\documentclass[11pt]{article}
\pdfoutput=1
\usepackage{dcolumn}
\usepackage{bm}

\usepackage{graphicx}
\usepackage{amssymb,amsmath}
\usepackage{multirow}
\usepackage{cite,color,url}
\usepackage[colorlinks=true
,urlcolor=blue
,anchorcolor=blue
,citecolor=blue
,filecolor=blue
,linkcolor=blue
,menucolor=blue
,linktocpage=true
,pdfproducer=medialab
,pdfa=true
]{hyperref}

\usepackage{slashed}
\usepackage{epsfig,psfrag,rotating,soul}
\usepackage{rotfloat}

\newcommand{\be}{\begin{equation}}
\newcommand{\ee}{\end{equation}}
\newcommand{\ba}{\begin{eqnarray}}
\newcommand{\ea}{\end{eqnarray}}
\def\stilde{\widetilde}
\def\conj{{{\rm c.c.}}}
\def\half{{1\over 2}}
\newcommand{\VL}{\left( \begin{array}{c}}
\newcommand{\VR}{\end{array} \right)}
\evensidemargin \oddsidemargin
\allowdisplaybreaks

\psfrag{lk}{$l_k$}
\psfrag{lm}{$l_m$}
\psfrag{blk}{$\bar {l}_k$}
\psfrag{blm}{$\bar {l}_m$}

\def\thefootnote{\fnsymbol{footnote}}

\let\OLDthebibliography\thebibliography
\renewcommand\thebibliography[1]{
  \OLDthebibliography{#1}
  \setlength{\parskip}{0pt}
  \setlength{\itemsep}{0pt plus 0.3ex}
}

\begin{document}
\thispagestyle{empty}

\begin{flushright}
IFT-UAM/CSIC-15-108\\
FTUAM-15-32\\
\end{flushright}

\vspace{0.5cm}

\begin{center}

\begin{Large}
\textbf{\textsc{Analysis of the $h, H, A \to \tau \mu$ decays induced from SUSY loops within the Mass Insertion Approximation}}
\end{Large}

\vspace{1cm}

{\sc
E. Arganda$^{1,3}$%
\footnote{\tt \href{mailto:ernesto.arganda@unizar.es}{ernesto.arganda@unizar.es}}%
, M.J. Herrero$^{2}$%
\footnote{\tt \href{mailto:maria.herrero@uam.es}{maria.herrero@uam.es}}%
, R. Morales$^{3}$%
\footnote{\tt \href{mailto:roberto.morales@fisica.unlp.edu.ar}{roberto.morales@fisica.unlp.edu.ar}}%
, A. Szynkman$^{3}$%
\footnote{\tt \href{mailto:szynkman@fisica.unlp.edu.ar}{szynkman@fisica.unlp.edu.ar}}%
}

\vspace*{.7cm}

{\sl
$^1$Departamento de F\'{\i}sica Te\'orica, Facultad de Ciencias,\\
Universidad de Zaragoza, E-50009 Zaragoza, Spain

\vspace*{0.1cm}

$^2$Departamento de F\'{\i}sica Te\'orica and Instituto de F\'{\i}sica Te\'orica, IFT-UAM/CSIC,\\
Universidad Aut\'onoma de Madrid, Cantoblanco, 28049 Madrid, Spain

\vspace*{0.1cm}

$^3$IFLP, CONICET - Dpto. de F\'{\i}sica, Universidad Nacional de La Plata, \\ 
C.C. 67, 1900 La Plata, Argentina

}

\end{center}

\vspace*{0.1cm}

\begin{abstract}
\noindent
In this paper we study the lepton favor violating decay channels of the neutral Higgs bosons of the Minimal Supersymmetric Standard Model into a lepton and an anti-lepton of different flavor. We work in the context of the most general flavor mixing scenario in the slepton sector, in contrast to the minimal flavor violation assumption more frequently used. Our analytic computation is a one-loop diagrammatic one, but in contrast to the full one-loop computation which is usually referred to the physical slepton mass basis, we use here instead the Mass Insertion Approximation (MIA) which uses the electroweak interaction slepton basis and treats perturbatively the mass insertions changing slepton flavor. 
By performing an expansion in powers of the external momenta in the relevant form factors, we will be able to separate explicitly in the analytic results the leading non-decoupling (constant at asymptotically large sparticle masses) and the next to leading decoupling contributions (decreasing with the sparticle masses). Our final aim is to provide a set of simple analytic formulas for the form factors and the associated effective vertices, that we think may be very useful for future phenomenological studies of the lepton flavor violating Higgs boson decays, and for their comparison with data. The accuracy of the numerical results obtained with the MIA are also analyzed and discussed here in comparison with the full one-loop results. Our most optimistic numerical estimates for the three neutral Higgs boson decays channels into $\tau$ and $\mu$ leptons, searching  for their maximum rates that are allowed by present constraints from $\tau \to \mu \gamma$ data and beyond Standard Model Higgs boson searches at the LHC, are also included.
\end{abstract}

\def\thefootnote{\arabic{footnote}}
\setcounter{page}{0}
\setcounter{footnote}{0}

\newpage

\section{Introduction}
\label{intro}

After the discovery of a new scalar particle  at the LHC~\cite{Aad:2012tfa,Chatrchyan:2012ufa}, identified  with the so long expected Higgs boson, and once its mass, now being set at 
$m_h =$ 125.09 $\pm$ 0.21 (stat.) $\pm$ 0.11 (syst.) GeV~\cite{Aad:2015zhl}, and other properties like some of its couplings to the Standard Model (SM) particles  have been measured (see~\cite{HiggsCouplingsCombined} for a recent review), one of the most challenging open questions still to be solved is to disentangle if this is an elemental or a composite particle and if there is new physics beyond the SM that could be hidden in the Higgs sector. In this regard, it is clear that the future ambitious experimental program, both at the CERN Large Hadron Collider (LHC) and future linear colliders, which will determine all the Higgs couplings with higher precision than at present, will play a central role. 
Among the most clear signals of Higgs physics beyond the SM, would be undoubtedly the discovery of new Higgs scalar bosons, and the discovery of new Higgs decay channels, both subjects being intensively explored at present at the LHC. We will focus here on these two possibilities by considering, on the one hand, the existence of new Higgs bosons, concretely those predicted in the Minimal Supersymmetric Standard Model (MSSM) and, on the other hand,  their potential new decay channels into leptons with different flavor, therefore violating lepton flavor number, which would be certainly very exotic Higgs decay channels, totally inhibited for the SM Higgs boson case.  

The subject of Lepton Flavor Violating Higgs Decays (LFVHD)  is actually a very active research field, being explored at present at the LHC.  
The first direct search of the particular decay $h \to \mu \tau$ (from now on we will refer in this short way to both $h \to \mu \bar \tau$ and $h \to \tau \bar \mu$ decays), has been performed by the
CMS Collaboration~\cite{Khachatryan:2015kon}, and an upper limit of BR($h \to \mu \tau$) $< 1.51 \times 10^{-2}$ at 95\% C.L. has been
set. Besides, CMS has also observed a slight excess with a significance of 2.4 standard deviations at $m_h =$ 125 GeV, whose best fit branching
ratio, if interpreted as a signal, is BR($h \to \mu \tau$) $= (8.4_{-3.7}^{+3.9}) \times 10^{-3}$. The ATLAS collaboration has recently released their results for the
same $h \to \mu \tau$ decay~\cite{Aad:2015gha} as well, focusing on hadronically decaying $\tau$ leptons. ATLAS has reported an upper limit of BR($h \to \mu \tau$) $< 1.85 \times 10^{-2}$ at 95\% C.L.
in agreement with the previous CMS result. It is worth mentioning that a small excess appears in one of the signal regions considered, even though it is not statistically significant.

On the theoretical side, the subject of  LFVHD has been studied for a long time in the literature within various models beyond the SM (for recent works see, for instance,~\cite{Blankenburg:2012ex,Harnik:2012pb,Davidson:2012ds,Arroyo:2013kaa,Celis:2013xja,Diaz-Cruz:2014pla,Arganda:2014dta,Bressler:2014jta,Dery:2014kxa,Sierra:2014nqa,Heeck:2014qea,Crivellin:2015mga,deLima:2015pqa,Dorsner:2015mja,Das:2015kea,Crivellin:2015lwa,Das:2015zwa,Yue:2015dia,Bhattacherjee:2015sia,He:2015rqa,Crivellin:2015hha,Cheung:2015yga,Botella:2015hoa,Baek:2015mea,Baek:2015fma}), but the most frequently explored ones are the supersymmetric (SUSY) models because the needed feature of flavor mixing among particles of different generations to produce these exotic decays is easily incorporated and well justified in the SUSY particles sector~\cite{DiazCruz:1999xe,DiazCruz:2002er,Brignole:2003iv,Brignole:2004ah,Kanemura:2004cn,Arganda:2004bz,Parry:2005fp,DiazCruz:2008ry,Crivellin:2010er,Crivellin:2011jt,Giang:2012vs,Arhrib:2012mg,Arhrib:2012ax,Arana-Catania:2013xma,Abada:2014kba,Arganda:2015naa}. More specifically, it is the flavor mixing among the three generations of the charged sleptons  
and/or sneutrinos, typically present in SUSY models, what  produces {\it via} their contributions at the one-loop level, these interesting Higgs decay channels with Lepton Flavor Violation (LFV).  In this work we will focus, in particular, on the LFVHD within the context of the MSSM and with the hypothesis of general flavor mixing in the  charged slepton and sneutrino sectors. This in contrast to the alternative and more restrictive Minimal Lepton Flavor Violation (MFV) hypothesis where the assumed unique origin of LFV comes from the Yukawa fermion couplings. Several examples where the neutrino Yukawa couplings, which can be large if neutrinos are Majorana fermions, are the responsible for generating these LFVHD have been explored in the literature. The issue of LFVHD being radiatively generated from loops with neutrinos was first explored in a non-SUSY context~\cite{Pilaftsis:1992st}, and later other cases were considered, including the case of the type-I seesaw model both with and without SUSY~\cite{Arganda:2004bz}, the inverse seesaw model~\cite{Arganda:2014dta} and its SUSY version~\cite{Arganda:2015naa}. 
The study of LFVHD within the more general context of Non-Minimal-Flavor Violation (NMFV) of the MSSM has also a long story.  The LFVHD rates of the neutral MSSM Higgs bosons into $\mu$ and $\tau$ leptons were computed  in the effective Lagrangian framework in~\cite{Brignole:2003iv} and a full-one loop diagrammatic computation in the physical SUSY particle basis was done in~\cite{Arganda:2004bz}. The issue of non-decoupling of the heavy SUSY particles in the LFVHD within this same MSSM context with NMFV was addressed numerically in~\cite{Arana-Catania:2013xma}.

In the present work we re-explore the LFV leptonic decays of the three neutral Higgs bosons, $h,H,A \to l_k \bar l_m$ ($k \neq m$), within the context of the MSSM with NMFV, and calculate their partial widths at the one-loop level with general slepton flavor mixings.  These mixings are parametrized by means of a complete set of slepton flavor mixing 
dimensionless parameters, $\delta^{AB}_{mk}$ with $AB=LL$, $RR$, $LR$, $RL$, and flavor indices $m,k=1,2,3$, with $m \neq k$. These parameters take into account, in a model-independent way and without any assumption on their particular origin, all the possible flavor mixings among the SUSY partners of the leptons with either left-handed or right-handed chirality. The novelty of this new computation is that we use a different technique, the so-called Mass Insertion Approximation (MIA)~\cite{Hall:1985dx,Gabbiani:1996hi,Misiak:1997ei}, that works with sleptons in the electroweak basis instead of the physical basis of the full one-loop computation, and  treats the off-diagonal in flavor entries of the slepton squared mass matrices $\Delta^{AB}_{mk}$ perturbatively, i.e., by means of mass insertions inside the propagators of the electroweak interaction sleptons eigenstates, instead of performing the exact diagonalization of the mass basis involved in the full one-loop computation.  Recent studies have additionally shown that the MIA results can alternatively be also obtained if one expands properly the starting expressions in the mass basis~\cite{Dedes:2015twa,Rosiek:2015jua}. The main advantage of using the MIA for the one-loop computation of the $\Gamma (H_x \to l_k \bar l_m)$ partial widths ($H_x=h,H,A$) is clear: it provides very simple analytic formulas for the form factors involved which after a proper expansion, to be valid in the case of heavy sparticle masses of our interest here, 
say $m_{\rm SUSY} \gtrsim {\cal O}(1\,{\rm TeV})$, can be recast in simple LFV effective vertices $V^{\rm eff}_{H_x l_m l_k}$, and these in turn are very useful for a simplified phenomenological study of the LFVHD rates in terms of the generic $\delta^{AB}_{mk}$'s and their comparison with data. In this work, by applying the MIA at the first (linear) order in the off-diagonal mass insertions $\Delta^{AB}_{mk}$ ($m\neq k$), we will compute  analytically all the relevant diagrams that contribute at the one-loop level to the LFV partial widths $\Gamma (h,H,A \to l_k \bar l_m)$. Furthermore, the MIA will also allow us to perform an analytic expansion of the involved form factors in powers of the external momenta and, in consequence, we will be able to capture analytically for the first time both the leading non-decoupling contributions of ${\cal O}((m_{h,H,A}/m_{\rm SUSY})^0)$, i.e., those that go to a constant value in the asymptotically large SUSY masses limit, and the next-to-leading decoupling contributions of ${\cal O}(m_{h,H,A}^2/m_{\rm SUSY}^2)$, which are numerically much smaller than the leading ones but they turn out to play an important role for some of the studied cases of the flavor mixings.  A few comments and estimates will also be done for the next-to-leading decoupling contributions of ${\cal O}(M_W^2/m_{\rm SUSY}^2)$, which are numerically very tiny. In this work we will also include a numerical computation of the LFVHD rates with the MIA for the case of most interest at present, $h,H,A \to \tau \bar \mu$, which will be systematically  compared with the full one-loop results to be able to conclude on the goodness of this approximation, the MIA, and its range of applicability. Finally we will conclude with simple analytic formulas for the useful LFV effective vertices, $V^{\rm eff}_{H_x \tau \mu}$, and with a numerical estimate of the maximum expected BR$(h,H,A \to \tau \bar \mu)$ rates that are allowed by the present experimental constraints from $\tau \to \mu \gamma$~\cite{Aubert:2009ag} and by the ATLAS and CMS searches for neutral Higgs bosons beyond the SM~\cite{Khachatryan:2014wca,Aad:2014vgg}. This numerical study will be performed in terms of the most relevant model parameters, emphasizing which flavor mixing parameters will be most efficiently tested at future colliders.
     
The paper is organized as follows. In section \ref{th-framework} we summarize the relevant aspects of the MSSM with general sfermion flavor mixing and present the chosen scenarios for our numerical estimates. Section~\ref{MIA-expressions} contains our analytic computation of the LFVHD widths within the MIA. We select and compute the relevant one-loop diagrams and derive the  form factors for LFVHD, their proper expansions, and the corresponding effective vertices. Section~\ref{results} contains all our numerical results for BR$(h,H,A \to \tau \bar \mu)$ and the comparison with the full one-loop predictions. The conclusions are summarized in section~\ref{conclusions}. The technicalities, including the relevant Feynman rules for the interaction vertices in the MSSM with NMFV, the analytic expressions of the form factors for each diagram, and the proper expansions of the loop integrals are collected in the appendices~\ref{FeynRules}, \ref{AnalyticFormFactors}, and~\ref{LoopIntegrals}, respectively.    

\section{The MSSM with general flavor mixing in the charged slepton and sneutrino sectors}
\label{th-framework}

In order to describe the MSSM with general sfermion mixing, the relevant model pieces are the superpotential and the soft SUSY-breaking Lagrangian.
The superpotential of the MSSM in terms of the relevant superfields is given by: \be
W_{\rm MSSM} =
\hat U Y^u \hat Q \hat H_2 -
\hat D Y^d \hat Q \hat H_1 -
\hat E Y^e \hat L \hat H_1 +
\mu \hat H_1 \hat H_2 \> \,,
\label{MSSMsuperpot}
\ee
where the Yukawa couplings $Y^{u,d,e}$ are $3 \times 3$ matrices in flavor space. All indices, including the flavor ones, have been omitted in eq.~(\ref{MSSMsuperpot}) for simplicity. 

The relevant soft SUSY-breaking MSSM Lagrangian for generic sfermion mixing is: 
\ba
{\cal L}_{\rm soft}^{\rm MSSM} &=& -\half\left ( M_3 \stilde g\stilde g
+ M_2 \stilde W \stilde W + M_1 \stilde B\stilde B 
+\conj \right )
\nonumber
\\
&&
-\left ( \tilde {\cal Q}_i {{\cal  A}}^{u}_{ij} \tilde {\cal U}^{*}_j {\cal H}_2
- \tilde {\cal Q}_i { {\cal A}}^{d}_{ij} \tilde {\cal D}^{*}_j {\cal H}_1  
- \tilde {\cal L}_i { {\cal A}}^{e}_{ij} \tilde {\cal E}^{*}_j {\cal H}_1  
+ \conj \right ) 
\nonumber
\\
&&
- \tilde {\cal Q}^{\dagger}_i m_{\tilde Q_{ij}}^2 \tilde {\cal Q}_j 
- \tilde {\cal U}^{*}_i m_{\tilde U_{ij}}^2 \, \tilde {\cal U}_j
- \tilde {\cal D}^{*}_i m_{\tilde D_{ij}}^2 \, \tilde {\cal D}_j
- \tilde {\cal L}^{\dagger}_i m_{\tilde L_{ij}}^2 \tilde {\cal L}_j 
- \tilde {\cal E}^{*}_i m_{\tilde R_{ij}}^2 \, \tilde {\cal E}_j
\nonumber \\
&&
- \, m_{H_1}^2 {\cal H}_1^* {\cal H}_1 - m_{H_2}^2 {\cal H}_2^* {\cal H}_2
- \left ( b {\cal H}_2 {\cal H}_1 + \conj \right ) \,,
\label{MSSMsoft}
\ea
where we use calligraphic capital letters for the sfermion fields in the interaction basis with generation indices, varying from 1 to 3,
\ba
\tilde {\cal U}_{1,2,3}=\tilde u_R,\tilde c_R,\tilde t_R \,;\quad 
\tilde {\cal D}_{1,2,3}=\tilde d_R,\tilde s_R,\tilde b_R &;& \quad 
\tilde {\cal Q}_{1,2,3}=(\tilde u_L \, \tilde d_L)^T, (\tilde c_L\, \tilde s_L)^T, (\tilde t_L \, \tilde b_L)^T \,, \\
\tilde {\cal E}_{1,2,3}=\tilde e_R,\tilde \mu_R,\tilde \tau_R &;& \quad  
\tilde {\cal L}_{1,2,3}=(\tilde \nu_{eL} \, \tilde e_L)^T, (\tilde \nu_{\mu L}\, \tilde \mu_L)^T, (\tilde \nu_{\tau L} \, \tilde \tau_L)^T \,,
\ea
and all the gauge indices have been omitted. All the trilinear couplings, 
${{\cal  A}}^f_{ij}$, and the soft squared masses of sfermions, $m_{ij}^2$, are $3\times 3$ matrices in the space of flavor. 

The two Higgs doublets of the MSSM are given by:
\ba
{\cal H}_1 &=& \VL  {\cal H}_1^0 \\[0.5ex] {\cal H}_1^- \VR \; = \; \VL v_1 
        + \frac{1}{\sqrt{2}} (\phi_1^0 - i\chi_1^0) \\[0.5ex] -\phi_1^- \VR~,  
        \nonumber \\
{\cal H}_2 &=& \VL {\cal H}_2^+ \\[0.5ex] {\cal H}_2^0 \VR \; = \; \VL \phi_2^+ \\[0.5ex] 
        v_2 +  \frac{1}{\sqrt{2}} (\phi_2^0 + i\chi_2^0) \VR~ ,
\label{higgsfeldunrot}
\ea
where $v_1$ and $v_2$ are the vacuum expectation values (VEV) of the neutral Higgs fields, $v_1=\langle{\cal H}_1^0\rangle$ and $v_2=\langle{\cal H}_2^0\rangle$, and the ratio between the two VEVs is defined as $\tan \beta = v_2/v_1$. In the present work, 
we focus on the three physical neutral Higgs bosons, which are built from the previous Higgs doublet components as:
\ba
H&=& \cos \alpha \, \phi_1^0 +  \sin \alpha  \, \phi_2^0 \,, \nonumber \\
h&=& -\sin \alpha \, \phi_1^0 +  \cos \alpha \, \phi_2^0 \,, \nonumber \\
A&=& -\sin \beta \, \chi_1^0  +  \cos  \beta \,   \chi_2^0 \,, 
\ea 
and use $m_A$ and $\tan \beta$ as input model parameters of the MSSM Higgs sector.  

Since we are interested here in the Lepton Flavor Violating Higgs decays of these three neutral MSSM Higgs bosons, $H_x \to l_k \bar l_m$ with $H_x=h,H,A$, we will focus on sfermion mixing in the slepton sector and we will ignore the possible sfermion mixing in the squark sector. Furthermore,   
we will work within a general flavor mixing context at the low energies, i.e., without assuming any high-energy hypothesis for the generation of the relevant soft-breaking terms producing this slepton flavor mixing. Therefore, we will work within a NMFV framework which goes beyond the more frequently used MFV hypothesis in which the sfermion mixing is always induced by the Yukawa couplings.

 The most general hypothesis for flavor
mixing among sleptons assumes a mass matrix in the interaction basis that is not diagonal
in the space of flavor, both for charged sleptons and sneutrinos. 
In the charged slepton sector the mass matrix is $6 \times 6$,
since there are six electroweak interaction eigenstates, 
${\tilde l}_{L,R}$ with $l=e, \mu, \tau$. For the sneutrinos 
 the mass matrix is $3 \times 3$, since within the MSSM there are
only three electroweak interaction eigenstates, ${\tilde \nu}_{L}$ with
$\nu=\nu_e, \nu_\mu, \nu_\tau$.

The non-diagonal $6 \times 6$ squared
mass matrix of sleptons when expressed in the electroweak interaction basis,  
 that we order here as   $(\tilde e_L, \tilde \mu_L, \tilde \tau_L, \tilde e_R, \tilde \mu_R ,  \tilde \tau_R)$, is written
 in terms of left- and right-handed
 blocks $M^2_{\tilde l \, AB}$  
 ($A,B=L,R$), which are non-diagonal $3\times 3$ matrices, as follows:
\begin{equation}
{\mathcal M}_{\tilde l}^2 =\left( \begin{array}{cc}
M^2_{\tilde l \, LL} & M^2_{\tilde l \, LR} \\ 
M_{\tilde l \, LR}^{2 \, \dagger} & M^2_{\tilde l \,RR}
\end{array} \right) \,,
\label{eq:slep-6x6}
\end{equation} 
 where:
 \ba
M_{\tilde l \, LL \, ij}^2 
  &= &  m_{\tilde L \, ij}^2 + \left( m_{l_i}^2
     + (-\frac{1}{2}+ \sin^2 \theta_W ) M_Z^2 \cos 2\beta \right) \delta_{ij}  \,, \notag\\
 M^2_{\tilde l \, RR \, ij}
 & = &  m_{\tilde R \, ij}^2 + \left( m_{l_i}^2
     -\sin^2 \theta_W M_Z^2 \cos 2\beta \right) \delta_{ij} \,, \notag \\
  M^2_{\tilde l \, LR \, ij}
 & = &  v_1 {\cal A}_{ij}^l- m_{l_{i}} \mu \tan \beta \, \delta_{ij} \,,
\label{eq:slep-matrix}
\ea
with flavor indices $i,j=1,2,3$ running by the three generations, respectively; and $(m_{l_1},m_{l_2},
m_{l_3})=(m_e,m_\mu,m_\tau)$ are the lepton masses.
It is worth recalling that the non diagonality in flavor comes
exclusively from the soft SUSY-breaking parameters, that could be
non vanishing for $i \neq j$. Specifically: the masses $m_{\tilde L \, ij}$
for the slepton $SU(2)$ doublets, $(\tilde \nu_{Li}\,\,\, \tilde
l_{Li})$, the masses $m_{\tilde R \, ij}$ for the slepton $SU(2)$
singlets, $(\tilde l_{Ri})$, and the trilinear couplings, ${\cal
  A}_{ij}^l$.   

In the sneutrino sector there is a $3\times
3$ squared mass matrix that, when expressed in the $(\tilde \nu_{eL}, \tilde \nu_{\mu L},
\tilde \nu_{\tau L})$ electroweak interaction basis, is given by: 
\begin{equation}
{\mathcal M}_{\tilde \nu}^2 =\left( \begin{array}{c}
M^2_{\tilde \nu \, LL}   
\end{array} \right) \,,
\label{eq:sneu-3x3}
\end{equation} 
 where
\begin{equation} 
  M_{\tilde \nu \, LL \, ij}^2 
  =   m_{\tilde L \, ij}^2 + \left( 
      \frac{1}{2} M_Z^2 \cos 2\beta \right) \delta_{ij} \,.  
\label{eq:sneu-matrix}
\end{equation} 
 
As a consequence of the $SU(2)_L$ gauge invariance,
the same soft masses $m_{\tilde L \, ij}$ enter in both the slepton and
sneutrino $LL$ mass matrices. It should be noted that
if the neutrino masses and neutrino flavor mixings (oscillations) were taken into account, the soft SUSY-breaking parameters in the
sneutrino sector would differ from the corresponding ones for the charged
slepton sector by a rotation with the PMNS matrix. This would be somehow equivalent to what happens in the squark sector where the soft masses for the squarks of down type and those of up type differ by a rotation given by the CKM matrix. However, due to the smallness of the neutrino masses, 
we do not expect
large effects from the inclusion of neutrino masses in the present computation and consequently we will neglect them in this work, as it is usually done in the context of the MSSM. 

The general flavor mixing in the slepton sector is introduced via the
non-diagonal terms in the soft breaking slepton mass matrices and
trilinear coupling matrices, and these are defined here in a model-independent way in terms of a set of 
dimensionless parameters $\delta^{AB}_{ij}$ ($A,B=L,R$; $i,j=1,2,3$, 
$i \neq j$), where $L$ and $R$ denote the 
``left-'' and ``right-handed'' SUSY partners of the corresponding
leptonic degrees of freedom, and $i,j$
indices run over the three generations. We
assume here that the $\delta^{AB}_{ij}$'s  provide the unique origin
of LFV processes with potentially measurable rates. Specifically, we define: 
  
\noindent \begin{equation}  
m^2_{\tilde L}= \left(\begin{array}{ccc}
 m^2_{\tilde L_{1}} & \delta_{12}^{LL} m_{\tilde L_{1}}m_{\tilde L_{2}} & \delta_{13}^{LL} m_{\tilde L_{1}}m_{\tilde L_{3}} \\
 \delta_{21}^{LL} m_{\tilde L_{2}}m_{\tilde L_{1}} & m^2_{\tilde L_{2}}  & \delta_{23}^{LL} m_{\tilde L_{2}}m_{\tilde L_{3}}\\
\delta_{31}^{LL} m_{\tilde L_{3}}m_{\tilde L_{1}} & \delta_{32}^{LL} m_{\tilde L_{3}}m_{\tilde L_{2}}& m^2_{\tilde L_{3}} \end{array}\right) \,,\label{mLL}\end{equation}

\noindent \begin{equation}
v_1 {\cal A}^l  =\left(\begin{array}{ccc}
m_e A_e & \delta_{12}^{LR} m_{\tilde L_{1}}m_{\tilde R_{2}} & \delta_{13}^{LR} m_{\tilde L_{1}}m_{\tilde R_{3}}\\
\delta_{21}^{LR}  m_{\tilde L_{2}}m_{\tilde R_{1}} & m_\mu A_\mu & \delta_{23}^{LR} m_{\tilde L_{2}}m_{\tilde R_{3}}\\
\delta_{31}^{LR}  m_{\tilde L_{3}}m_{\tilde R_{1}} & \delta_{32}^{LR}  m_{\tilde L_{3}} m_{\tilde R_{2}}& m_{\tau}A_{\tau}\end{array}\right) \,,\label{v1Al}\end{equation}

\noindent \begin{equation}  
m^2_{\tilde R}= \left(\begin{array}{ccc}
 m^2_{\tilde R_{1}} & \delta_{12}^{RR} m_{\tilde R_{1}}m_{\tilde R_{2}} & \delta_{13}^{RR} m_{\tilde R_{1}}m_{\tilde R_{3}}\\
 \delta_{21}^{RR} m_{\tilde R_{2}}m_{\tilde R_{1}} & m^2_{\tilde R_{2}}  & \delta_{23}^{RR} m_{\tilde R_{2}}m_{\tilde R_{3}}\\
\delta_{31}^{RR}  m_{\tilde R_{3}} m_{\tilde R_{1}}& \delta_{32}^{RR} m_{\tilde R_{3}}m_{\tilde R_{2}}& m^2_{\tilde R_{3}} \end{array}\right) \,.\label{mRR}\end{equation}

Some comments are in order regarding our parametrization above. First, for simplicity, in all this work we are assuming that all $\delta_{ij}^{AB}$
parameters are real, hence, hermiticity of the squared mass matrices implies $\delta_{ij}^{AB}= \delta_{ji}^{BA}$. Second, the diagonal entries in eq.~(\ref{v1Al}) have been normalized as usually done in the literature, namely, by factorizing out the corresponding lepton Yukawa coupling: ${\cal A}^l_{ii}= y_{l_i} A_{l_i}$, with $A_{l_1}=A_e$, $A_{l_2}=A_\mu$,  $A_{l_3}=A_\tau$, and $y_{l_i}=m_{l_i}/v_1$. Third, it should be noted that the choice in eqs.~(\ref{mLL}), (\ref{v1Al}), and (\ref{mRR}) is to normalize the non-diagonal in flavor entries with respect to the geometric mean of the corresponding diagonal squared soft masses. Thus, the non-diagonal $LL$ and $RR$ terms, with $m \neq k$,   are normalized as:  
\be
\Delta^{LL}_{mk} \equiv (m^2_{\tilde L})_{mk}= \delta^{LL}_{mk}m_{\tilde L_{m}}m_{\tilde L_{k}} \,,
\label{DeltaLL}
\ee
and
\be
\Delta^{RR}_{mk} \equiv (m^2_{\tilde R})_{mk}= \delta^{RR}_{mk}m_{\tilde R_{m}}m_{\tilde R_{k}} \,.
\label{DeltaRR}
\ee
However, in the case of sfermion mixing of $LR$ (and $RL$) type, and taking into account that the origin of these off-diagonal mass entries is intrinsically connected to the value of the soft trilinear couplings, having dimension of mass, we find more appropriate for the purpose of this work, dealing with very large SUSY masses, to normalize them alternatively as follows:
\be
\Delta^{LR}_{mk} \equiv (v_1 {\cal A}^l)_{mk}= \tilde \delta^{LR}_{mk} v_1 \sqrt{m_{\tilde L_{m}}m_{\tilde R_{k}}} \,,
\label{DeltaLR}
\ee
and similarly,
\be
\Delta^{RL}_{mk} \equiv (v_1 {\cal A}^l)_{km}= \tilde \delta^{RL}_{mk} v_1 \sqrt{m_{\tilde R_{m}}m_{\tilde L_{k}}} \,.
\label{DeltaRL}
\ee
This implies an obvious relation between $\delta^{LR}_{mk}$ and $ \tilde \delta^{LR}_{mk}$ which should be kept in mind:
\be
\delta^{LR}_{mk}=\tilde \delta^{LR}_{mk} \frac{v_1}{\sqrt{m_{\tilde L_{m}}m_{\tilde R_{k}}}} \,,
\ee   
and similarly for the $RL$ case.

Besides, if one wishes to relate the previous electroweak interaction basis and the physical mass basis one must perform the corresponding rotations:
\begin{equation}
\VL  \tilde l_{1} \\ \tilde l_{2}  \\ \tilde l_{3} \\
                                    \tilde l_{4}   \\ \tilde l_{5}  \\\tilde l_{6}   \VR
  \; = \; R^{\tilde l}  \VL \tilde e_L \\ \tilde \mu_L \\\tilde \tau_L \\ 
  \tilde e_R\\ \tilde \mu_R  \\ \tilde \tau_R  \VR ~,~~~~
\VL  \tilde \nu_{1} \\ \tilde \nu_{2}  \\  \tilde \nu_{3}  \VR             \; = \; R^{\tilde \nu}  \VL \tilde \nu_{eL} \\ \tilde \nu_{\mu L}  \\  \tilde \nu_{\tau L}   \VR ~,
\label{newsleptons}
\end{equation} 
where $R^{\tilde l}$ and $R^{\tilde \nu}$ are the respective $6\times 6$ and
$3\times 3$ unitary rotating matrices that provide the diagonal
mass-squared matrices as follows, 
\begin{eqnarray}
{\rm diag}\{m_{\tilde l_1}^2, m_{\tilde l_2}^2, 
          m_{\tilde l_3}^2, m_{\tilde l_4}^2, m_{\tilde l_5}^2, m_{\tilde l_6}^2 
           \}  & = &
R^{\tilde l}   {\cal M}_{\tilde l}^2    
 R^{\tilde l \dagger}    ~, \label{sleptons}\\
{\rm diag}\{m_{\tilde \nu_1}^2, m_{\tilde \nu_2}^2, 
          m_{\tilde \nu_3}^2  
          \}  & = &
R^{\tilde \nu}     {\cal M}_{\tilde \nu}^2    
 R^{\tilde \nu \dagger}    ~.
\label{sneutrinos} 
\end{eqnarray}

Regarding the particle interactions that are involved in the present computation of the LFV Higgs decay widths,  $\Gamma(H_x \to l_k \bar l_m)$ with $k,m= 1,2,3$, $k \neq m$, and $H_x=h,H,A$, we have collected all the relevant Feynman rules in Appendix~\ref{FeynRules}, including all the needed insertions, vertices, and propagators, which we have expressed in the proper basis here. Concretely, we work with the mass basis for the external particles, $H_x$, $l_k$, and $\bar l_m$, and with the electroweak interaction basis for the internal sparticles in the loops, which from now on will be shortly denoted by: ${\tilde l}^{L,R}_i \,(i=1,2,3)$, ${\tilde \nu}_i \,(i=1,2,3)$, ${\tilde W}^{\pm}$, ${\tilde W}^3$, ${\tilde B}$, ${\tilde H}^{\pm}$, and ${\tilde H}_{1,2}$. This choice of basis is the most convenient one for the computation in the MIA, in contrast to the full one-loop computation where the physical mass basis is also usually set for the internal sleptons, sneutrinos, charginos, and neutralinos: ${\tilde l}_\alpha\, (\alpha=1,...,6)$, $ {\tilde \nu}_\alpha\, (\alpha=1,2,3)$, 
${\tilde \chi}^{\pm}_i\, (i=1,2)$, and ${\tilde \chi}^{0}_i\,(i=1,...,4)$.

Finally, to close this section of model specifications, we shortly summarize next the heavy SUSY scenarios that we work with for the estimates of this research.  In order to simplify our numerical analysis, and to reduce the number of independent parameters, we define here three simplified SUSY scenarios, where the relevant  parameters with mass dimensions are related to a single SUSY mass scale, $m_\text{SUSY}$:
\begin{itemize}
\item {\it Equal masses} scenario. In this scenario we choose the simplest case with all the relevant parameters involved set to be equal:
\begin{equation}
M_1 = M_2 = M_3 = \mu = m_{\tilde L} = m_{\tilde R} = A_\mu = A_\tau = m_\text{SUSY} \,.
\end{equation}
\item {\it GUT approximation} scenario. In this scenario we set an approximate GUT relation for the gaugino masses:
\begin{equation}
M_2 = 2 M_1 = M_3/4 \,.
\end{equation}
And, for simplicity, we also relate the soft parameters and the $\mu$ parameter to a common scale by choosing:
\begin{eqnarray}
m_{\tilde L} &=& m_{\tilde R} =M_2= A_\mu = A_\tau = m_\text{SUSY} \,, \\
\mu &=& a \,m_\text{SUSY} \,,
\end{eqnarray}
where $a$ is a constant coefficient that we will fix to two different values for comparison, namely, $a = \frac{3}{4}$ and $\frac{4}{3}$.
\item {\it Generic} scenario. In this scenario we wish to explore the non-equal mass generic case. Thus, we set different values for all the mass parameters involved. For the purpose of this work, the particular values of each parameter is not much relevant, but the important feature here is setting all of them to be heavy by a common $m_{\rm SUSY}$ scale. Concretely, we take:
\begin{eqnarray}
M_1 &=& 2.2 \, m_\text{SUSY} \,, M_2 = 2.4 \, m_\text{SUSY} \,, M_3 = 2.6 \, m_\text{SUSY} \,, \mu = 2.1 \, m_\text{SUSY} \,, \\
m_{\tilde L_{1}} &=& 2 \, m_\text{SUSY} \,, m_{\tilde L_{2}} = 1.8 \, m_\text{SUSY} \,, m_{\tilde L_{3}} = 1.6 \, m_\text{SUSY} \,, \\
m_{\tilde R_{1}} &=& 1.4 \, m_\text{SUSY} \,, m_{\tilde R_{2}} = 1.2 \, m_\text{SUSY} \,, m_{\tilde R_{3}} = \, m_\text{SUSY} \,, \\
A_\mu &=& 0.6 \, m_\text{SUSY} \,, A_\tau = 0.8 \, m_\text{SUSY} \,.
\end{eqnarray}
\end{itemize}
For the first two scenarios that are defined above, we use a short notation for the common soft masses, namely, $m_{\tilde L}$ for $m_{\tilde L}=m_{\tilde L_{1}}=m_{\tilde L_{2}}=m_{\tilde L_{3}}$, etc. For simplicity, in all the three scenarios we have also assumed a vanishing soft-trilinear coupling for the first generation in the charged slepton sector, i.e., $A_e=0$. Concerning the soft masses of the squark sector, they are indeed irrelevant for LFV processes. However, since we want to identify the discovered scalar boson with the lightest MSSM Higgs boson, we set these parameters to values which give a prediction of $m_h$ compatible with the LHC data in the mass range of 125 GeV $\pm$ 3 GeV, and fix them to the particular values $m_{\tilde Q} =$ $m_{\tilde U} =$ $m_{\tilde D} =$ $A_t =$ $A_b =$ 5 TeV in the three scenarios described just above. Besides, as already said, the other MSSM input parameters to be set in the numerical analysis are $m_A$ and $ \tan \beta$. Finally, regarding the $\delta^{AB}_{ij}$ parameters they will be taken in the conservative interval, $|\delta^{AB}_{ij}|<1$, since we wish to keep our computation in the perturbative regime. This computation will be reported in the next section.

\section{Analytic results of the LFVHD widths in the MIA}
\label{MIA-expressions}

Here we present our analytic computation of the partial widths for the LFVHD, 
$\Gamma(H_x \to l_k \bar l_m)$ with $k,m= 1,2,3$, $k \neq m$, and $H_x=h,H,A$. These can be written with full generality in terms of the two form factors $F_{L,R}^{(x)}$ involved in the decay amplitude of this $H_x(p_1) \to l_k(-p_2) \bar l_m (p_3)$ process,
\begin{equation}
i {\cal M} = -i g \bar{u}_{l_k} (-p_2) (F_L^{(x)} P_L + F_R^{(x)} P_R) v_{l_m}(p_3) \,,
\end{equation}
as follows:
\begin{eqnarray}
\Gamma (H_x \to {l_k} \bar{l}_m)& = &\frac{g^2}{16 \pi m_{H_x}} 
\sqrt{\left(1-\left(\frac{m_{l_k}+m_{l_m}}{m_{H_x}}\right)^2\right)
\left(1-\left(\frac{m_{l_k}-m_{l_m}}{m_{H_x}}\right)^2\right)} \nonumber\\
&& \times \left((m_{H_x}^2-m_{l_k}^2-m_{l_m}^2)(|F_L^{(x)}|^2+|F_R^{(x)}|^2)- 4 m_{l_k} m_{l_m} Re(F_L^{(x)} F_R^{(x)*})\right) \,,
\label{decay}
\end{eqnarray}
where $p_1$ is the ingoing Higgs boson momentum, $-p_2$ the outgoing momentum of the lepton $l_k$, and $p_3$ the outgoing momentum of the antilepton $\bar l_m$, with $p_1=p_3-p_2$.  
We focus here on the $H_x \to l_k \bar l_m$ channel, but due to the fact that we work with real parameters, the predictions for the $CP$-conjugate channel  
$H_x \to l_m \bar l_k$ will be equal.
 
The present computation of $\Gamma (H_x \to {l_k} \bar{l}_m)$ is performed by taking into account the following assumptions and considerations: 1) The amplitude is evaluated at the one-loop level, 2) only loops containing sleptons and sneutrinos contribute since they are the only particles propagating the LFV by means of the $\Delta^{AB}_{mk}$ entries with $m \neq k$, 3) the particle content assumed here is that of the MSSM, 4) the external particles $h,H,A$ and $l_k$, $\bar \l_m$ are expressed in the physical mass basis, 5)  the internal loop sparticles are expressed in the electroweak interaction basis, and 6) we use the Mass Insertion Approximation~\cite{Hall:1985dx,Gabbiani:1996hi,Misiak:1997ei} to describe the propagation of slepton mixing changing flavor, and work in the linear approximation for each insertion $\Delta^{AB}_{mk}$, with $AB=LL,RR,LR,RL$, and $m \neq k$, i.e, considering one single insertion at a time. 

\begin{figure}[t!]
\begin{center}
\begin{tabular}{llll}
\includegraphics[width=0.23\textwidth]{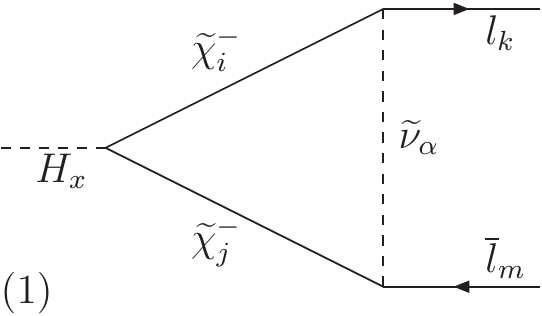} &
\includegraphics[width=0.23\textwidth]{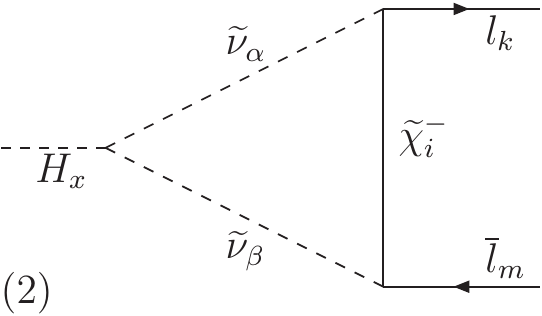} &
\includegraphics[width=0.23\textwidth]{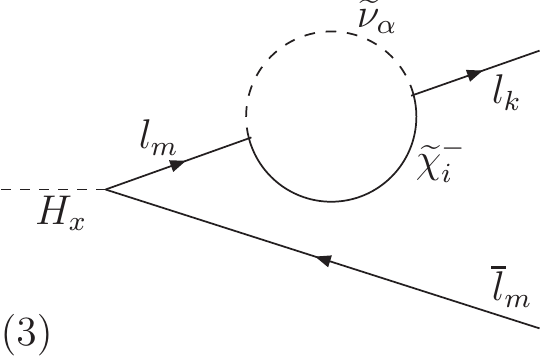} &
\includegraphics[width=0.23\textwidth]{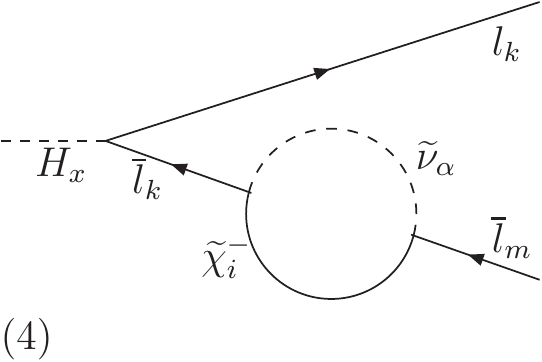} \\
\includegraphics[width=0.23\textwidth]{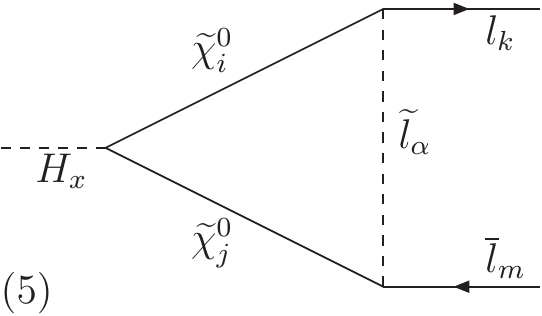} &
\includegraphics[width=0.23\textwidth]{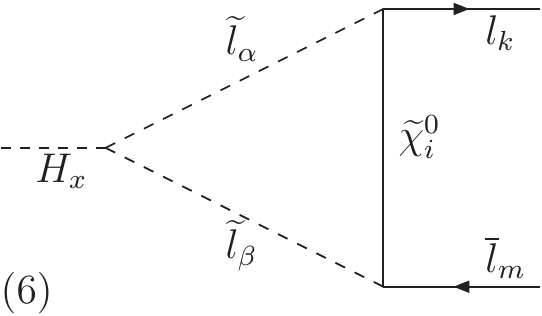} &
\includegraphics[width=0.23\textwidth]{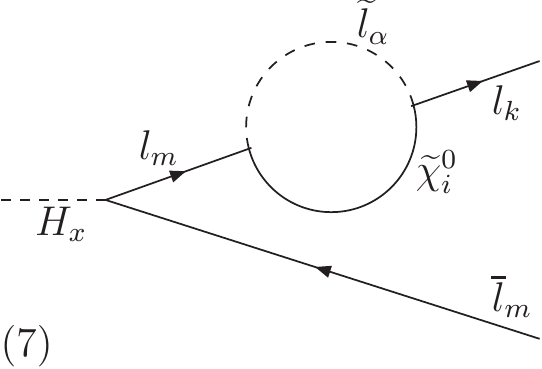} &
\includegraphics[width=0.23\textwidth]{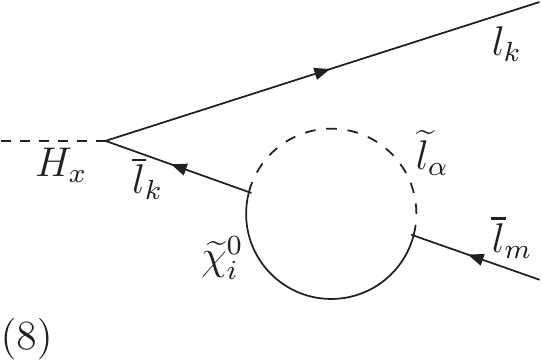}
\end{tabular}
\caption{Full one-loop diagrams for $H_x \to l_k \bar l_m$ decays in the MSSM mass basis.}
\label{fulldiag}
\end{center}
\end{figure}

In order to estimate the goodness of the MIA that we use here, we have systematically compared all our results with the full-one loop results which were firstly computed in~\cite{Arganda:2004bz}. In this case, all the particles involved in the $H_x \to l_m \bar l_k$ decay, both external and internal to the loops, are usually expressed in the mass basis. For completeness, we display in figure~\ref{fulldiag} the eight one-loop diagrams that contribute to the full one-loop result. For our posterior  numerical analysis and comparison with our computation in the MIA,  we have also implemented in our code the full one-loop formulas for each of these eight diagrams contributing to $F_{L,R}^{(x)}$, which we take from~\cite{Arganda:2004bz}. From now on, we will use the labels $(i)$, with $i=1,...,8$ associated to each of these diagrams according to figure~\ref{fulldiag}, in the comparison of the full versus MIA results.   

\begin{figure}[t!]
\begin{center}
\begin{tabular}{lll}
\includegraphics[width=0.3\textwidth]{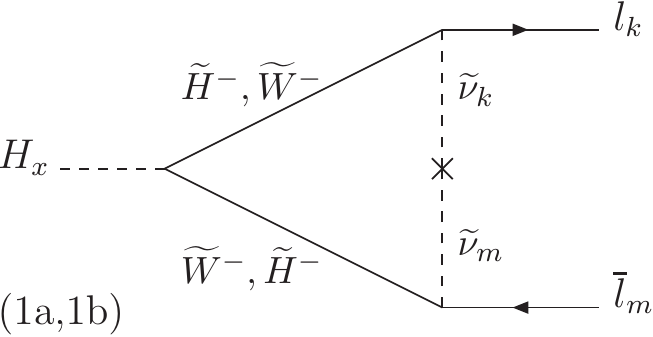} &
\includegraphics[width=0.3\textwidth]{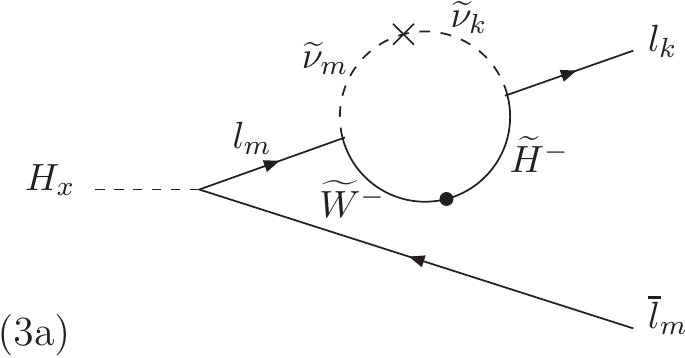} &
\includegraphics[width=0.3\textwidth]{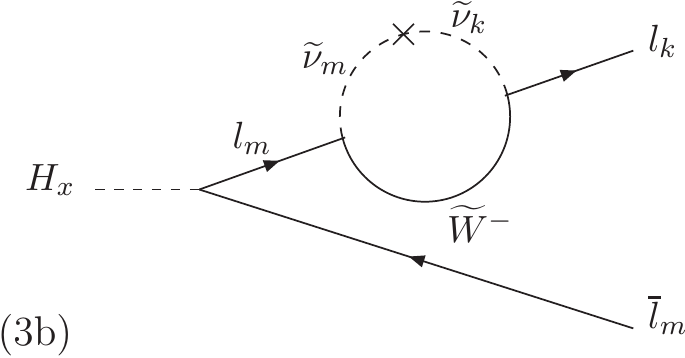} \\
\includegraphics[width=0.3\textwidth]{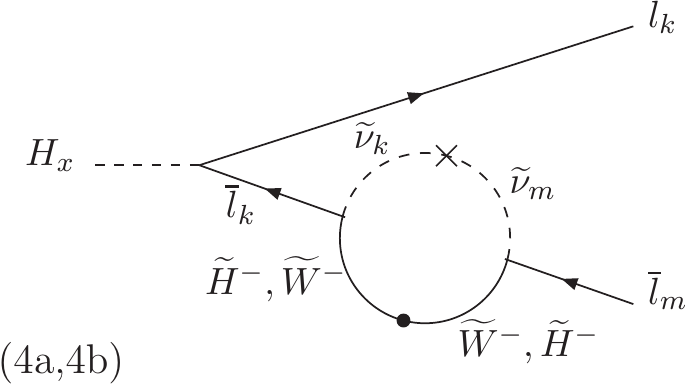} &
\includegraphics[width=0.3\textwidth]{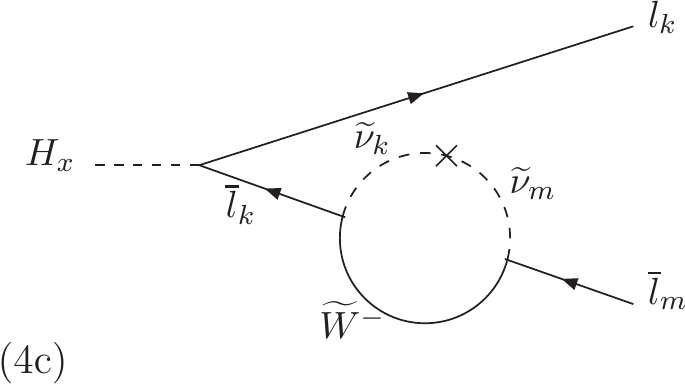} &
\includegraphics[width=0.3\textwidth]{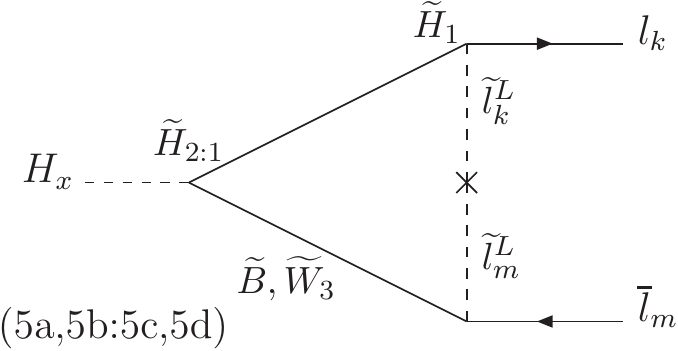} \\
\includegraphics[width=0.3\textwidth]{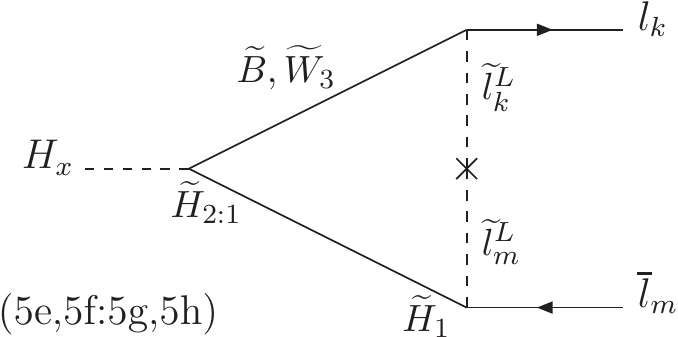} &
\includegraphics[width=0.3\textwidth]{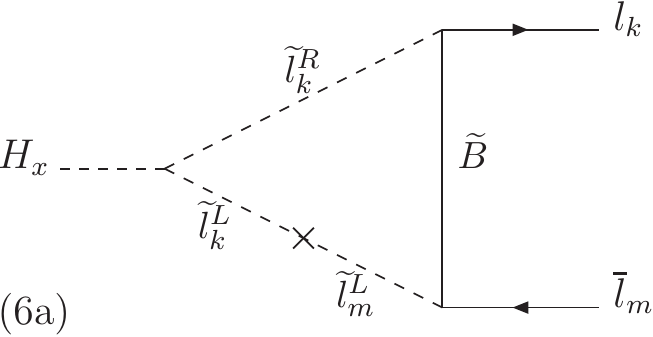} &
\includegraphics[width=0.3\textwidth]{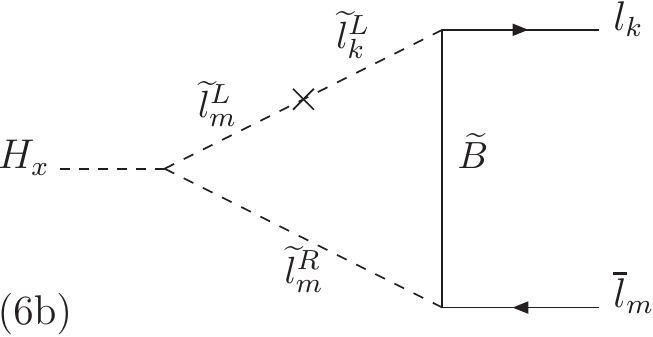} \\
\includegraphics[width=0.3\textwidth]{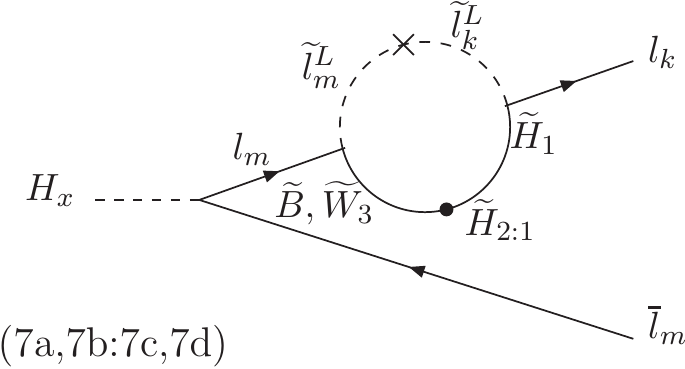} &
\includegraphics[width=0.3\textwidth]{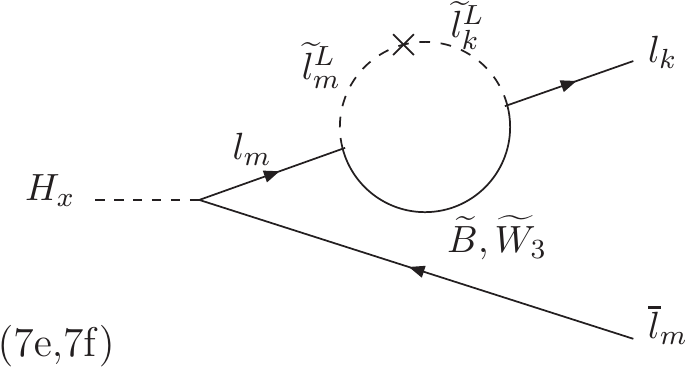} &
\includegraphics[width=0.3\textwidth]{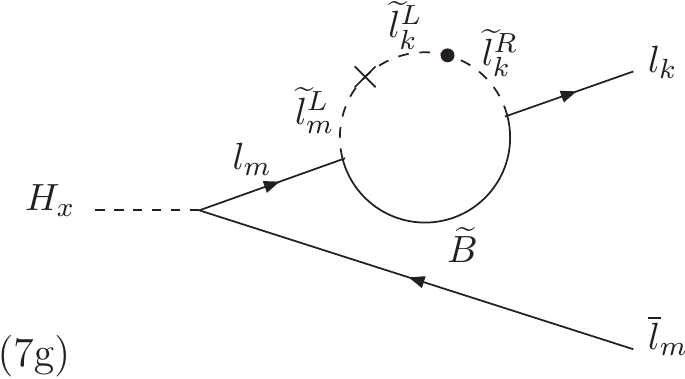} \\
\includegraphics[width=0.3\textwidth]{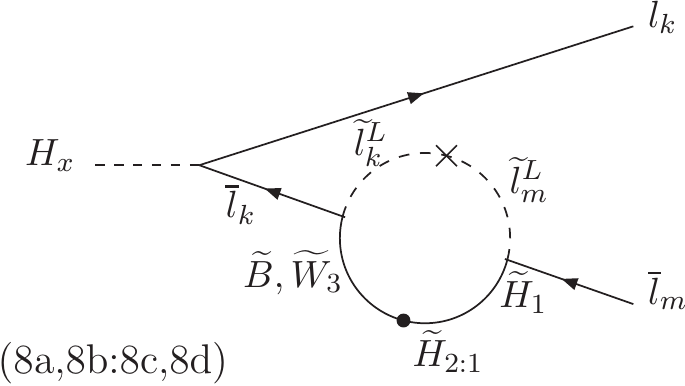} &
\includegraphics[width=0.3\textwidth]{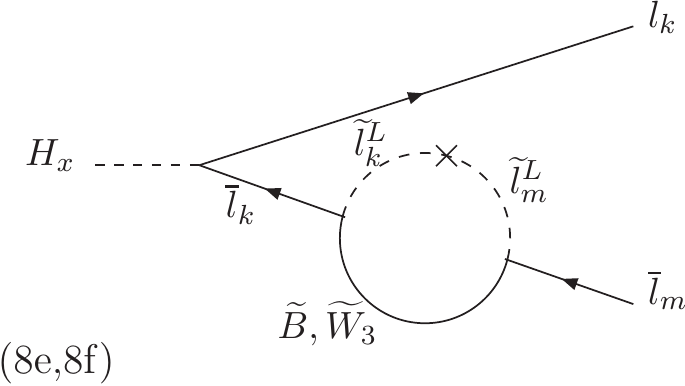} &
\includegraphics[width=0.3\textwidth]{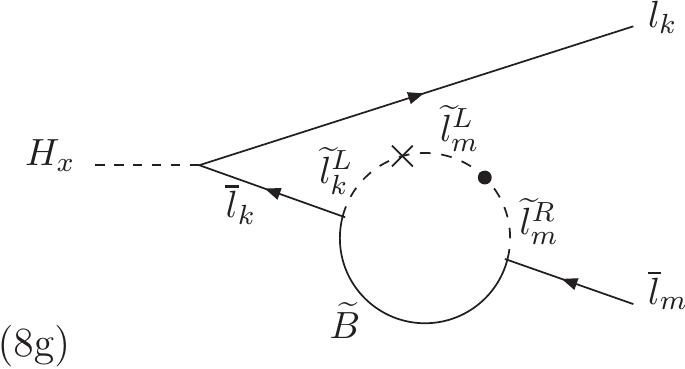} \\
\includegraphics[width=0.3\textwidth]{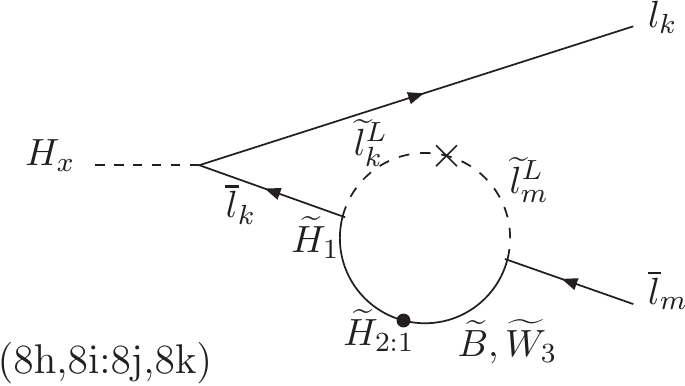} &
&
\includegraphics[width=0.3\textwidth]{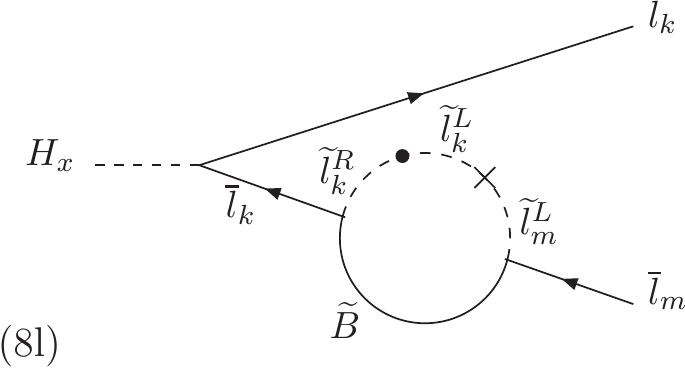} 
\end{tabular}
\caption{Relevant one-loop diagrams within the Mass Insertion Approximation for $H_x \to l_k \bar l_m$ decays in the MSSM electroweak interaction basis  for the internal SUSY particles, with one insertion changing flavor given by $\times=\Delta^{LL}_{mk}$.}
\label{diagsMIALL}
\end{center}
\end{figure}

\begin{figure}[t!]
\begin{center}
\begin{tabular}{ll}
\includegraphics[width=0.3\textwidth]{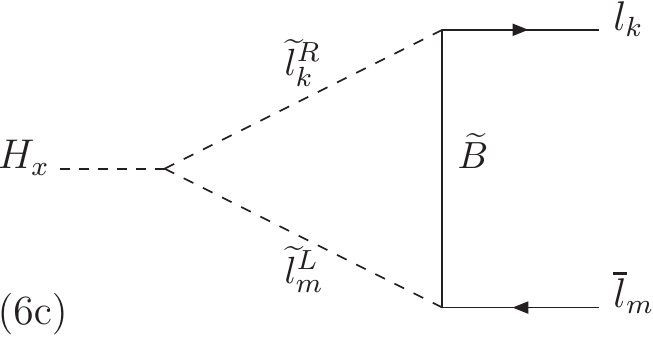} &
\includegraphics[width=0.3\textwidth]{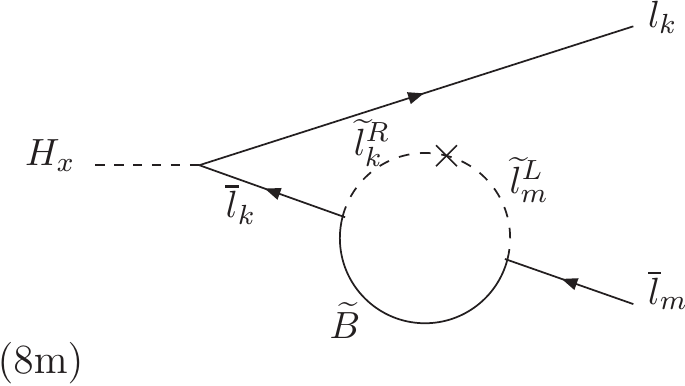} 
\end{tabular}
\caption{Relevant one-loop diagrams within the Mass Insertion Approximation for $H_x \to l_k \bar l_m$ decays in the MSSM electroweak interaction basis 
 for the internal SUSY particles, with one insertion changing flavor given by $\times=\Delta^{LR}_{mk}$.}
\label{diagsMIALR}
\end{center}
\end{figure}

\begin{figure}
\begin{center}
\begin{tabular}{ll}
\includegraphics[width=0.3\textwidth]{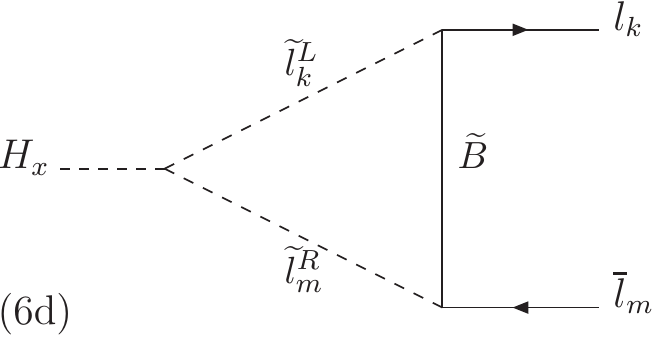} &
\includegraphics[width=0.3\textwidth]{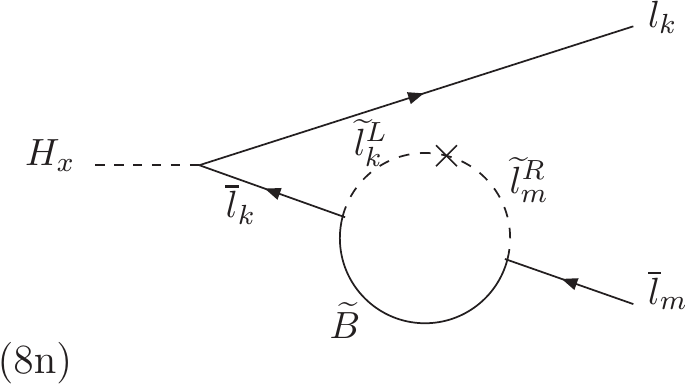} 
\end{tabular}
\caption{Relevant one-loop diagrams within the Mass Insertion Approximation for $H_x \to l_k \bar l_m$ decays in the MSSM electroweak interaction basis
 for the internal SUSY particles, with one insertion changing flavor given by $\times=\Delta^{RL}_{mk}$.}
\label{diagsMIARL}
\end{center}
\end{figure}

\begin{figure}[t!]
\begin{center}
\begin{tabular}{lll}
\includegraphics[width=0.3\textwidth]{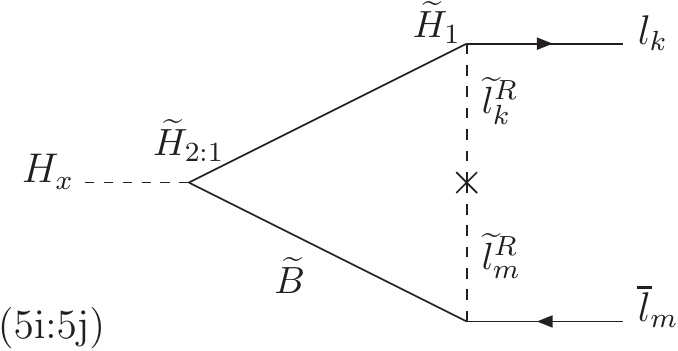} &
\includegraphics[width=0.3\textwidth]{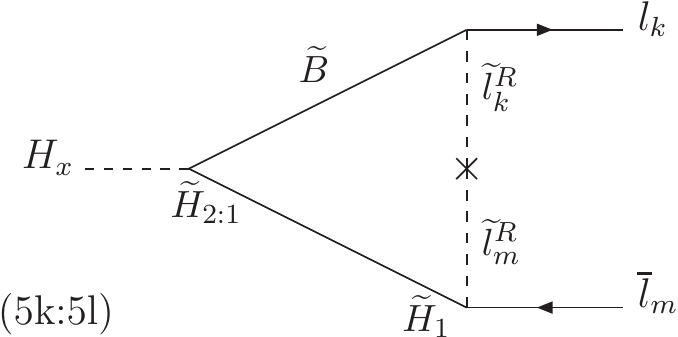} &
\includegraphics[width=0.3\textwidth]{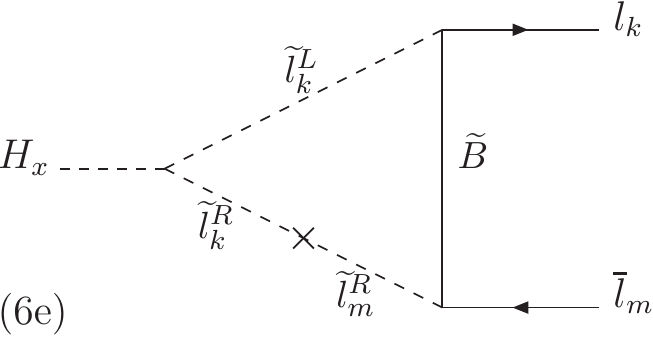} \\
\includegraphics[width=0.3\textwidth]{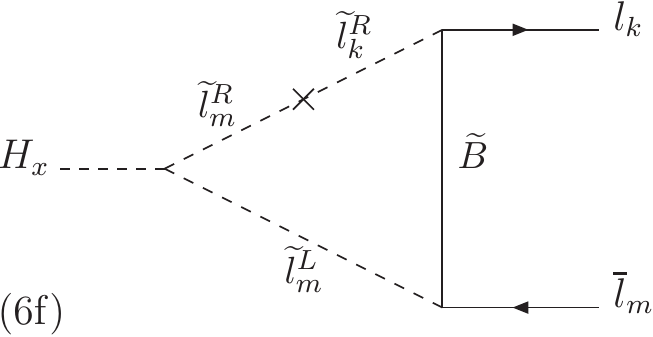} &
\includegraphics[width=0.3\textwidth]{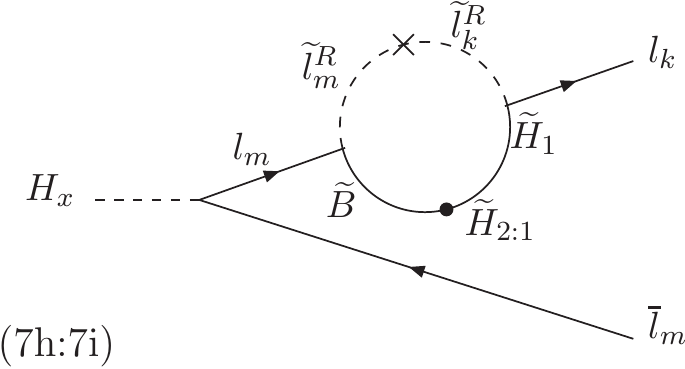} &
\includegraphics[width=0.3\textwidth]{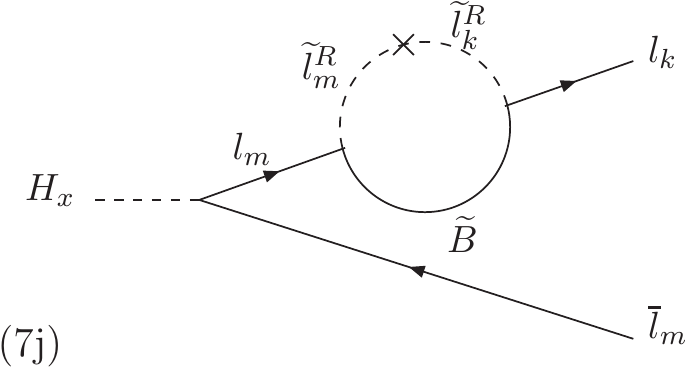} \\
\includegraphics[width=0.3\textwidth]{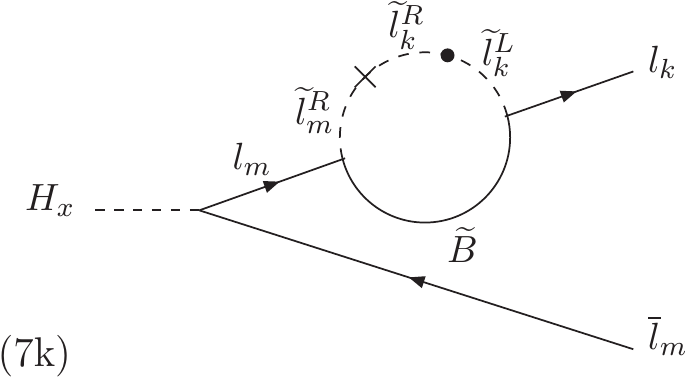} &
\includegraphics[width=0.3\textwidth]{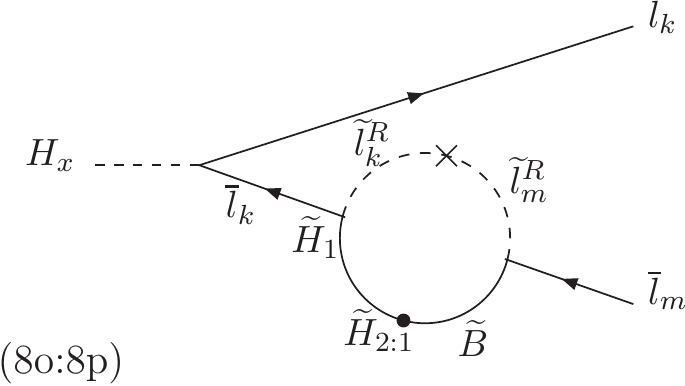} &
\includegraphics[width=0.3\textwidth]{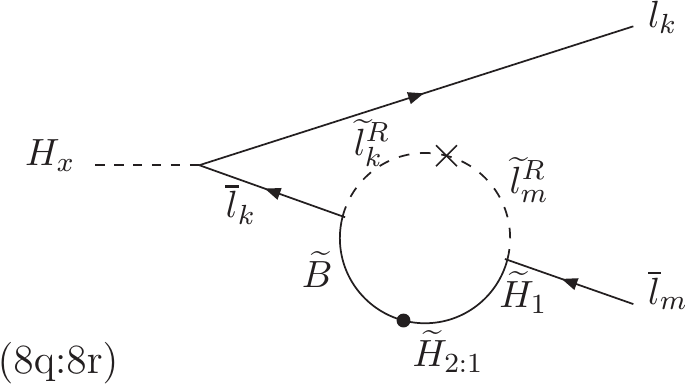} \\
\includegraphics[width=0.3\textwidth]{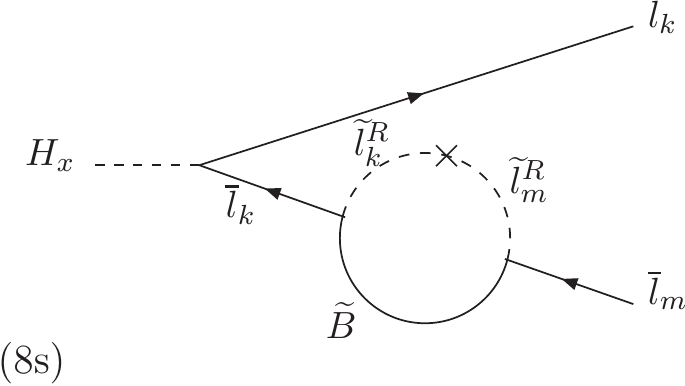} &
\includegraphics[width=0.3\textwidth]{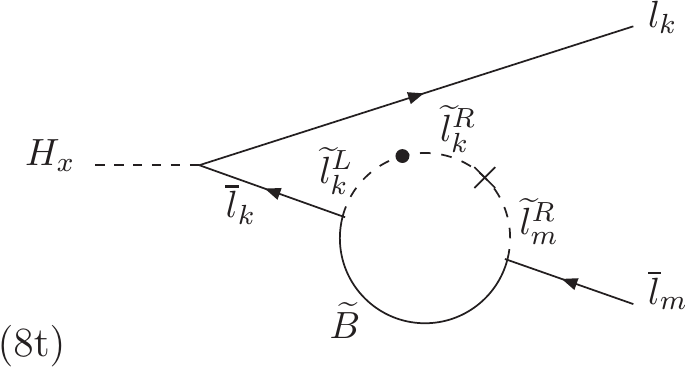} &
\includegraphics[width=0.3\textwidth]{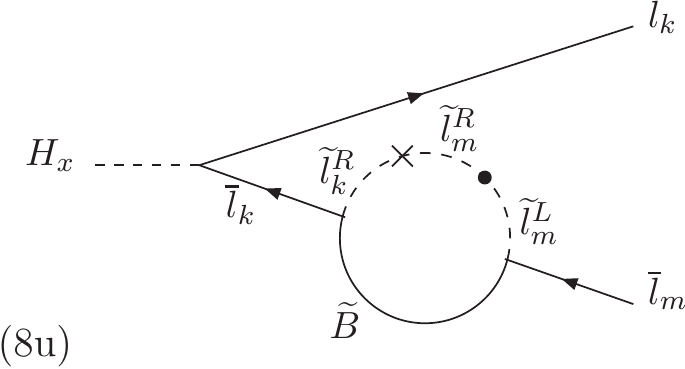} 
\end{tabular}
\caption{Relevant one-loop diagrams within the Mass Insertion Approximation for $H_x \to l_k \bar l_m$ decays in the MSSM electroweak interaction basis 
 for the internal SUSY particles, with one insertion changing flavor given by $\times=\Delta^{RR}_{mk}$.} 
\label{diagsMIARR}
\end{center}
\end{figure}

Next, we present our computation of the form factors $F_{L,R}^{(x)}$ within the MIA.  The results are presented in the following simple form, 
\be
 F_{L,R}^{(x)}= \Delta^{LL}_{mk} F_{L,R}^{(x)LL}+ {\Delta}^{LR}_{mk} F_{L,R}^{(x)LR}+ {\Delta}^{RL}_{mk} F_{L,R}^{(x)RL}+ \Delta^{RR}_{mk} F_{L,R}^{(x)RR} \,,
\label{formfactors}
\ee
where the contribution from each single insertion is explicitly separated. 
In order to extract all the relevant contributions in the MIA to each of these form factors, we have selected and computed, in a systematic way, all the diagrams that dominate the decay rates in the kinematic region of our interest here, namely, for very heavy internal sparticle masses as compared to the external particle masses: $m_{\rm SUSY} \gg m_{H_x}, m_{l_k}, m_{l_m}$. The set of contributing one-loop diagrams in the MIA are displayed in figures~\ref{diagsMIALL}, \ref{diagsMIALR}, \ref{diagsMIARL}, and \ref{diagsMIARR}, for each case with a non-vanishing insertion, $\Delta^{LL}_{mk}$, 
$\Delta^{LR}_{mk}$, $\Delta^{RL}_{mk}$, and $\Delta^{RR}_{mk}$, correspondingly. The labels assigned to these diagrams refer explicitly to the particular class of full one-loop diagram they should be compared with. For instance, the contributions from the MIA diagrams with labels $(1a)$, $(1b)$, when added, should be compared with the full diagram $(1)$, the ones with labels $(3a)$, $(3b)$, when added, should be compared with $(3)$, etc.
It should be noted that, in the scenarios that we are working with, all the sparticle masses are considered to be heavy by means of a unique common SUSY mass scale, generically called here $m_{\rm SUSY}$. In each of the three considered scenarios, the particular relation between each soft mass and $m_{\rm SUSY}$ varies, but in all scenarios the sparticle masses grow linearly with the common $m_{\rm SUSY}$ scale. Saying it in different words, we are integrating  to one-loop order all the internal SUSY particles, considering all of them very heavy, and without keeping any of them at low energies with a fixed mass. This feature allows us to classify the various contributions from the loop diagrams in the MIA into two categories, depending on their behavior in the asymptotic limit $m_{\rm SUSY} \to \infty$: 1) Contributions that go to a constant, which will be called from now on non-decoupling contributions, and 2) contributions that go to zero, which will be called from now on decoupling contributions. Furthermore, among these later 
we will distinguish between the dominant decoupling contributions, which  decrease with $m_{\rm SUSY}$ as powers of $(m_{H_x}/m_{\rm SUSY})$, and the subdominant 
decoupling contributions, which decrease with $m_{\rm SUSY}$ as powers of $(m_{\rm EW}/m_{\rm SUSY})$, with $m_{\rm EW}$ being any of the other electroweak masses involved, namely, $M_W$, $M_Z$, $m_{l_k}$, and $m_{l_m}$. Here we will not include these subdominant decoupling contributions. For instance, diagrams of type (2), that would be classified as (2a), (2b), with a 
$H_x \,{\tilde \nu}_{Lk} \, {\tilde \nu}_{Lm}$ vertex and one insertion of $\Delta^{LL}_{mk}$ type into one of the two sneutrino internal propagators, would be one of these cases, leading to contributions that are subdominant and decoupling by powers of $(m_{\rm EW}/m_{\rm SUSY})$, and consequently we have not included them into our selected diagrams. Although for all the cases studied in this work, we have checked that these corrections are not relevant numerically from a phenomenological point of view, in some specific cases in which important cancellations among the leading non-decoupling contributions occur, we have found that they may play some important role in order to obtain a better convergence between the full and the MIA results. This will be commented later in our numerical analysis.

The analytic results of the form factors in eq.~(\ref{formfactors}), $F_{L,R}^{(x)AB}$ with $AB=LL,LR,RL,RR$, from all the diagrams in figures~\ref{diagsMIALL}, \ref{diagsMIALR}, \ref{diagsMIARL}, and \ref{diagsMIARR} are collected in Appendix~\ref{AnalyticFormFactors}. The contributions from each diagram are presented separately and expressed in terms of the relevant one-loop functions, $C_0$, $C_2$, $D_0$, and ${\tilde D}_0$, which are given in Appendix~\ref{LoopIntegrals}. Some comments are in order regarding these analytic results. First of all, it is immediate to learn that within the MIA each diagram by itself is ultraviolet finite, since the contributing loop functions, $C_0$, $C_2$, $D_0$, and ${\tilde D}_0$, are all convergent. This is in contrast to the full one-loop computation, where there are some diagrams that are ultraviolet divergent, more specifically, all diagrams in figure~\ref{fulldiag} except $(2)$ and $(6)$ and, of course, the total full one-loop result given by the sum of the eight diagrams is ultraviolet convergent~\cite{Arganda:2004bz}. Second, according to our previously explained classification into non-decoupling and decoupling contributions, we can already conclude from these analytic results which particular terms will dominate. In particular, by selecting just the contributions from the loop functions at zero external momenta, we are capturing all the non-decoupling terms, and we can already conclude that these only appear in the $LL$ and $RR$ form factors but not in the $LR$ and $RL$ ones.
 
By considering zero external momenta in eqs.~(\ref{FLLL}) through (\ref{FRRR}), neglecting  $m_\mu$, and after some algebraic simplifications due to the symmetry properties of the loop functions,  we obtain for the case of our main interest here, $H_x \to \tau \bar \mu$ with $k=3$ and $m=2$, the following simple results for the non-decoupling (ND) part of the form factors, which is by far the dominant part:
\ba
\left(\Delta^{LL}_{23} F_{L}^{(x)LL}  \right)_{\rm ND}&=& \left(\frac{g^2}{16 \pi^{2}} \frac{m_{\tau}}{2 M_{W}}\right) \left[\frac{\sigma_{2}^{(x)}+\sigma_{1}^{(x)^{*}}t_{\beta}}{c_{\beta}} \right] 
(\delta^{LL}_{23} m_{\widetilde{L}_{2}}  m_{\widetilde{L}_{3}}) \nonumber \\
&& \times \left[ \frac{3}{2} \mu M_{2} D_{0}({ {0,0,0,m_{\widetilde{L}_{2}},m_{\widetilde{L}_{3}},\mu,M_{2}}})  \right. \nonumber \\
&&\left. -\frac{t_{W}^{2}}{2} \mu M_{1}D_{0}({ {0,0,0,m_{\widetilde{L}_{2}},m_{\widetilde{L}_{3}},\mu,M_{1}}}) \right. \nonumber \\
&&\left. -t_{W}^{2} \mu M_{1} D_{0}({ {0,0,0,m_{\widetilde{L}_{2}},m_{\widetilde{L}_{3}},m_{\widetilde{R}_{3}}}},M_{1}) \right] \,,
\label{FLLLND}  
\ea 
 where only eight diagrams contribute: $(1a)$, $(4a)$, $(5a)$, $(5b)$, $(6a)$,
 $(8h)$, $(8i)$, and $(8l)$, and
 \ba
\left(\Delta^{RR}_{23} F_{R}^{(x)RR} \right)_{\rm ND}&=& \left(\frac{g^2 t_{W}^{2}}{16 \pi^{2}} \frac{m_\tau}{2 M_{W}}\right) \left[\frac{\sigma_{2}^{(x)^{*}}+\sigma_{1}^{(x)}t_{\beta}}{c_{\beta}}\right]
(\delta^{RR}_{23} m_{\widetilde{R}_{2}}  m_{\widetilde{R}_{3}}) \nonumber \\
&& \times \left[ \mu M_{1} D_{0}({ {0,0,0,m_{\widetilde{R}_{2}},m_{\widetilde{R}_{3}},\mu,M_{1}}}) \right. \nonumber \\
&&\left. - \mu M_{1} D_{0}({ {0,0,0,m_{\widetilde{R}_{2}},m_{\widetilde{R}_{3}},
m_{\widetilde{L}_{3}},M_{1}}}) \right] \,, 
\label{FRRRND}
\ea
where only four diagrams contribute: $(5i)$, $(6e)$, $(8o)$, and $(8t)$.
The rest of form factors have a vanishing ND part. The coefficients $\sigma_{1}^{(x)}$ and $\sigma_{2}^{(x)}$ are defined in eq.~(\ref{sigma1y2}), and $t_W = \tan\theta_W$, $s_\beta = \sin\beta$, $c_\beta = \cos\beta$, and $t_\beta = \tan\beta$.
 
It is interesting to notice that only the loop function $D_0$ at zero external momenta is involved in these simple expressions for the ND parts. The definition of this $D_0$ for arbitrary masses is given in eq.~(\ref{D0zeromom}).  It is clear that if one considers all mass parameters to be asymptotically heavy, the two functions in eqs.~(\ref{FLLLND}) and (\ref{FRRRND}) tend to a constant value, meaning that the integration out of the heavy SUSY particles does leave as a remnant a non-vanishing value of the $\Gamma(H_x \to \tau {\bar \mu})$ partial widths that is constant with $m_{\rm SUSY}$ if either $\delta^{LL}_{23}$ or $\delta^{RR}_{23}$ are non vanishing.  We also wish to emphasize that for the particular choice of $\mu=m_{\widetilde{L}_{3}}$ there is an important cancellation in the $RR$ form factor  between the two contributing terms in eq.~(\ref{FRRRND}), leading to a vanishing of the ND part in this case.  It is also worth mentioning  that the above analytic results at zero external momenta of eqs.~(\ref{FLLLND}) and (\ref{FRRRND}) are in agreement with previous results obtained in the alternative  framework of the effective Lagrangian approach~\cite{Brignole:2003iv}.  

On the other hand, the above simple expressions also tell us about how large can be this constant value as a function of the other relevant parameters, namely, $\tan \beta$ and $m_A$. Indeed, these two dependencies are fully contained in the factor inside the big squared parenthesis, which can be easily derived using eq.~(\ref{sigma1y2}) and setting $s_\alpha$ and $c_\alpha$ in terms of the input parameters $m_A$ and $\tan \beta$, namely, 
$s_\alpha =\sin\alpha = -c_\beta+{\cal O}(M_Z^2/m_A^2)$ and $c_\alpha = \cos\alpha = s_\beta+{\cal O}(M_Z^2/m_A^2)$. This simple exercise gives the relevant dependence with $m_A$ and $\tan \beta$ in the two form factors above. We find that for the case of $\delta^{LL}_{23}$ and $\delta^{RR}_{23}$ mixings, and for generic masses, the modulo of the form factors go at large $\tan \beta$ as: 
\be
\left\vert \frac{\sigma_{2}^{(h)}+\sigma_{1}^{(h)^{*}}t_{\beta}}{c_{\beta}} \right\vert  \propto \left(\frac{M_{Z}}{m_{A}}\right)^{2}t_{\beta}   \ \ \ \ {\rm and} \ \ \ \ \left\vert \frac{\sigma_{2}^{(H,A)}+\sigma_{1}^{(H,A)^{*}}t_{\beta}}{c_{\beta}} \right\vert \propto t_{\beta}^{2} \,.
\label{tb-ma-dep}
\ee
By collecting all findings together, we can summarize this general power counting with all the relevant factors in the case of $\delta^{LL}_{23}$ and $\delta^{RR}_{23}$ as follows:
 
\ba
\left(\Delta^{LL}_{23} F_{L}^{(h)LL} \right)_{\rm ND}
 &\sim & {\cal O}\left(\delta^{LL}_{23} \left(\frac{g^2}{16 \pi^{2}}\right) \left(\frac{m_{\tau}}{ M_{W}}\right)^1 
\left(\frac{m_{h}}{m_{\rm SUSY}}\right)^0
 \left(\frac{M_{Z}}{m_{A}}\right)^{2} (t_{\beta})^1 \right) \,, \label{h-LLpowercounting} \\
\left(\Delta^{LL}_{23} F_{L}^{(H,A)LL}   \right)_{\rm ND}
 &\sim & {\cal O}\left(\delta^{LL}_{23} \left(\frac{g^2}{16 \pi^{2}}\right) \left(\frac{m_{\tau}}{M_{W}}\right)^1 
\left(\frac{m_{H,A}}{m_{\rm SUSY}}\right)^0
 \left(\frac{M_{Z}}{m_{A}}\right)^{0} (t_{\beta})^2 \right) \,, \label{HA-LLpowercounting} \\
\left(\Delta^{RR}_{23} F_{R}^{(h)RR}   \right)_{\rm ND}
 &\sim & {\cal O}\left(\delta^{RR}_{23} \left(\frac{g^2 t_W^2}{16 \pi^{2}}\right) \left(\frac{m_{\tau}}{ M_{W}}\right)^1 
\left(\frac{m_{h}}{m_{\rm SUSY}}\right)^0
 \left(\frac{M_{Z}}{m_{A}}\right)^{2} (t_{\beta})^1 \right) \,, \label{h-RRpowercounting} \\
\left(\Delta^{RR}_{23} F_{R}^{(H,A)RR}   \right)_{\rm ND}
 &\sim & {\cal O}\left(\delta^{RR}_{23} \left(\frac{g^2 t_W^2}{16 \pi^{2}}\right) \left(\frac{m_{\tau}}{M_{W}}\right)^1 
\left(\frac{m_{H,A}}{m_{\rm SUSY}}\right)^0
 \left(\frac{M_{Z}}{m_{A}}\right)^{0} (t_{\beta})^2 \right) \,, \label{HA-RRpowercounting}
  \ea
which show, on the one hand, the expected decoupling behavior with $m_A$ in the lightest Higgs boson $h$ case, recovering the well known feature of vanishing LFVHD rates within a SM Higgs-like scenario, and, on the other hand, the also well known feature of the enhanced heavy $H$ and $A$ LFVHD rates at large $\tan\beta$, which grow as $(\tan\beta)^4$. 

In the case of $\delta^{LR}_{23}$ and $\delta^{RL}_{23}$ mixings, the effective form factors, as we have said, decouple with the large sparticle masses, since the potential non-decoupling terms coming from the evaluation of the loop functions at zero external momenta vanish when adding the two relevant diagrams: $(6c)$ and $(8m)$ in the $LR$ case and $(6d)$ and $(8m)$ in the $RL$ one.
In these two cases, the leading contribution then comes from the decoupling 
(D) terms of 
${\cal O}(m_{H_x}^2/m_{\rm SUSY}^2)$ in the $C_0$ loop functions expansions:
\ba
\left(\Delta^{LR}_{23}F_{L}^{(x)LR} \right)_{\rm D}&=& \frac{g^2 t_{W}^{2}}{16 \pi^{2}}  
(\tilde \delta^{LR}_{23} v_1 \sqrt{m_{\tilde L_{2}}m_{\tilde R_{3}}}) \frac{M_{1} \sigma_{1}^{(x)^{*}}}{2 M_{W} c_{\beta}} \nonumber \\
&& \times \left( 
- C_{0}(p_{2},p_{1},M_{1},m_{\widetilde{R}_{3}},m_{\widetilde{L}_{2}})  
+ C_{0}(p_{3},0,M_{1},m_{\widetilde{L}_{2}},m_{\widetilde{R}_{3}})
  \right) \,, 
\ea
and 
\ba
\left(\Delta^{RL}_{23}F_{R}^{(x)RL}\right)_{\rm D} &=& \frac{g^2 t_{W}^{2}}{16 \pi^{2}} 
(\tilde \delta^{RL}_{23} v_1 \sqrt{m_{\tilde R_{2}}m_{\tilde L_{3}}}) \frac{M_{1} \sigma_{1}^{(x)}}{2 M_{W} c_{\beta}} \nonumber \\
&& \times \left( 
- C_{0}(p_{2},p_{1},M_{1},m_{\widetilde{L}_{3}},m_{\widetilde{R}_{2}}) 
+ C_{0}(p_{3},0,M_{1},m_{\widetilde{R}_{2}},m_{\widetilde{L}_{3}})  \right) \,.   
\ea   
It is remarkable that these results above for the $LR$ and $RL$ cases are not dependent on the lepton masses nor on $\mu$.
We also see the factorized dependence in $\tan \beta$, this time inside 
$\sigma_{1}^{(x)}$. Thus, we can summarize the general power counting with all the relevant factors in the case of $\tilde\delta^{LR}_{23}$ as follows:
\ba
\left(\Delta^{LR}_{23} F_{L}^{(h)LR} \right)_{\rm D}
 &\sim & {\cal O}\left(\tilde \delta^{LR}_{23} \left(\frac{g^2 t_W^2}{16 \pi^{2}}\right) \left(\frac{v}{ M_{W}}\right)^1 
\left(\frac{m_{h}}{m_{\rm SUSY}}\right)^2
 \left(\frac{M_{Z}}{m_{A}}\right)^{0} (t_{\beta})^{-1} \right) \,, \label{h-LRpowercounting} \\
\left(\Delta^{LR}_{23} F_{L}^{(H,A)LR} \right)_{\rm D}
 &\sim & {\cal O}\left(\tilde \delta^{LR}_{23} \left(\frac{g^2 t_W^2}{16 \pi^{2}}\right) \left(\frac{v}{ M_{W}}\right)^1 
\left(\frac{m_{H,A}}{m_{\rm SUSY}}\right)^2
 \left(\frac{M_{Z}}{m_{A}}\right)^{0} (t_{\beta})^0 \right) \,, \label{HA-LRpowercounting}
\ea
and similarly for $\tilde\delta^{RL}_{23}$, by interchanging $L$ by $R$ in the formulas above.  

Finally, to finish this section we find illustrative to include the analytic results in the simplest scenario where all soft mass parameters are equal, i.e, the {\it Equal masses} scenario with just one SUSY scale: $m_{\rm SUSY}=m_S$. In this case the formulas can be greatly simplified and they could be useful both as a reference benchmark scenario to compare with and to perform an easy phenomenological analysis. First, the form factors are expressed as: 
\be
 F_{L,R}^{(x)}= \delta^{LL}_{23} {\hat F}_{L,R}^{(x)LL}+ \tilde{\delta}^{LR}_{23} 
 {\hat F}_{L,R}^{(x)LR}+  \tilde{\delta}^{RL}_{23} {\hat F}_{L,R}^{(x)RL}+ \delta^{RR}_{23} {\hat F}_{L,R}^{(x)RR} \,.
\label{hatFF} 
\ee
Then, by using the formulas of the loop functions in eq.~(\ref{eqmass}) of Appendix~\ref{LoopIntegrals}, we have found the results collected at the end of Appendix~\ref{AnalyticFormFactors}, where we explicit the contributions from each diagram. After adding the contributions from all the diagrams, the total results of the form factors, which can be interpreted as effective LFV interaction vertices, are the following: 
\ba
{\hat F}_{L}^{(x)LL} 
&=& \frac{g^2}{16 \pi^{2}}  \frac{m_{\tau}}{2 M_{W} c_{\beta}} \left[ 
\left(
\sigma_{2}^{(x)}+\sigma_{1}^{(x)^{*}}t_{\beta}
\right) \frac{1-t_{W}^{2}}{4}   \right. \nonumber \\ 
&&\left.  \ \ 
+ \frac{m_{H_{x}}^{2}}{m_{S}^{2}} \left( \sigma_{2}^{(x)} 
\frac{3-5 t_{W}^{2}}{120} + \sigma_{1}^{(x)^{*}} \frac{9-11t_{W}^{2}}{240}\right)   \right] \,,
\label{hatFLLL}
\ea
\ba
 {\hat F}_{R}^{(x)LL} 
&=&  \frac{g^2}{16 \pi^{2}}  \frac{m_{\mu}}{2 M_{W} c_{\beta}} \left[ 
\left(
\sigma_{2}^{(x)^{*}}+\sigma_{1}^{(x)}t_{\beta}
\right) \frac{1-t_{W}^{2}}{4}  \right.  \nonumber\\
&& \left. \ \  
 + \frac{m_{H_{x}}^{2}}{m_{S}^{2}} \left( \sigma_{2}^{(x)^{*}} \frac{3-5t_{W}^{2}}{120} + \sigma_{1}^{(x)} \frac{9-11t_{W}^{2}}{240}\right)  \right] \,,
\label{hatFRLL}
 \ea 
\ba
{\hat F}_{L}^{(x)LR} 
&=&  \frac{g t_W^2}{16 \pi^{2}} \frac{1}{24\sqrt{2}}  \frac{m_{H_{x}}^{2}}{m_{S}^{2}}\left[ \sigma_{1}^{(x)^{*}} \right] \,,
\label{hatFLLR}
\ea
\ba
{\hat F}_{R}^{(x)RL}& =&  {+}{\hat F}_{L}^{(x)LR*}  \ \ ; \ \ \   {\hat F}_{R}^{(x)LR}={\hat F}_{L}^{(x)RL}=0 \,,
\label{hatFRRL}
\ea
\ba
 {\hat F}_{L}^{(x)RR} 
  &=&  -\frac{g^2 t_W^2}{16 \pi^{2}} \frac{m_{\mu}}{2 M_{W} c_{\beta}} \frac{m_{H_{x}}^{2}}{m_{S}^{2}} \left[\frac{2 \sigma_{2}^{(x)} + \sigma_{1}^{(x)^{*}}}{120} \right] \,,
\label{hatFLRR}
\ea
\ba
{\hat F}_{R}^{(x)RR} 
  &=&  -\frac{g^2 t_W^2}{16 \pi^{2}}  \frac{m_{\tau}}{2 M_{W} c_{\beta}} \frac{m_{H_{x}}^{2}}{m_{S}^{2}} \left[ \frac{2 \sigma_{2}^{(x)^{*}} + \sigma_{1}^{(x)}}{120}  \right] \,.
\label{hatFRRR} 
\ea 
We can see clearly in the total results above how relevant are the strong cancellations that occur in this {\it Equal masses} scenario among the various contributing diagrams. In fact, the behavior of the $RR$ case at large $m_S$ changes qualitatively with respect to a generic scenario with heavy sparticles, since we find in contrast a decoupling behavior, with the form factor going as $(m_{H_x}^2/m_{S}^2)$, due to an exact cancellation of the non-decoupling terms in this particular case. Regarding the $LL$ case, we find again non decoupling, and for the $LR$ and $RL$ cases we find decoupling as in the generic case.

Interestingly, if we keep just the leading non-decoupling terms and neglect $m_\mu$ in the previous formulas for the {\it Equal masses} scenario, we are left with only one relevant form factor, ${\hat F}_{L}^{(x)LL}$, and therefore the total effect of the heavy SUSY particles can be summarized in terms of a very simple effective LFV vertex given by $(-ig V^{\rm eff}_{H_x \tau \mu}P_L)$ with:
\be
V^{\rm eff}_{H_x \tau \mu}=\frac{g^2}{16 \pi^{2}}\frac{m_{\tau}}{2 M_W}
\left[\frac{\sigma_{2}^{(x)}+\sigma_{1}^{(x)^{*}}t_{\beta}}{c_{\beta}} \right]
\left(\frac{1-t_W^2}{4}\right) \delta^{LL}_{23} \,. 
\label{veff}
\ee
It should be noted that this is valid for all $\tan \beta$ values. We have further checked that when the large $\tan \beta$ limit is considered in this eq.~(\ref{veff}), we find agreement with the results found from the full one-loop computation in~\cite{Arganda:2004bz}. Concretely, using eq.~(\ref{tb-ma-dep}) for the lightest Higgs boson vertex we find the expected decoupling behavior in the large $m_A \gg M_Z$ limit going as $(M_Z/m_A)^2$, which then makes this $h$ boson to resemble as the SM Higgs boson. We also get agreement with the expected $(\tan\beta)^2$ enhanced LFV vertex~\cite{Arganda:2004bz} in the case of the $H$ and $A$ Higgs bosons:
 
\ba
 V^{\rm eff}_{h \tau \mu}|_{t_\beta \gg 1}&=&-\frac{g^2}{16 \pi^{2}}\frac{m_{\tau}}{ M_W} \frac{M_Z^2}{m_A^2} t_\beta 
\left(\frac{1-t_W^2}{4}\right) \delta^{LL}_{23}\,,\nonumber \\
V^{\rm eff}_{H \tau \mu}|_{t_\beta \gg 1}&=&-iV^{\rm eff}_{A \tau \mu}|_{t_\beta \gg 1}=-\frac{g^2}{16 \pi^{2}}\frac{m_{\tau}}{2 M_W} t_\beta^2 
\left(\frac{1-t_W^2}{4}\right) \delta^{LL}_{23} \,.  
\label{vefflargetbeta}
\ea 

\section{Numerical results of the $h,H,A \to \tau \bar \mu$  decay rates}
\label{results}

In this section we analyze the behavior of the radiative corrections from SUSY loops to the LFV neutral Higgs bosons decays $h, H, A \to \tau \bar \mu$, comparing numerically the predictions of the full one-loop calculation~\cite{Arganda:2004bz} with the MIA results, calculated for the first time here. The SUSY mass spectra for the three scenarios considered along this work are computed numerically with the code {\tt SPheno}~\cite{SPheno}. The LFVHD rates are computed with our private FORTRAN code in which we have implemented both the analytic results of the MIA of eqs.~(\ref{FLLL}) through (\ref{FRRR}) and also the complete one-loop formulas of~\cite{Arganda:2004bz}. The masses of the three neutral MSSM Higgs bosons, with two-loop corrections included, and their corresponding total decay widths are computed by means of the code {\tt FeynHiggs}~\cite{FeynHiggs}. We have explicitly checked that all the numerical results for BR$(H\to \tau \bar \mu)$ are nearly equal to those of BR$(A\to \tau \bar \mu)$ and, for shortness, we will show in this section only the latter.

\begin{figure}[t!]
\begin{center}
\begin{tabular}{cc}
\includegraphics[width=0.49\textwidth]{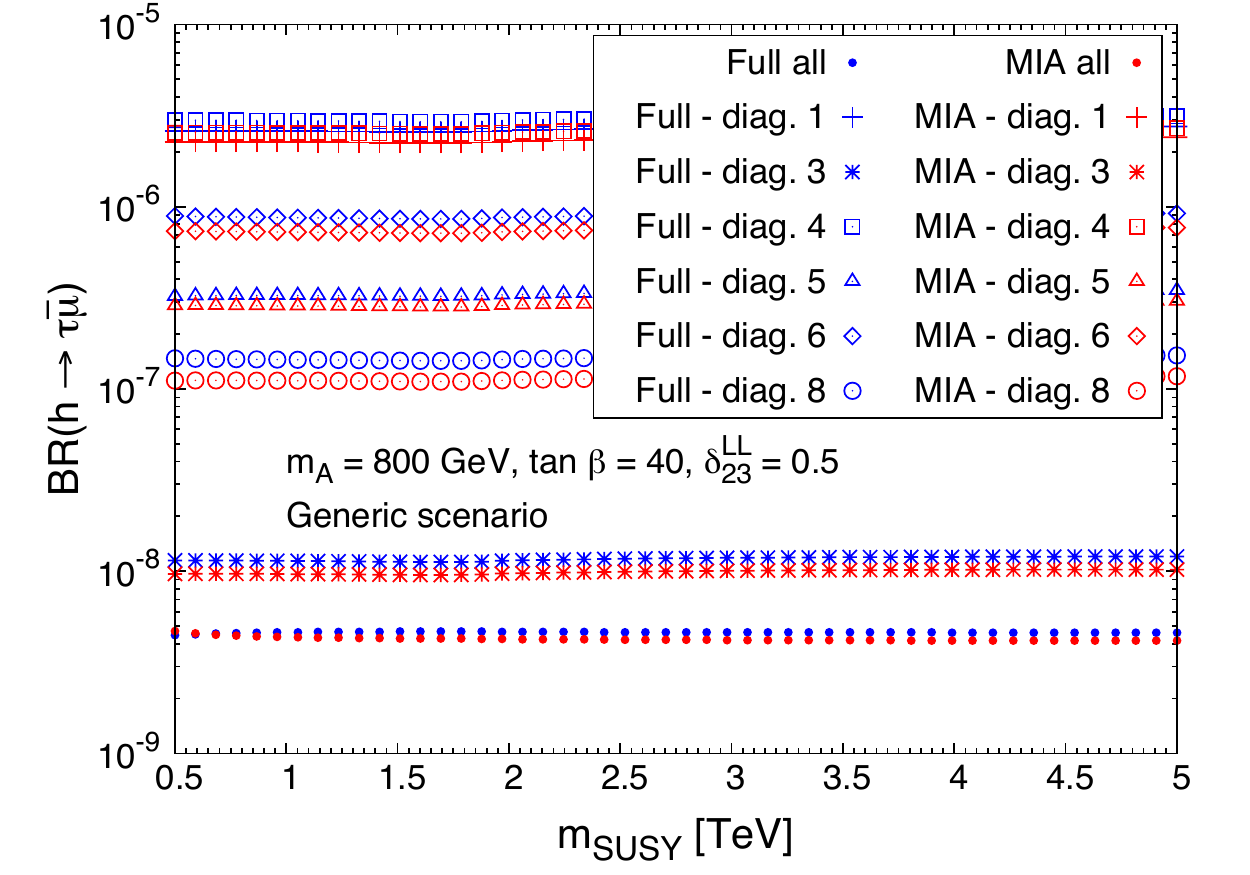} &
\includegraphics[width=0.49\textwidth]{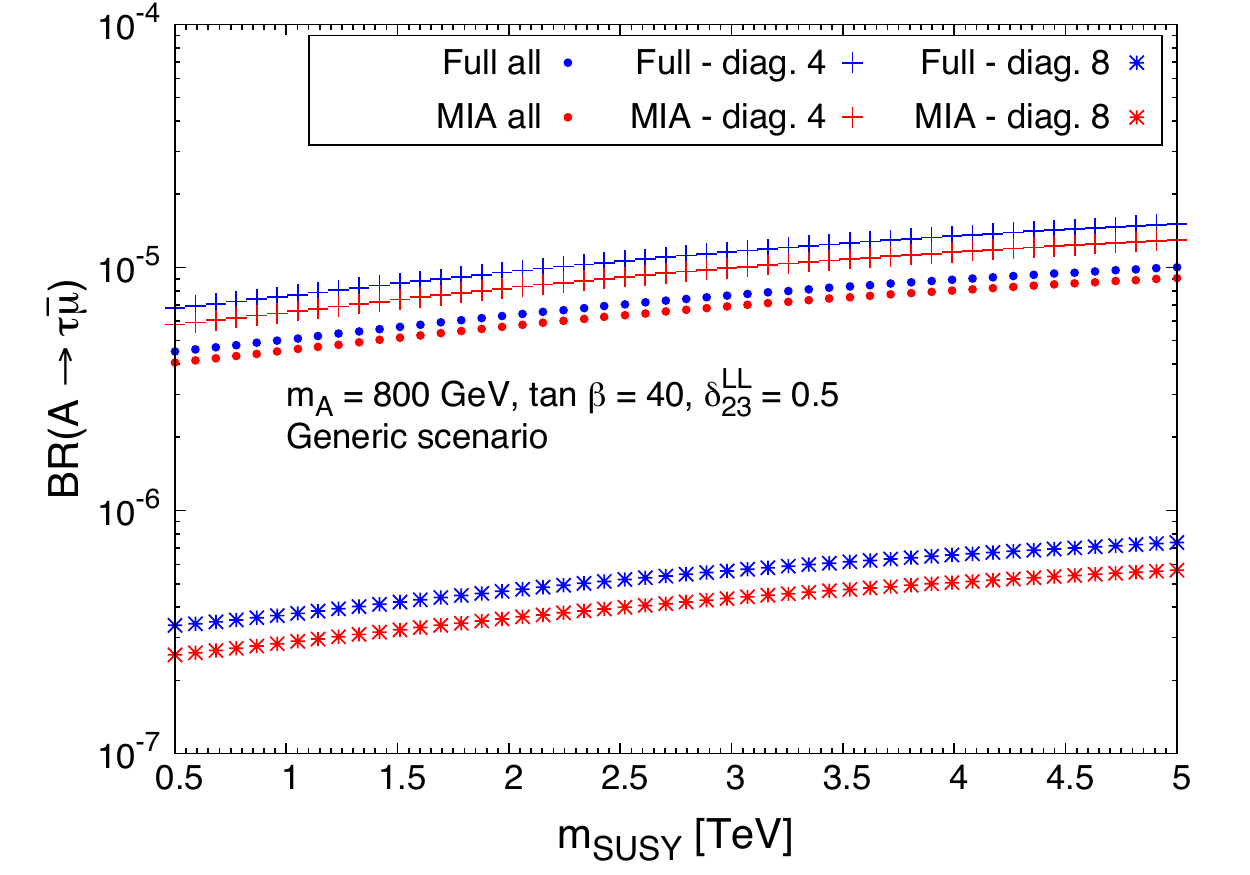} \\
\includegraphics[width=0.49\textwidth]{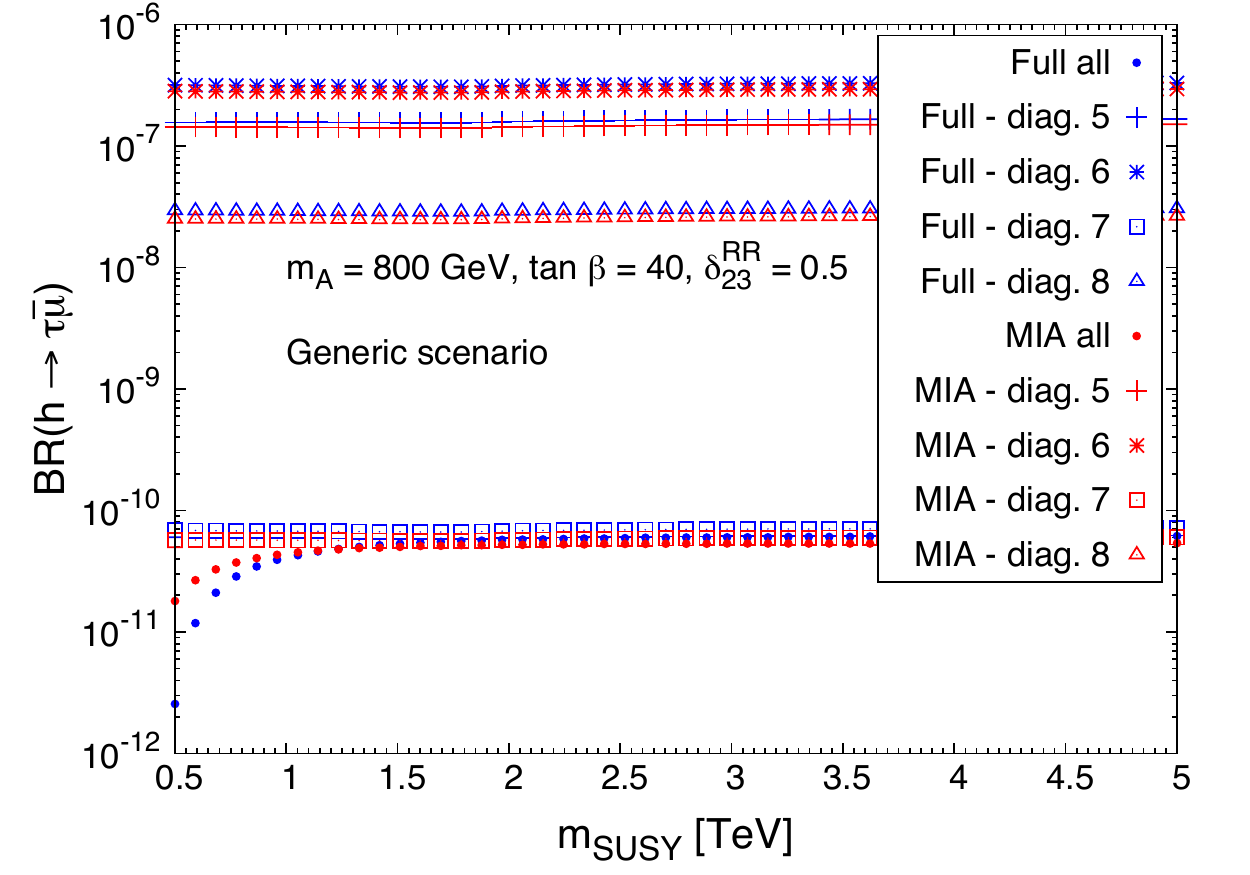} &
\includegraphics[width=0.49\textwidth]{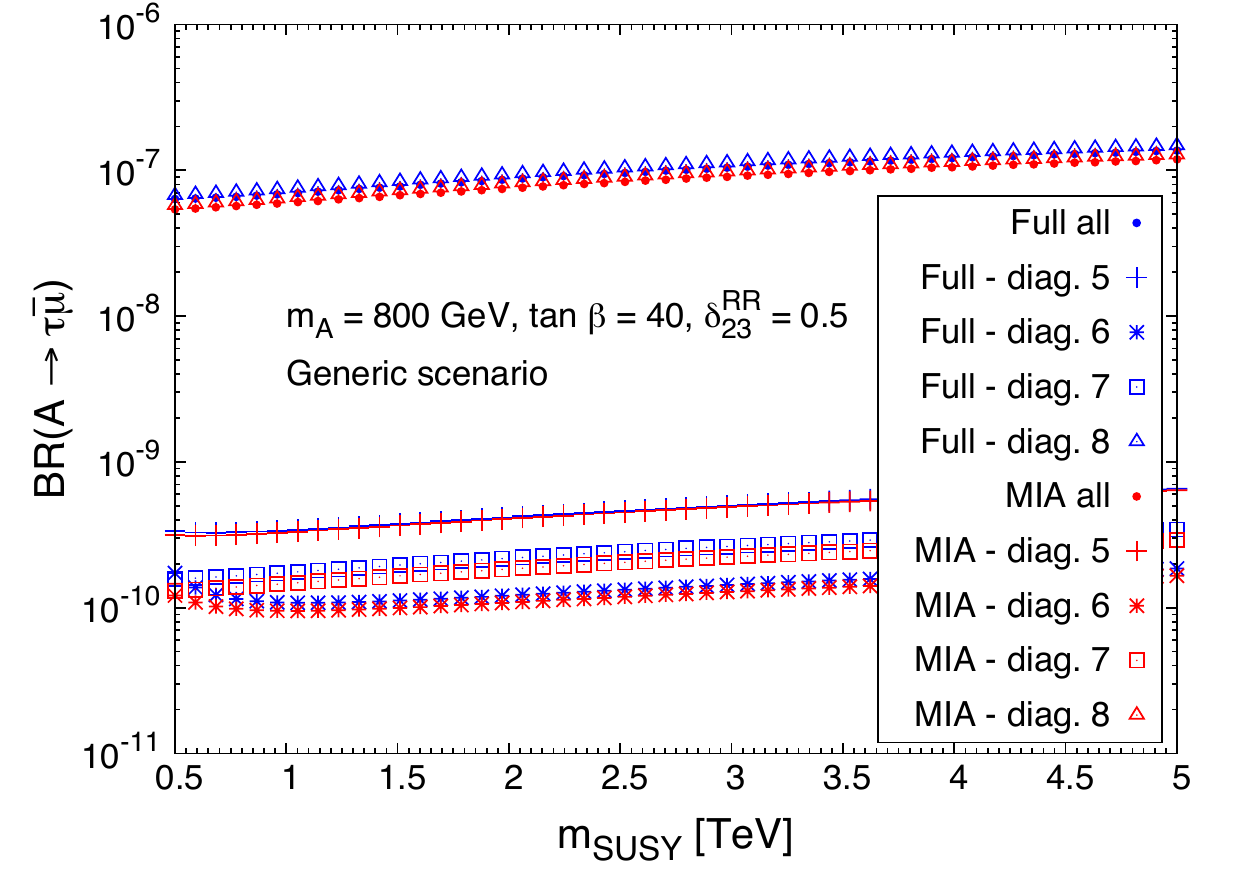} \\
\includegraphics[width=0.49\textwidth]{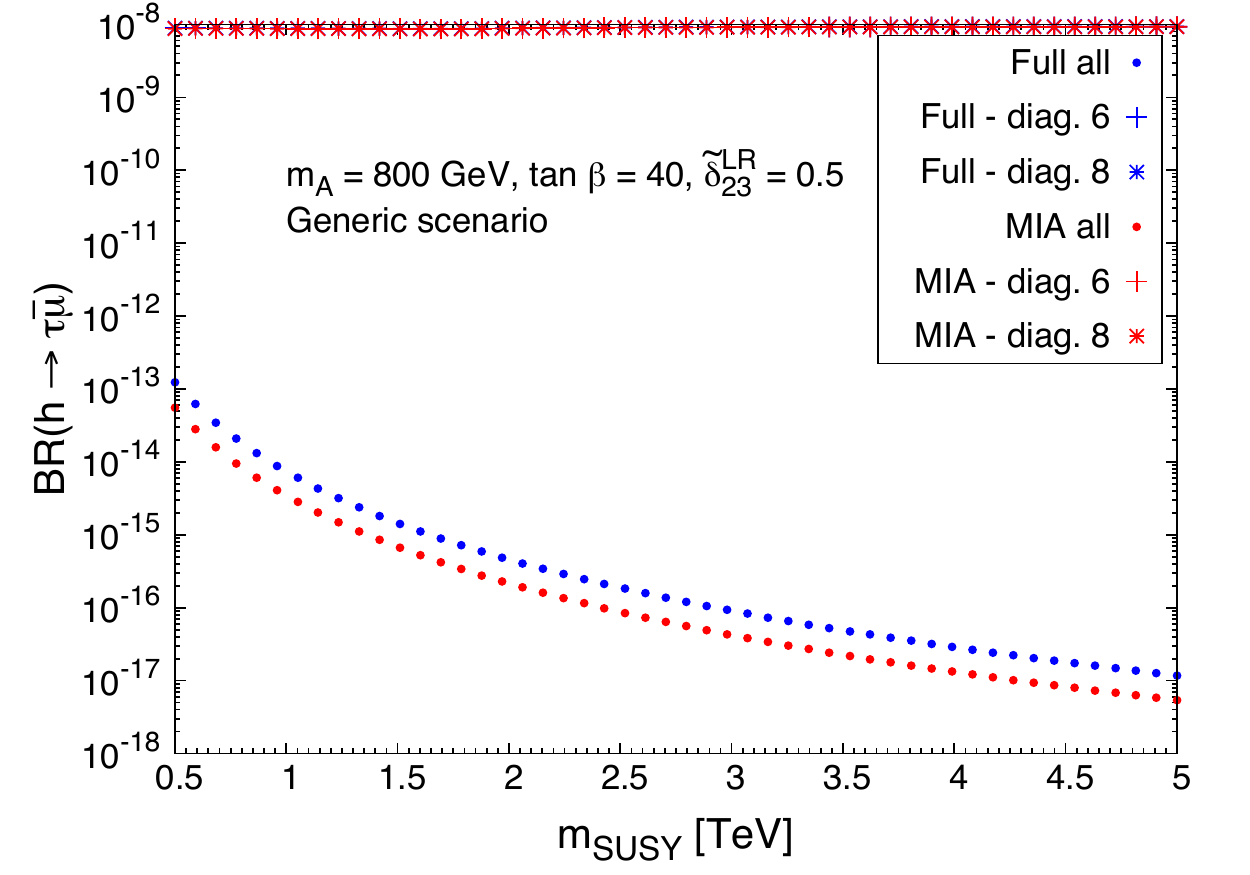} &
\includegraphics[width=0.49\textwidth]{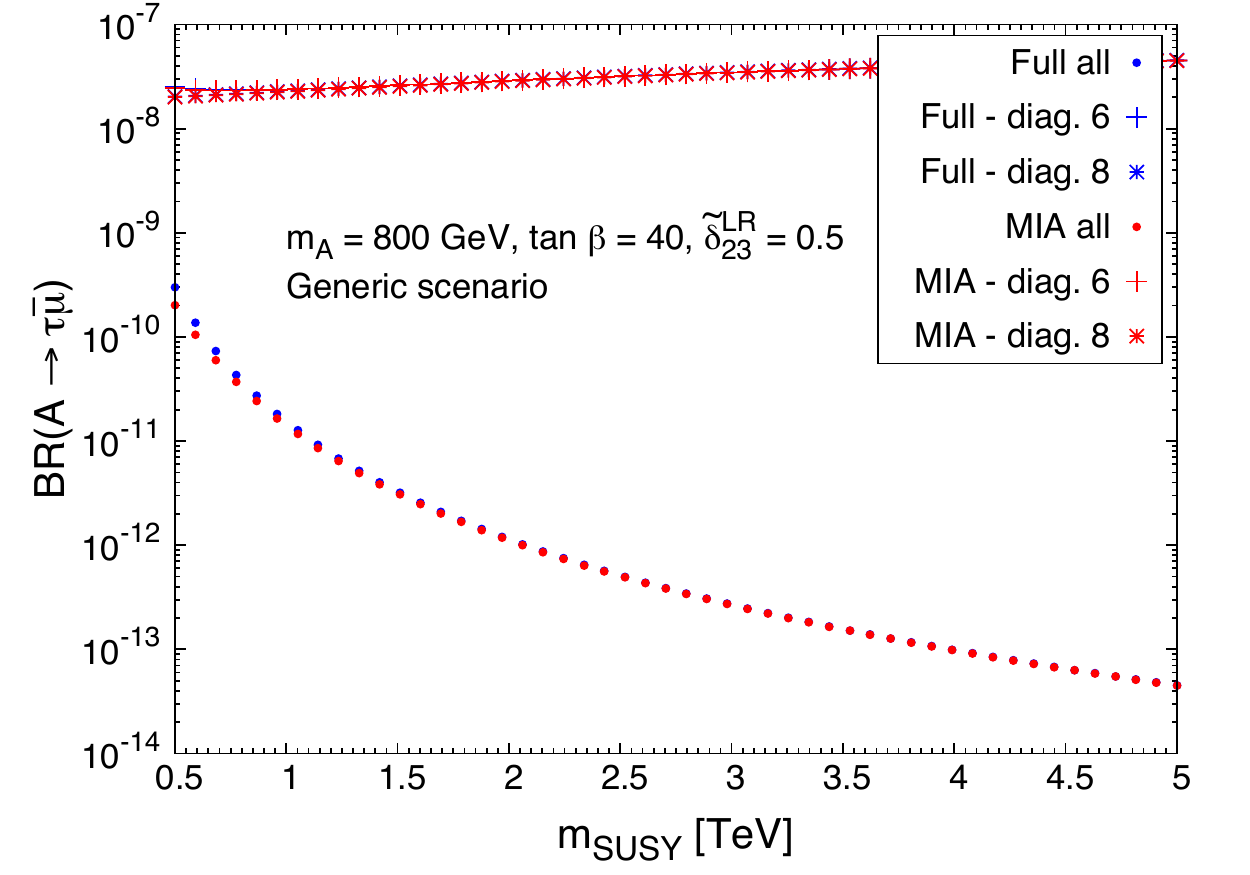}
\end{tabular}
\caption{Contributions of the dominant diagrams and the total one to BR($h \to \tau \bar \mu$) (left panels) and BR($A \to \tau \bar \mu$) (right panels) as functions of $m_\text{SUSY}$ in the {\it Generic} scenario with $m_A =$ 800 GeV and $\tan\beta =$ 40, for $\delta_{23}^{LL} = 0.5$ (upper panels), $\delta_{23}^{RR} = 0.5$ (middle panels), and $\tilde \delta_{23}^{LR} = 0.5$ (lower panels). The results for $\tilde \delta_{23}^{RL} = 0.5$ (not shown) are identical to those of $\tilde \delta_{23}^{LR} = 0.5$. In each case, the other flavor changing deltas are set to zero. The results for the heavy scalar $H$ (not shown) are nearly equal to these ones for the pseudoscalar $A$.
}\label{LFVHD-mSUSY_generic}
\end{center}
\end{figure}

We start the presentation of the numerical results with the most general scenario considered along this work, the {\it Generic} scenario, in which all the SUSY mass parameters are different. We show in figure~\ref{LFVHD-mSUSY_generic} the contributions of the dominant diagrams and the total one to BR($h \to \tau \bar \mu$) (left panels) and BR($A \to \tau \bar \mu$) (right panels) as functions of $m_\text{SUSY}$, within this scenario with $m_A =$ 800 GeV and $\tan\beta =$ 40, for $\delta_{23}^{LL} = 0.5$ (upper panels), $\delta_{23}^{RR} = 0.5$ (middle panels), and $\tilde \delta_{23}^{LR} = 0.5$ (lower panels). In each case, the other flavor changing deltas are set to zero. Since the results for $\tilde \delta_{23}^{RL} = 0.5$ are identical to those of $\tilde \delta_{23}^{LR} = 0.5$, they are not shown here. 

The BR($h \to \tau \bar \mu$) for the $LL$ case is displayed on the upper left panel of figure~\ref{LFVHD-mSUSY_generic}. First of all, we can see that each diagram contribution and the total prediction present the expected non-decoupling behavior with $m_\text{SUSY}$, with a very good agreement between the full one-loop results and the MIA ones. The agreement is found for each diagram contribution and for the total result. It should be noted that although the non-decoupling behavior of the partial width manifests in that it goes to a  constant value at large $m_\text{SUSY}$, in the plots we see however a slight increase of the branching ratios due to the slight decrease of the total width with $m_\text{SUSY}$. Regarding the dominant contributions, they come from diagrams 1 and 4, and we have found that they nearly cancel between each other. The rest of subdominant diagrams (3, 5, 6, and 8) are indeed important, since the remnant contributions of diagrams 1 and 4 interfere negatively with their contributions and fall down the total contribution below the diagram 3 one, what is the lowest one. Therefore, it is clear that there is in the $LL$ case a strong cancellation among diagrams of the BR($h \to \tau \bar \mu$) that reduces the rates around three orders of magnitude, from the dominant contributions (diagrams 1 and 4) to the total one. This strong cancellation does not occur for BR($A \to \tau \bar \mu$) as we can observe on the right panel of figure~\ref{LFVHD-mSUSY_generic}. The dominant contribution to this process in the $LL$ case comes from the diagram 4, followed by far by the diagram 8. There is indeed a small negative interference between these two diagrams, resulting in total contributions slightly lower than the diagram 4 ones. The non-decoupling behavior with $m_\text{SUSY}$ of all the contributions is also manifest, and the results in the MIA are very close to the full one-loop ones again. 
 
Now we move our attention to the $RR$ case. The dominant and total contributions to BR($h \to \tau \bar \mu$) are depicted on the middle left panel of figure~\ref{LFVHD-mSUSY_generic}, in which we can see that the diagram 6 is the dominant one, followed by the diagram 5 and secondly by the diagram 8. In this case there is again a very strong cancellation among the contributions of these three diagrams, and the surviving contribution comes from the diagram 7, which reproduces very well the total result for BR($h \to \tau \bar \mu$). As in the previous cases, the agreement between the full one-loop calculation and the MIA one is very good, and all the contributions to the LFVHD partial width show a constant behavior as $m_\text{SUSY}$ grows. On the other hand, we observe on the middle right panel that for BR($A \to \tau \bar \mu$) the dominant contribution comes from the diagram 8, reproducing extremely well the total result. In this $RR$ case there cannot be any class of interference among diagrams because the rest of contributions are at the most two orders of magnitude smaller than dominant one. All of them present also the expected non-decoupling behavior with $m_\text{SUSY}$. We obtain again a very good agreement between the full and the MIA calculations. As we have already said in the $LL$ case, it happens also here for the $RR$ case that the slight increase of both branching ratios, BR($h \to \tau \bar \mu$) and BR($A \to \tau \bar \mu$), with $m_\text{SUSY}$ has not its origin in the LFV Higgs partial decay widths, since they are constant as $m_\text{SUSY}$ grows, but it is due to a small reduction of the total decay widths with $m_\text{SUSY}$. 
 
To end up with this {\it Generic} scenario, the results of the $h \to \tau \bar \mu$ and $A \to \tau \bar \mu$ rates for the $LR$ case are displayed on the lower panels of figure~\ref{LFVHD-mSUSY_generic}. Both LFVHD rates can be understood by means of the contributions from the most relevant diagrams that are diagrams 6 and 8 in this case. The dominant non-decoupling terms, constant with $m_\text{SUSY}$, of these two diagrams are identical in the MIA but with opposite sign. Thus, they exactly cancel and the remaining dominant decoupling contributions in the branching ratios are proportional to $(m_{H_x}/m_\text{SUSY})^4$, what explains the final decoupling behavior of these rates with $m_\text{SUSY}$ observed in the plots. A good agreement between the full result and the MIA one is again achieved.

As main conclusions of the figure~\ref{LFVHD-mSUSY_generic}, we could say that in the {\it Generic} scenario the MIA approximates very well the full one-loop results, diagram by diagram and the total contributions. The LFVHD rates present a clear non-decoupling behavior with $m_\text{SUSY}$ if $\delta_{23}^{LL}$ or $\delta_{23}^{RR}$ is the responsible for the flavor mixing, whilst these rates have a strong decoupling behavior with the SUSY mass scale when the $\tilde \delta_{23}^{LR}$ or $\tilde \delta_{23}^{RL}$ is connected.

\begin{figure}[t!]
\begin{center}
\begin{tabular}{cc}
\includegraphics[width=0.49\textwidth]{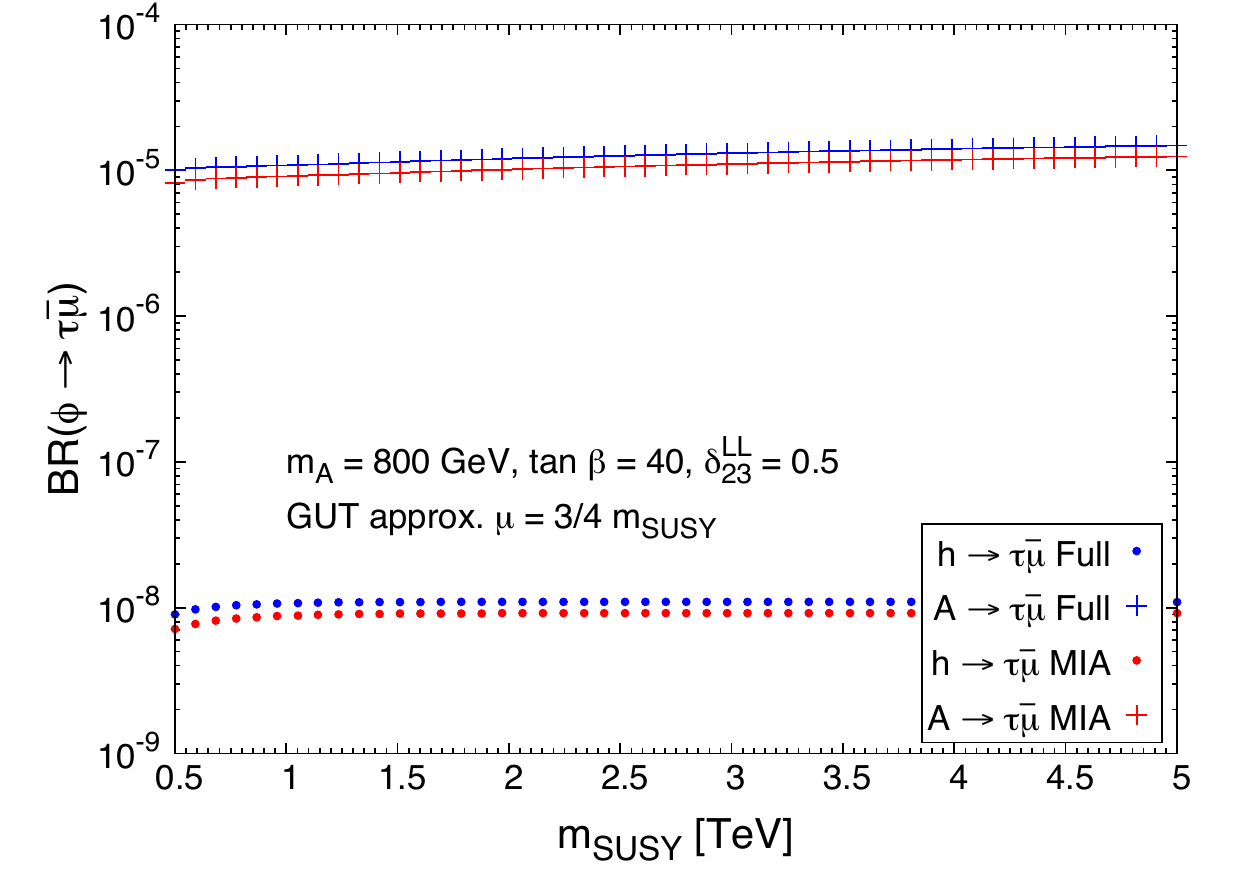} &
\includegraphics[width=0.49\textwidth]{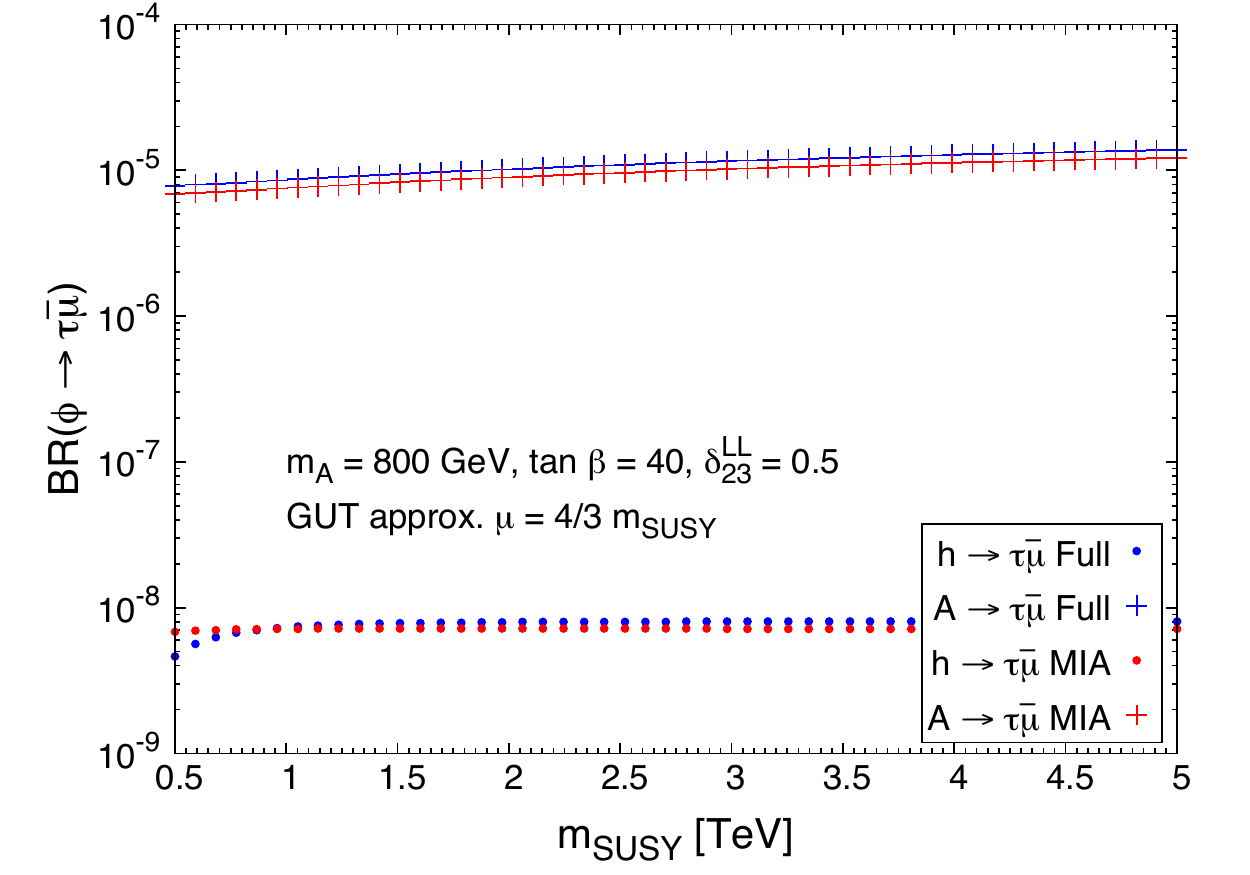} \\
\includegraphics[width=0.49\textwidth]{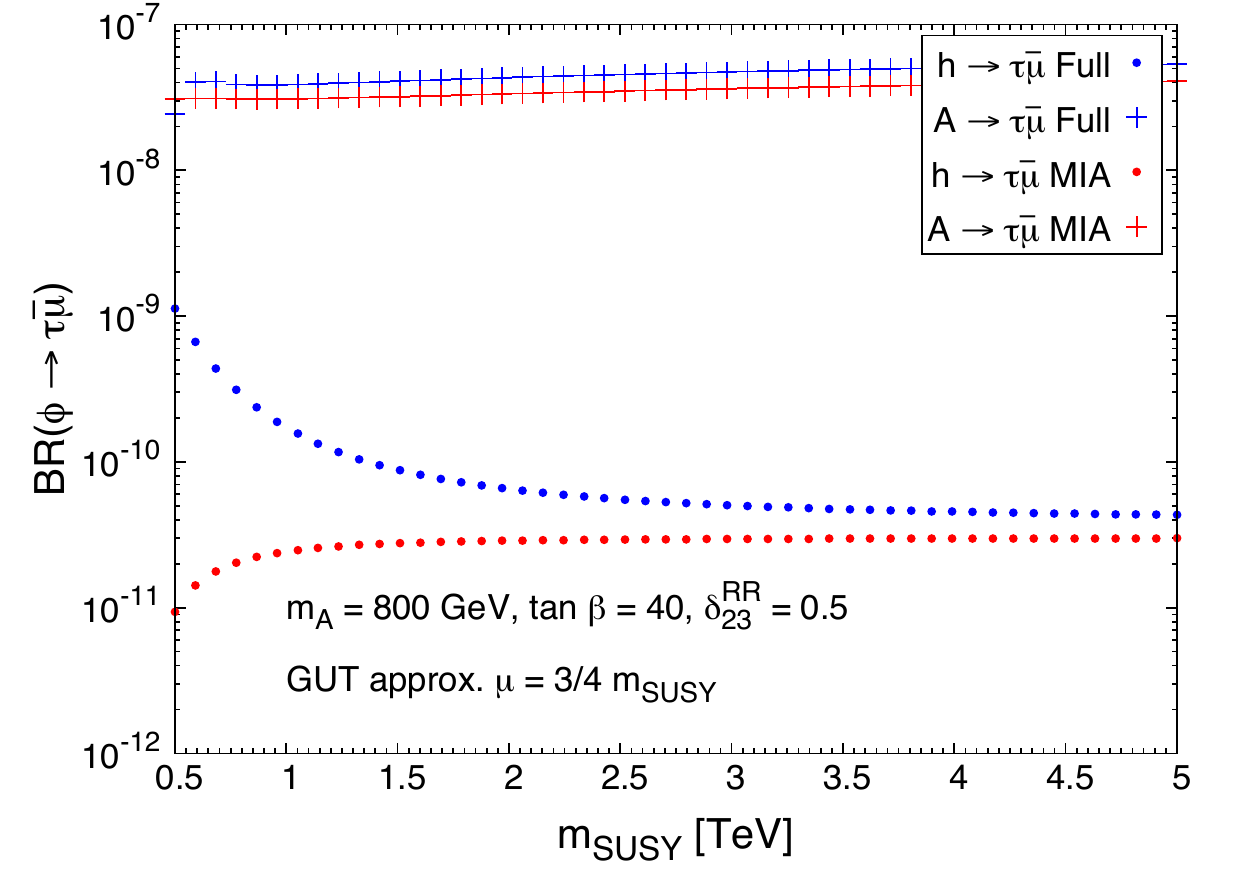} &
\includegraphics[width=0.49\textwidth]{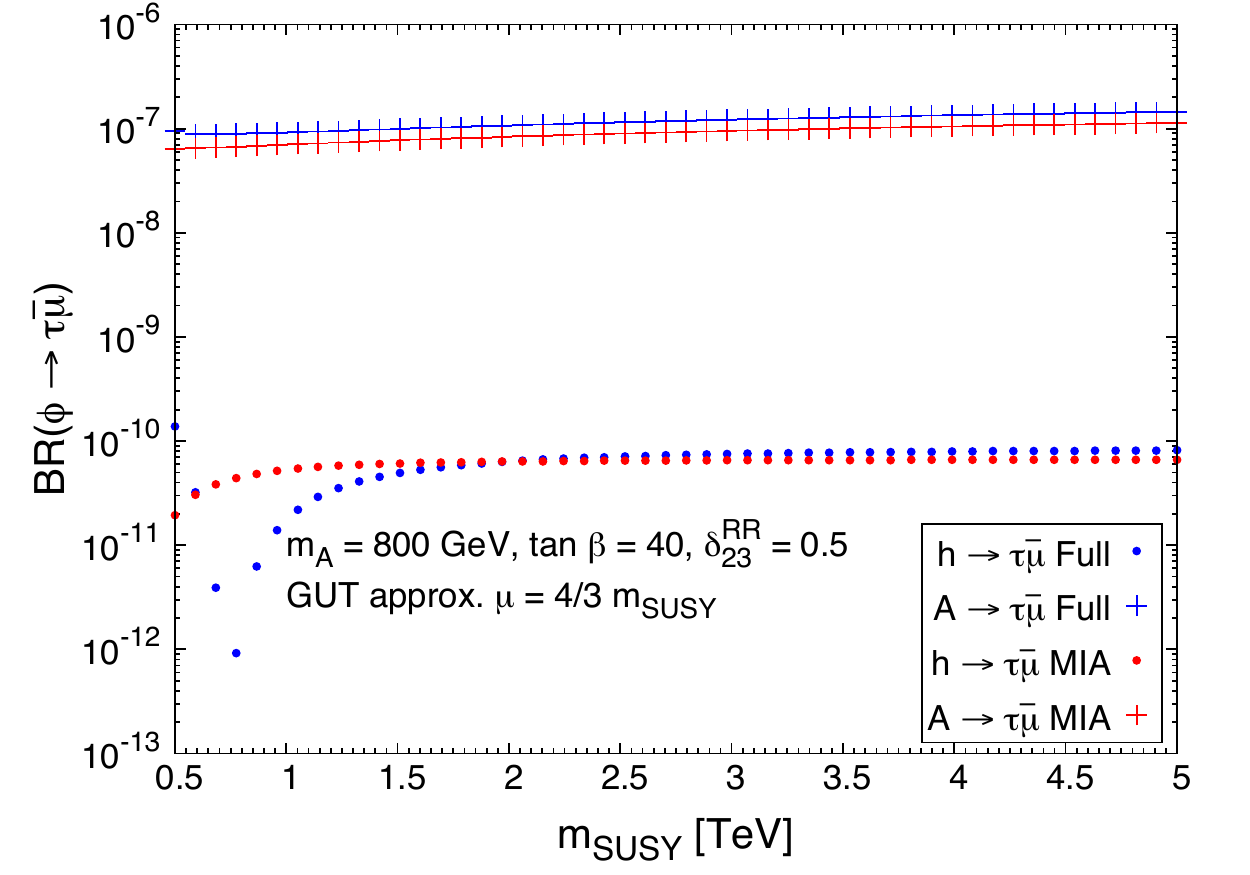} \\
\includegraphics[width=0.49\textwidth]{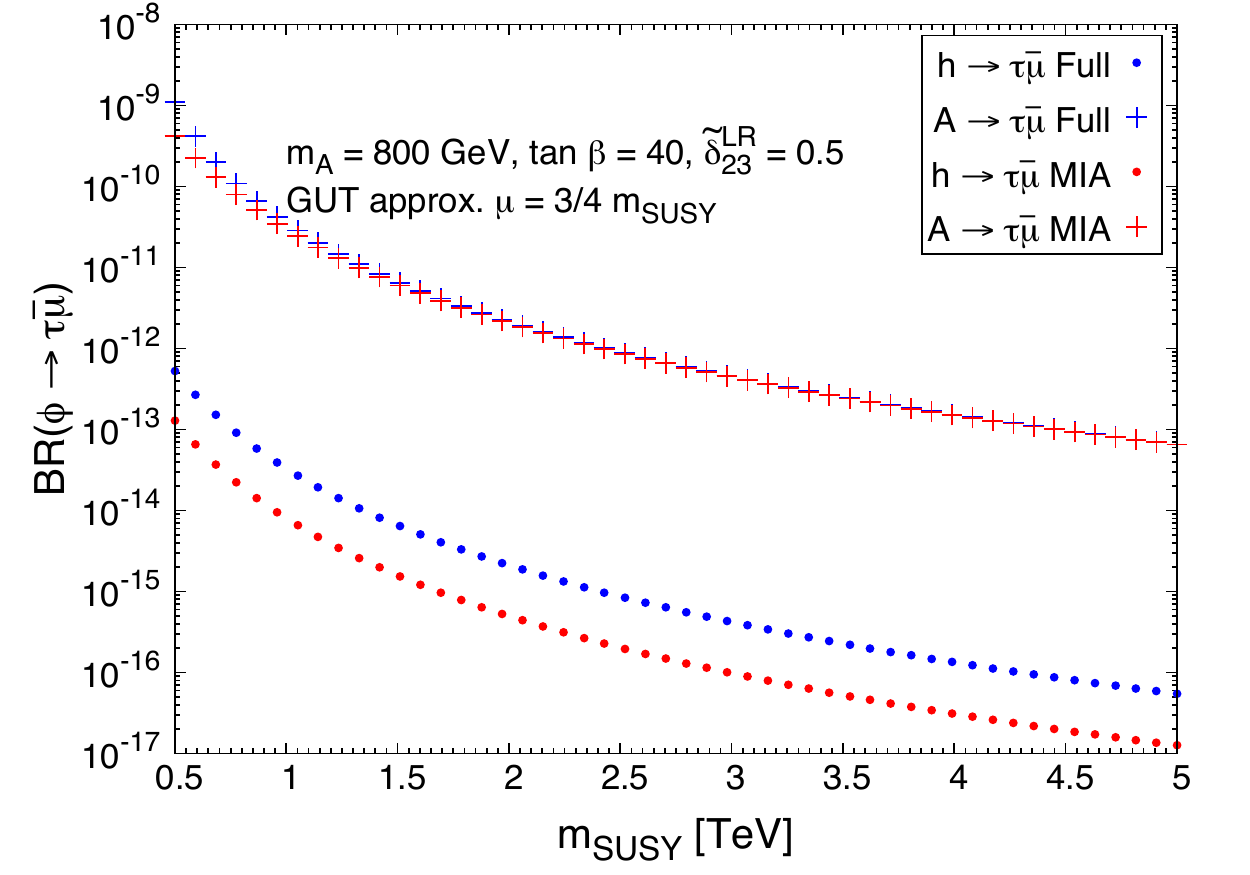} &
\includegraphics[width=0.49\textwidth]{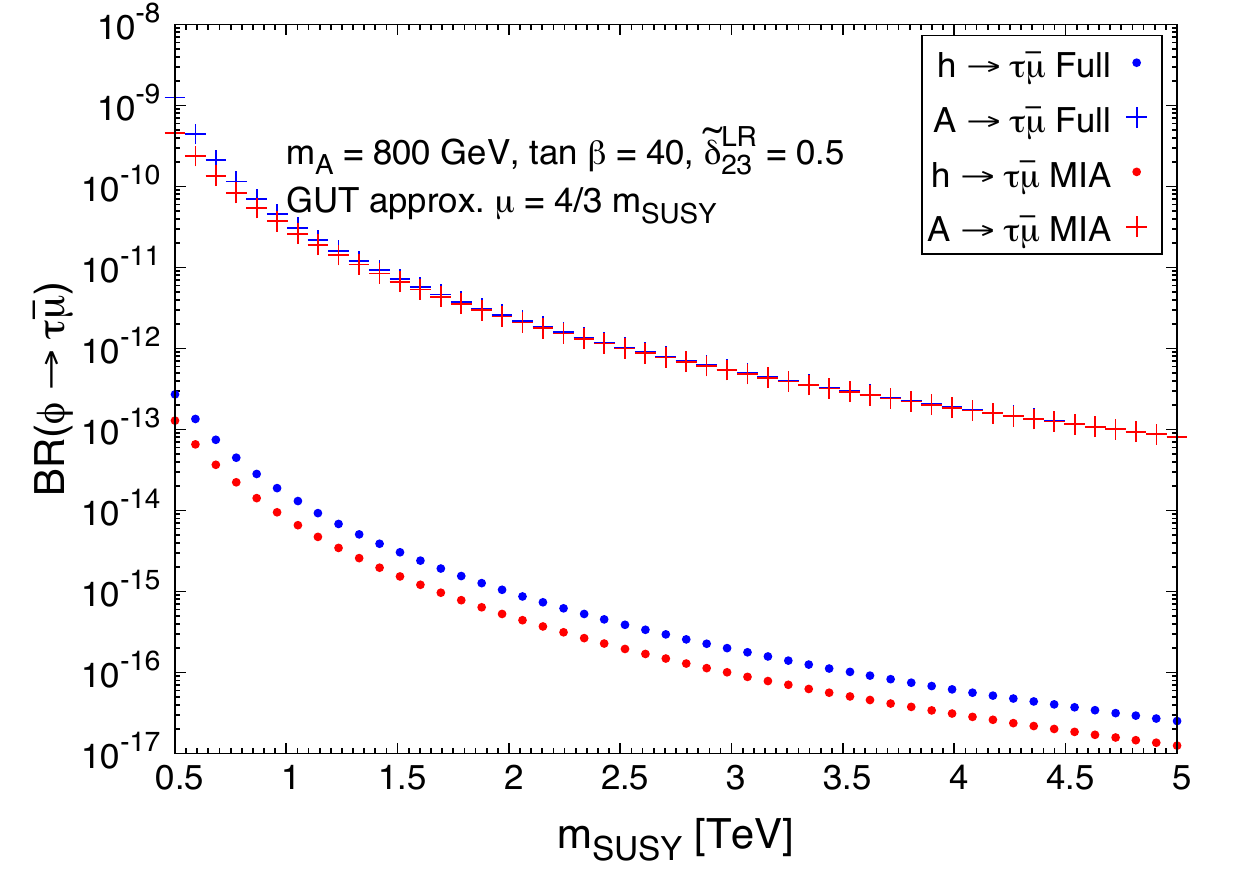}
\end{tabular}
\caption{BR($h \to \tau \bar \mu$) and BR($A \to \tau \bar \mu$) as functions of $m_\text{SUSY}$ in the {\it GUT approximation} scenario with $\mu =3/4 \, m_\text{SUSY}$ (left panels) and $\mu =4/3 \, m_\text{SUSY}$ (right panels), for $\delta_{23}^{LL} = 0.5$ (upper panels), $\delta_{23}^{RR} = 0.5$ (middle panels), and $\tilde \delta_{23}^{LR} = 0.5$ (lower panels). The results for $\tilde \delta_{23}^{RL} = 0.5$ (not shown) are identical to those of $\tilde \delta_{23}^{LR} = 0.5$. In each case, the other flavor changing deltas are set to zero. In all the panels we have set $m_A =$ 800 GeV and $\tan\beta =$ 40. The results for the heavy scalar $H$ (not shown) are nearly equal to these ones for the pseudoscalar $A$.}
\label{LFVHD-mSUSY_GUTapprox}
\end{center}
\end{figure}

In figure~\ref{LFVHD-mSUSY_GUTapprox} the results for BR($h \to \tau \bar \mu$) and BR($A \to \tau \bar \mu$) as functions of $m_\text{SUSY}$ are displayed in the {\it GUT approximation} scenario with $\mu =3/4 \, m_\text{SUSY}$ (left panels) and $\mu =4/3 \, m_\text{SUSY}$ (right panels), for $\delta_{23}^{LL} = 0.5$ (upper panels), $\delta_{23}^{RR} = 0.5$ (middle panels), and $\tilde \delta_{23}^{LR} = 0.5$ (lower panels). In both scenarios we have set $m_A =$ 800 GeV and $\tan\beta =$ 40. The first conclusion from this figure is that the LFVHD rates in this {\it GUT approximation} scenario show again a non-decoupling behavior with $m_\text{SUSY}$ in the $LL$ and $RR$ cases and a decoupling behavior with $m_\text{SUSY}$ in the $LR$ case, as in the {\it Generic} scenario. We also see that the MIA works very well in all the cases $LL$, $RR$, and $LR$ cases, reproducing accurately the results of the full one-loop computation at large $m_{\rm SUSY}$. The only exception is the prediction of BR($h \to \tau \bar \mu$), where we have found some discrepancies between the MIA and the full results in the $RR$ case and also a little one in the $LR$ case, being these differences larger for $\mu= 3/4 \, m_\text{SUSY}$ than for $\mu=4/3 \, m_\text{SUSY}$. We have also detected that these discrepancies are due to the fact that, in the light Higgs boson case, the missing decoupling terms in our MIA computation of the form factors of ${\cal O}(M_W^2/m_{\rm SUSY}^2)$ compete with the leading decoupling terms of ${\cal O}(m_h^2/m_{\rm SUSY}^2)$ and, for some particular cases in which there are strong cancellations among the dominant non-decoupling contributions, they may play some important role in order to obtain a better convergence between the MIA and the full results. We have also checked that this divergence appears more pronounced where there is some degree of degeneracy among the mass parameters, as it happens partially in the {\it GUT approximation} scenario and totally in the {\it Equal masses} one. Indeed, for this latter scenario with $RR$ mixing, we have checked that, by means of an explicit analytic computation in the MIA of these decoupling ${\cal O}(M_W^2/m_{\rm SUSY}^2)$ contributions from the most relevant additional diagrams  (see at the end of Appendix~\ref{AnalyticFormFactors}), we achieve a better convergence between the MIA and the full results. However, we believe that is not worth including those extra terms in our general estimates here, since they are numerically extremely tiny and therefore irrelevant for the associated phenomenology. 
 
\begin{figure}[t!]
\begin{center}
\begin{tabular}{cc}
\includegraphics[width=0.49\textwidth]{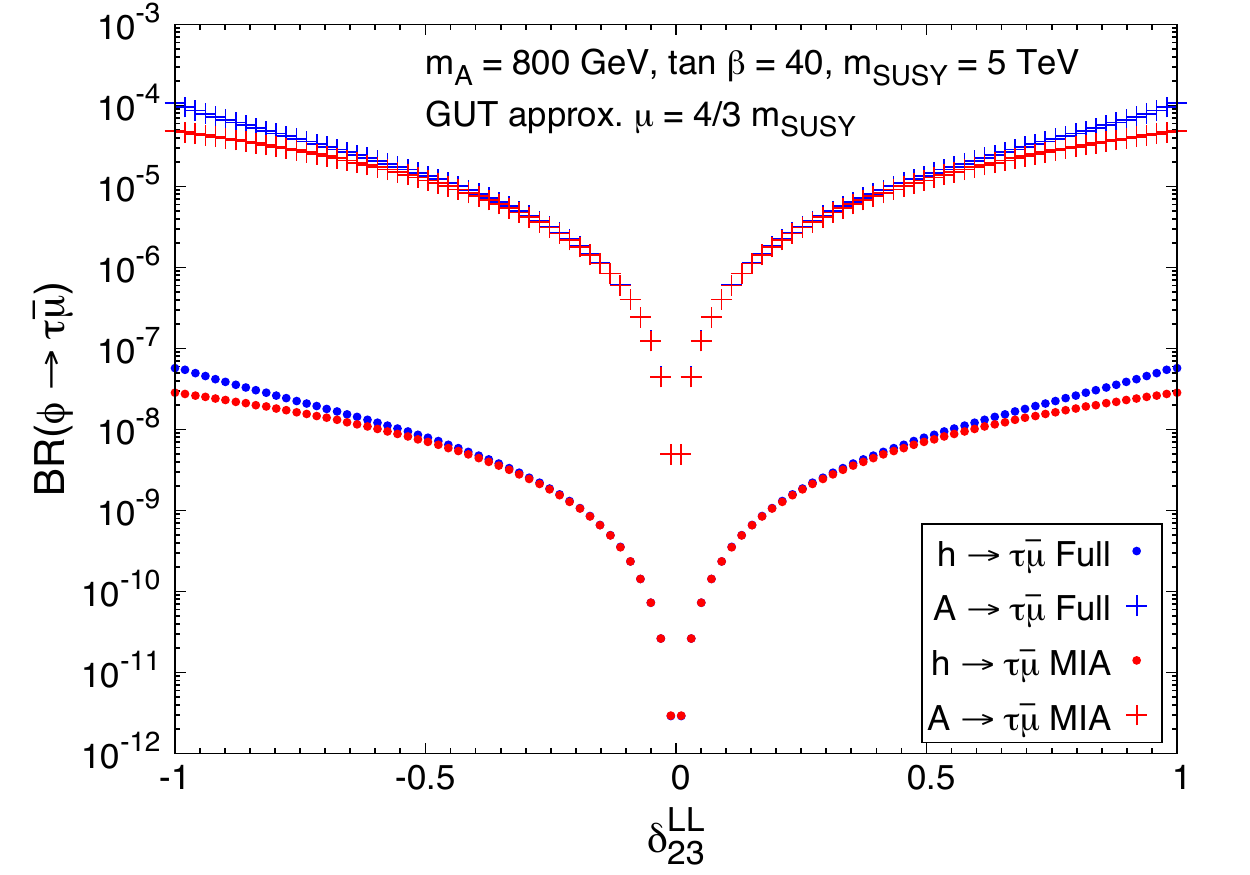} &
\includegraphics[width=0.49\textwidth]{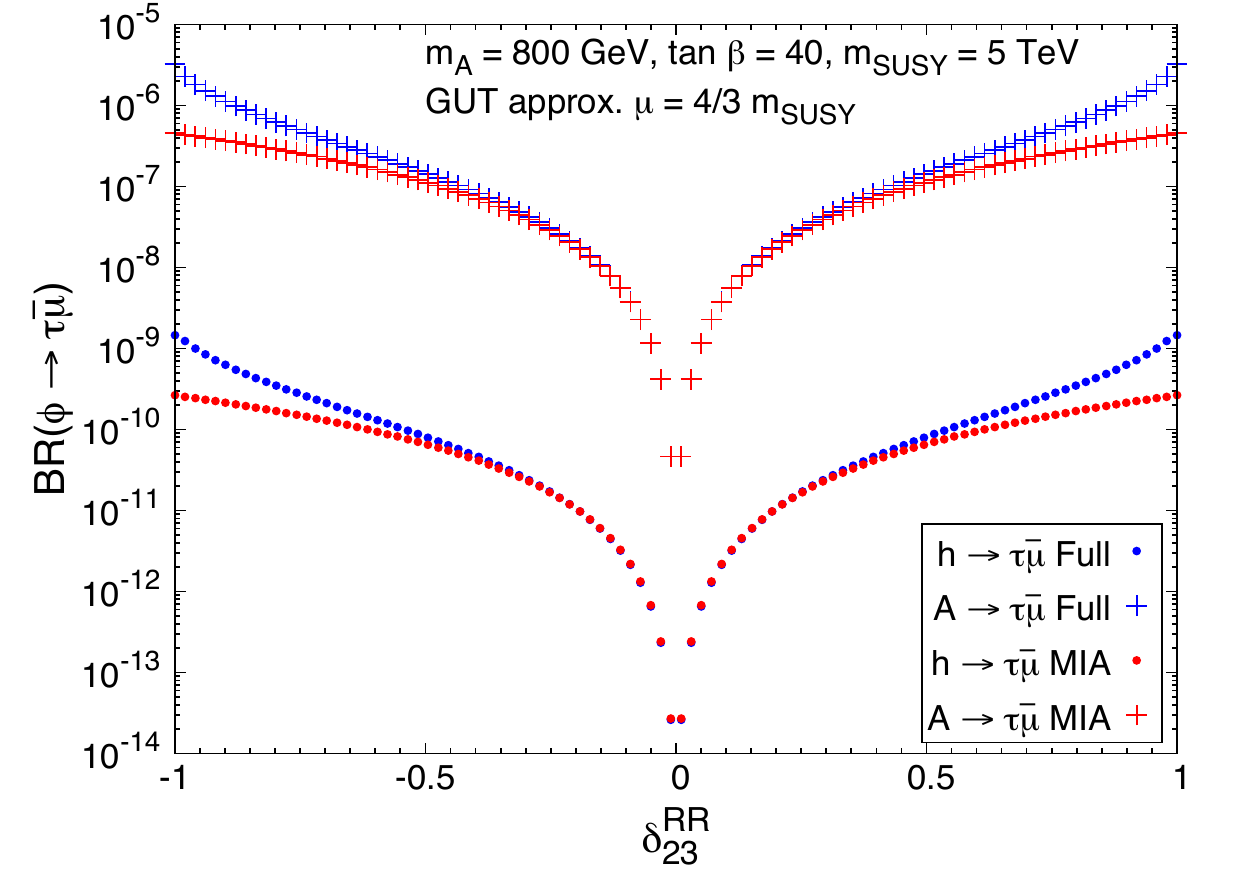} \\
\includegraphics[width=0.49\textwidth]{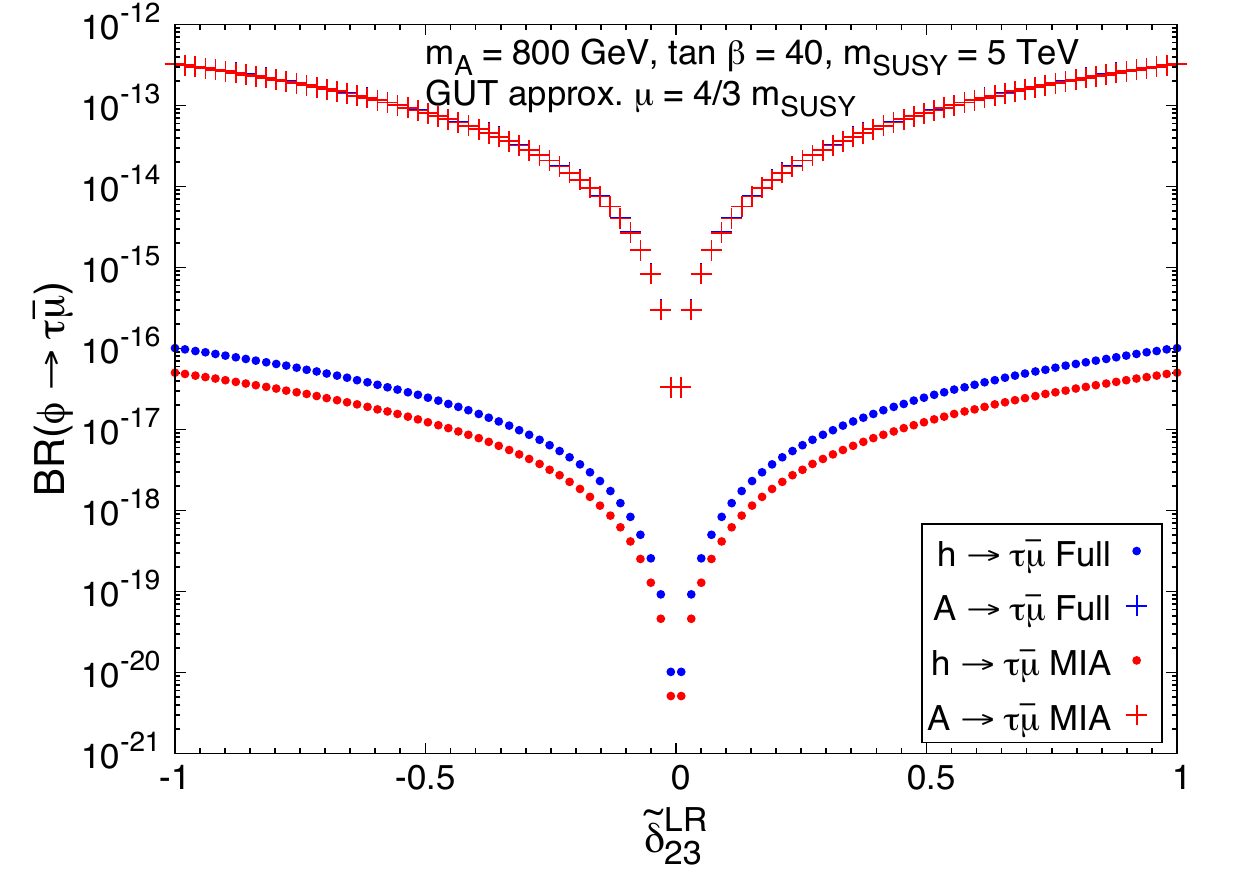} &
\includegraphics[width=0.49\textwidth]{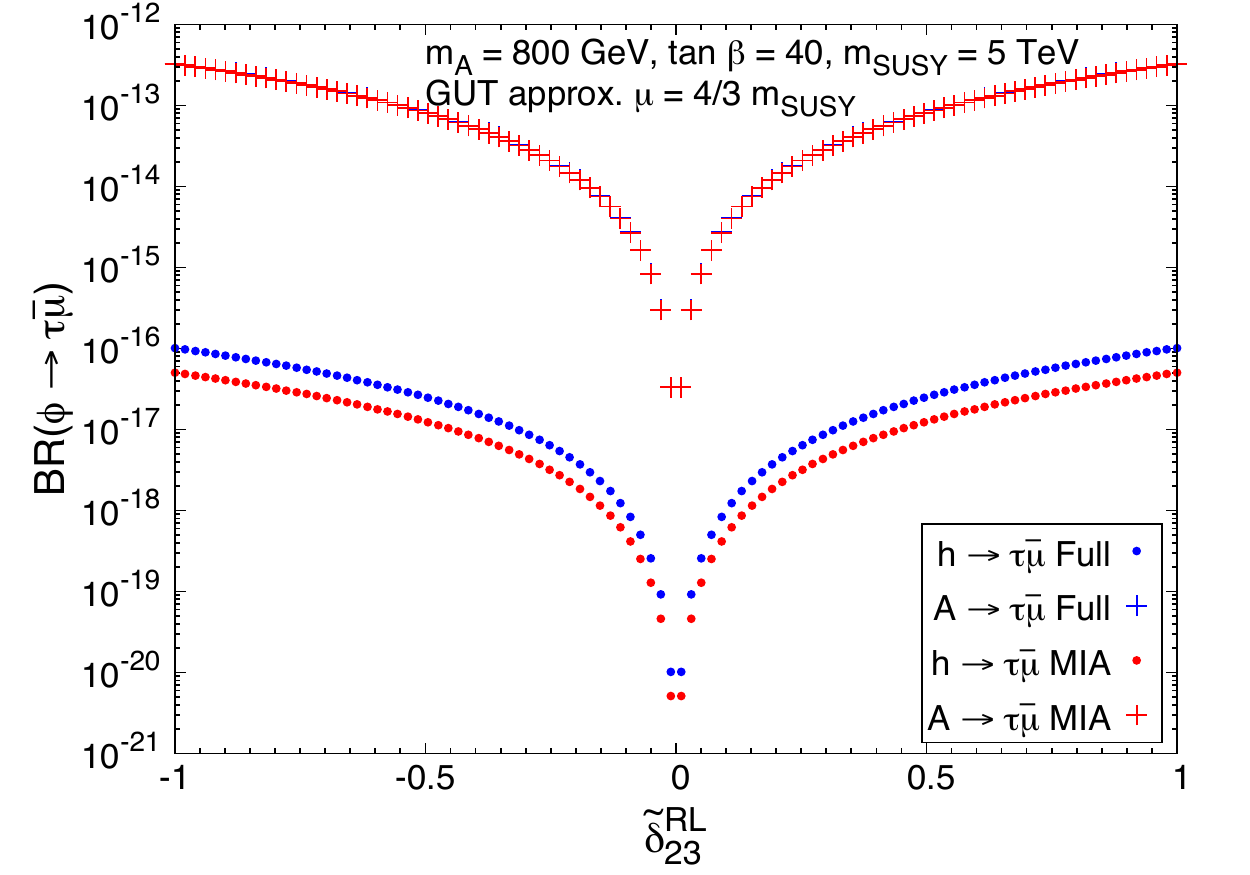}
\end{tabular}
\caption{BR($h \to \tau \bar \mu$) and BR($A \to \tau \bar \mu$) as functions of $\delta_{23}^{LL}$ (upper left panel), $\delta_{23}^{RR}$ (upper right panel), $\tilde \delta_{23}^{LR}$ (lower left panel), and $\tilde \delta_{23}^{RL}$ (lower right panel), in the {\it GUT approximation} scenario with $\mu =4/3 \, m_\text{SUSY}$, $m_\text{SUSY} =$ 5 TeV, $m_A =$ 800 GeV, and $\tan\beta =$ 40. In each case, the other flavor changing deltas are set to zero. The results for the heavy scalar $H$ (not shown) are nearly equal to these ones for the pseudoscalar $A$.
}\label{LFVHD-deltas_GUTapprox}
\end{center}
\end{figure}

Next we study in figure~\ref{LFVHD-deltas_GUTapprox} the dependence of the LFVHD rates on the four flavor changing deltas considered in this work, $\delta_{23}^{LL}$ (upper left panel), $\delta_{23}^{RR}$ (upper right panel), $\tilde \delta_{23}^{LR}$ (lower left panel), and $\tilde \delta_{23}^{RL}$ (lower right panel), within the {\it GUT approximation} scenario with $\mu =4/3 \, m_\text{SUSY}$, $m_\text{SUSY} =$ 5 TeV, $m_A =$ 800 GeV, and $\tan\beta =$ 40. First of all, it is clear that the behaviors of the branching ratios are symmetric with respect to positive and negative values of the deltas and we observe the expected increase of the LFVHD rates in the MIA with each delta, as $|\delta_{23}^{XY}|^2$. On the upper panels we observe a very good agreement between the MIA and the full one-loop results for BR($h \to \tau \bar \mu$) and BR($A \to \tau \bar \mu$) in the $LL$ and $RR$ cases, up to values of $|\delta_{23}^{LL,RR}| \simeq$ 0.6. From this value, the predictions of the full results start to separate from the MIA ones, showing the expected departure from the quadratic ${\cal O}(\delta^2)$ dependence. Anyway, the discrepancy between the MIA and the full calculation is not large, at the most of a factor of 3 for $|\delta_{23}^{LL}| =$ 1 and of 6 for $|\delta_{23}^{RR}| =$ 1. The results of the $LR$ and $RL$ cases are identical and we comment together. The full/MIA agreement for the $A \to \tau \bar \mu$ rates is almost exact and the predictions of both calculations do not separate for values of $\tilde \delta_{23}^{LR(RL)}$ close to 1, since they are still perturbative (remember eqs.~(\ref{DeltaLR}) and~(\ref{DeltaRL})). Again, the observed small discrepancies in BR($h \to \tau \bar \mu$) between the MIA and the full results are due to the missing subdominant decoupling contributions of ${\cal O}(M_W^2/m_{\rm SUSY}^2)$ in our MIA calculation.

\begin{figure}[t!]
\begin{center}
\begin{tabular}{cc}
\includegraphics[width=0.49\textwidth]{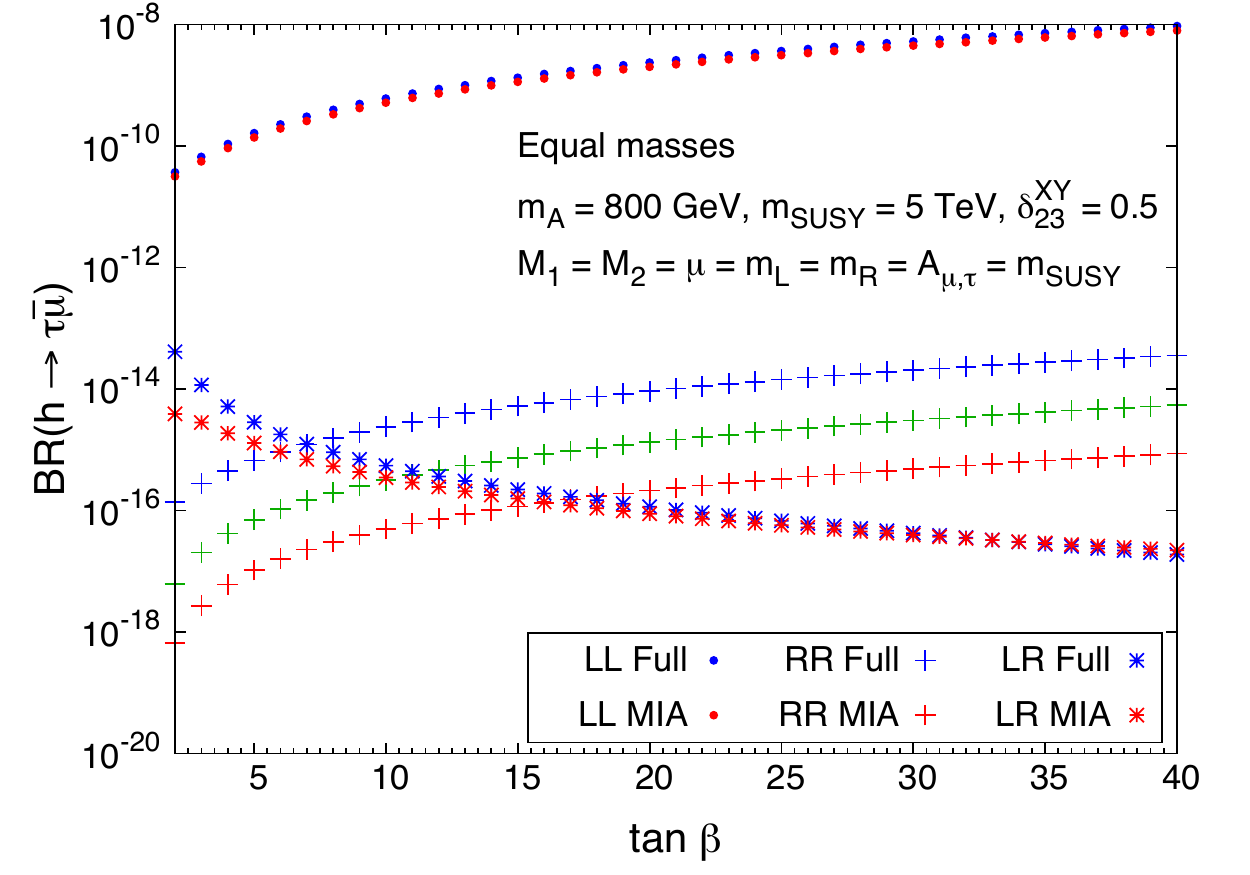} &
\includegraphics[width=0.49\textwidth]{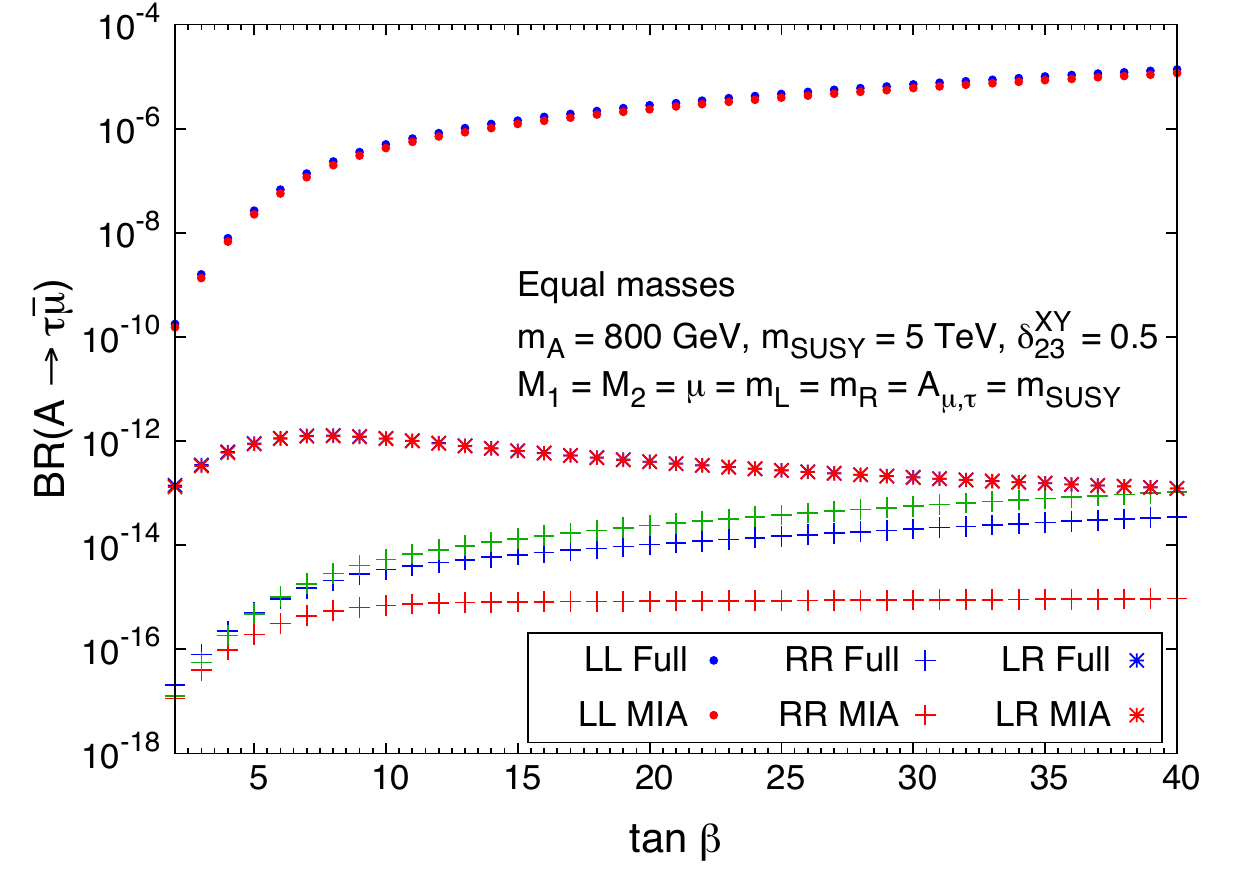}
\end{tabular}
\caption{BR($h \to \tau \bar \mu$) (left panel) and BR($A \to \tau \bar \mu$) (right panel) as functions of $\tan\beta$ in the {\it Equal masses} scenario with $m_\text{SUSY} =$ 5 TeV, $m_A =$ 800 GeV, and $\delta_{23}^{XY} =$ 0.5, with $XY =$ $LL$, $RR$, $LR$  ($\tilde \delta$, for the latter), in each case.  The green crosses are the MIA predictions in the $RR$ case after including the ${\cal O}(M_W^2/m_\text{SUSY}^2)$ corrections. The results for the heavy scalar $H$ (not shown) are nearly equal to these ones for the pseudoscalar $A$.
}\label{LFVHD-tanb_eqM}
\end{center}
\end{figure}

The dependence of the LFVHD rates as functions of $\tan\beta$ is depicted in figure~\ref{LFVHD-tanb_eqM} within the {\it Equal masses} scenario with $m_\text{SUSY} =$ 5 TeV, $m_A =$ 800 GeV, and $\delta_{23}^{XY} =$ 0.5, with $XY =$ $LL$, $RR$, $LR$, in each case. The full/MIA agreement in the $LL$ and $LR$ cases is very accurate for both LFVHD branching ratios, BR($h \to \tau \bar \mu$) and BR($A \to \tau \bar \mu$), while there is an appreciable disagreement for the $RR$ predictions, of up to two orders of magnitude. The main reason to explain these discrepancies is that in this {\it Equal masses} scenario the cancellation among diagrams is even stronger than in the previous ones, since all of the SUSY mass parameters are identical. This strong cancellation makes that the non-decoupling dominant terms of all the diagrams completely cancel. The remnant terms in the form factors proportional to $(m_{H_x}/m_\text{SUSY})^2$ are not sufficient to reproduce the full one-loop results and then, to obtain a better convergence in this $RR$ case, one should include the MIA subdominant decoupling contributions, proportional to $(M_W/m_\text{SUSY})^2$.  In order to check this expected better convergence, we have computed the most relevant diagrams providing the most important ${\cal O}(M_W^2/m_\text{SUSY}^2)$ corrections in the MIA for this particular $RR$ case in the {\it Equal masses} scenario. We include these analytic results at the end of Appendix~\ref{AnalyticFormFactors}. Our numerical estimates of the LFVHD rates for this $RR$ case after including these 
additional ${\cal O}(M_W^2/m_\text{SUSY}^2)$ corrections are also displayed (in green) in figure~\ref{LFVHD-tanb_eqM}, for comparison. We can clearly see that there is indeed a better convergence to the full result. However, as we have already said in all those cases where the disagreement MIA/full is clearly manifest, the predicted rates are very tiny and irrelevant for phenomenological purposes.

On the other hand, the different behaviors with $\tan\beta$ of the full LFVHD rates depending on each delta are well reproduced by the MIA predictions. They can be understood, in the case of generic SUSY masses,  from eqs.~(\ref{h-LLpowercounting})-(\ref{HA-RRpowercounting}) and~(\ref{h-LRpowercounting})-(\ref{HA-LRpowercounting}), and in the case of equal SUSY masses from eqs.~(\ref{hatFLLL})-(\ref{hatFRRR}) and~(\ref{FRRReqmassesMW2})-(\ref{FRRReqmassesMW2largetanb}),
and knowing that, at large $\tan\beta$, the total Higgs decay widths go as $\Gamma_\text{tot}(H, A) \sim$ $(\tan\beta)^2$ and $\Gamma_\text{tot}(h)$ is approximately constant with $\tan\beta$. The partial widths of the $h \to \tau \bar \mu$ decay in the $LL$ and $RR$ cases, for generic SUSY masses, go as $(\tan\beta)^2$ and the $H, A \to \tau \bar \mu$ decay widths are proportional to $(\tan\beta)^4$, therefore all the corresponding branching ratios grow as $(\tan\beta)^2$. By contrast, in the $LR$ case, $\Gamma(h \to \tau \bar \mu) \sim$ $(\tan\beta)^{-2}$ and $\Gamma(H, A \to \tau \bar \mu)$ are independent of $\tan\beta$, thus BR($h, H, A \to \tau \bar \mu$) $\sim (\tan\beta)^{-2}$.

\begin{figure}[t!]
\begin{center}
\begin{tabular}{cc}
\includegraphics[width=0.48\textwidth]{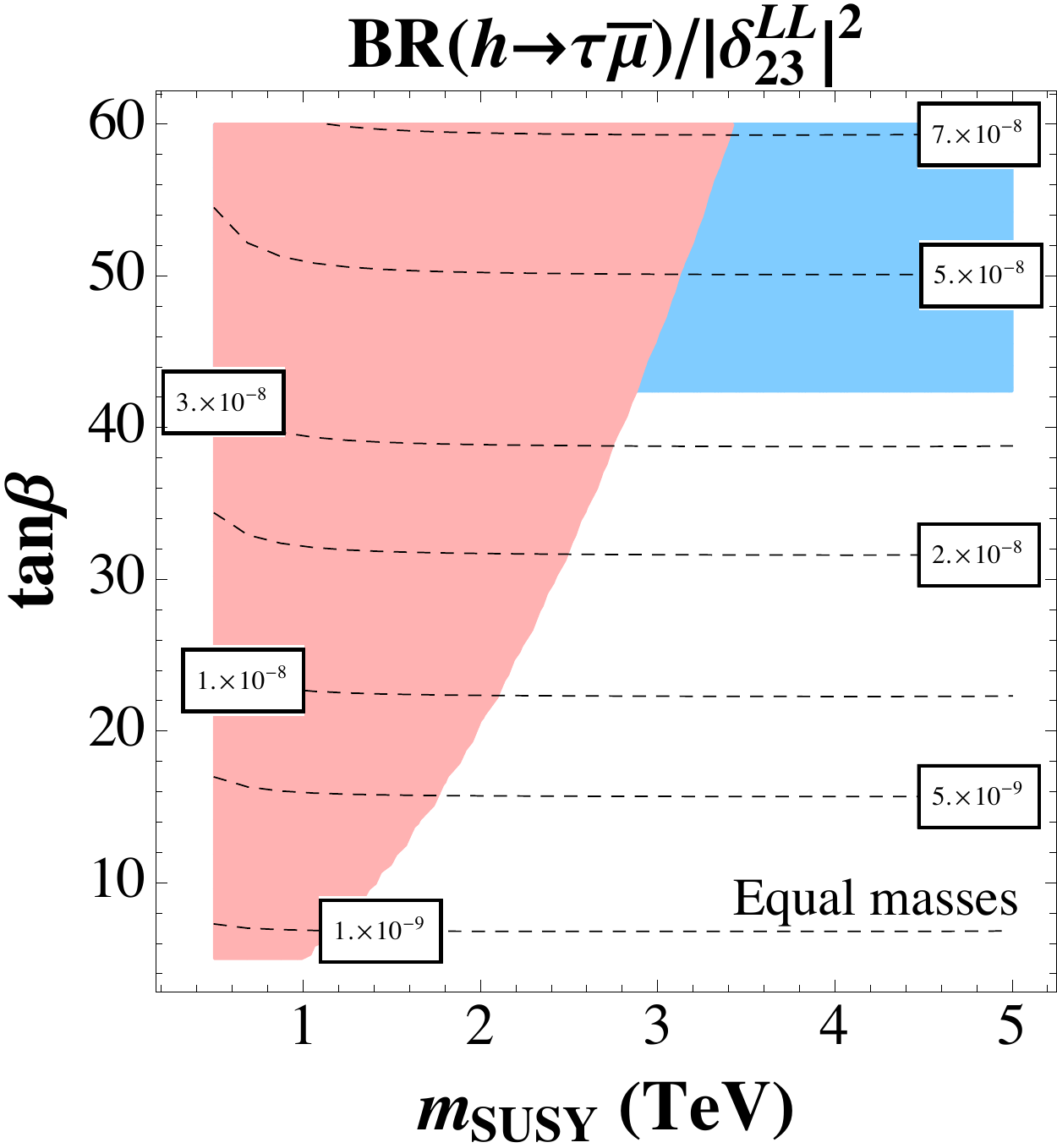} &
\includegraphics[width=0.48\textwidth]{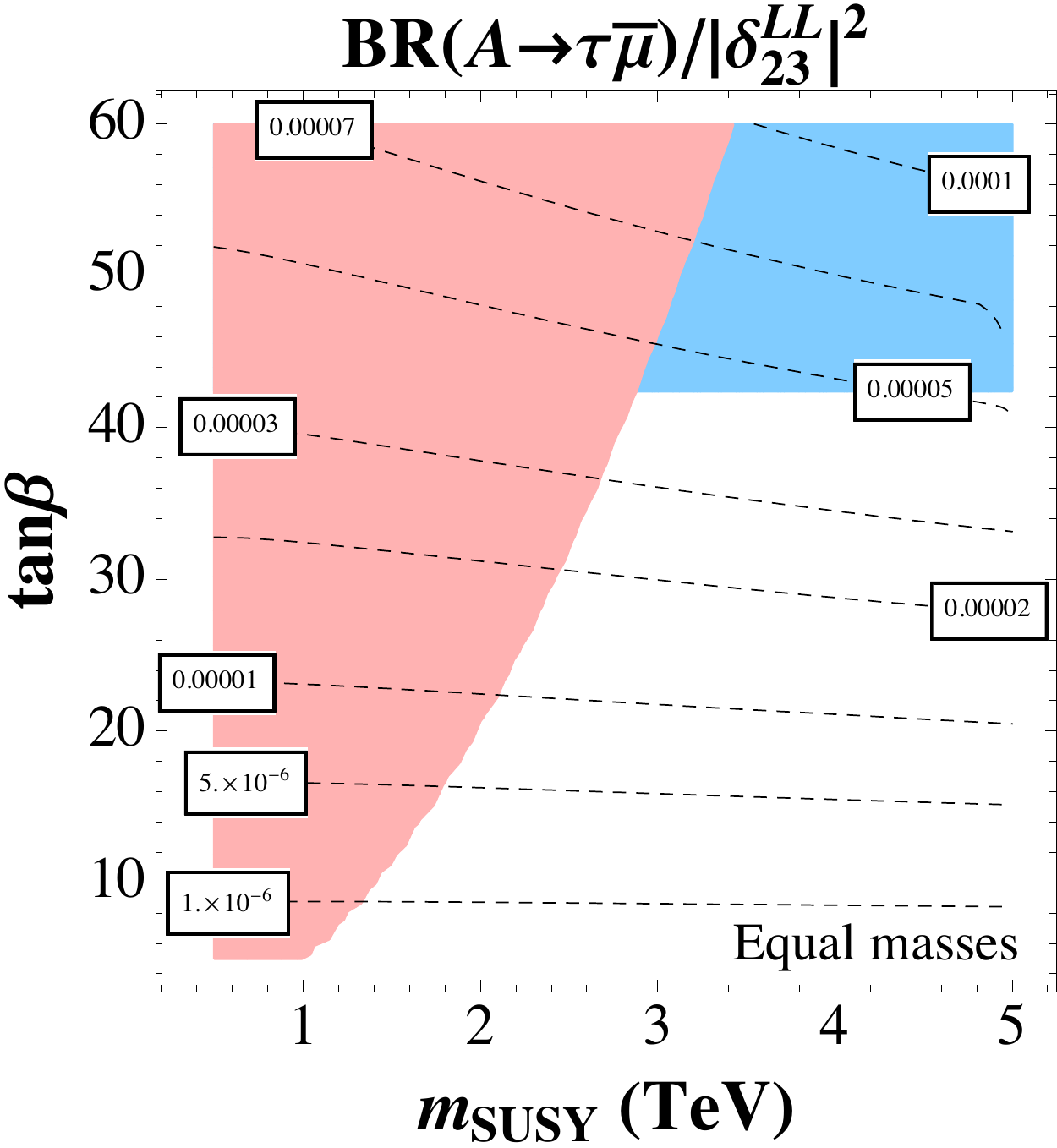}
\end{tabular}
\caption{Contour lines of BR($h \to \tau \bar \mu$)/$|\delta_{23}^{LL}|^2$ (left panel) and BR($A \to \tau \bar \mu$)/$|\delta_{23}^{LL}|^2$ (right panel) in the $[m_\text{SUSY}, \tan\beta]$ plane within the {\it Equal masses} scenario with $m_A =$ 800 GeV. The shaded red area is excluded by the current experimental upper limit for $\tau \to \mu \gamma$ channel, BR($\tau \to \mu \gamma$) $< 4.4 \times 10^{-8}$~\cite{Aubert:2009ag}. The shaded blue area represents the 95\% C.L. excluded regions by the negative searches by ATLAS and CMS for neutral MSSM Higgs bosons decaying to a pair of $\tau$ leptons~\cite{Khachatryan:2014wca,Aad:2014vgg}. The results for the heavy scalar $H$ (not shown) are nearly equal to these ones for the pseudoscalar $A$.
}\label{LFVHDdelta23LL2-mSUSYtanb_eqM}
\end{center}
\end{figure}

From figures~\ref{LFVHD-mSUSY_generic}-\ref{LFVHD-tanb_eqM} we learn that the only delta that can lead us to phenomenologically interesting LFVHD rates is $\delta_{23}^{LL}$. In order to try to find the largest LFV Higgs branching ratios, we are going to investigate the quantities BR($h \to \tau \bar \mu$)/$|\delta_{23}^{LL}|^2$ and BR($H, A \to \tau \bar \mu$)/$|\delta_{23}^{LL}|^2$ that are delta-independent when computed with the MIA. First, the contour lines of these two observables in the $[m_\text{SUSY}, \tan\beta]$ plane are displayed in figure~\ref{LFVHDdelta23LL2-mSUSYtanb_eqM}, within the {\it Equal masses} scenario with $m_A =$ 800 GeV. In both contour plots, the shaded red area is excluded by the current experimental upper limit for $\tau \to \mu \gamma$ channel, BR($\tau \to \mu \gamma$) $< 4.4 \times 10^{-8}$~\cite{Aubert:2009ag}, and the shaded blue area represents the 95\% C.L. excluded regions by the negative searches by ATLAS and CMS for neutral MSSM Higgs bosons decaying to a pair of $\tau$ leptons~\cite{Khachatryan:2014wca,Aad:2014vgg}. It is clear again the non-decoupling behavior with $m_\text{SUSY}$ of the LFVHD rates and their growth with $\tan\beta$. The largest values obtained for BR($h \to \tau \bar \mu$)/$|\delta_{23}^{LL}|^2$ and BR($H, A \to \tau \bar \mu$)/$|\delta_{23}^{LL}|^2$ are $7 \times 10^{-8}$ and $1 \times 10^{-4}$, respectively, but unfortunately they are excluded by the $\tau \to \mu \gamma$ upper limit and/or the ATLAS and CMS searches for MSSM Higgs bosons. The maximum values for these delta-independent rates, allowed by data, are BR($h \to \tau \bar \mu$)/$|\delta_{23}^{LL}|^2$ $\sim$ $3 \times 10^{-8}$ and BR($H, A \to \tau \bar \mu$)/$|\delta_{23}^{LL}|^2$ $\sim$ $5 \times 10^{-5}$, very far away both from the current LHC sensitivity to these LFV processes~\cite{Khachatryan:2015kon,Aad:2015gha}.

\begin{figure}[t!]
\begin{center}
\begin{tabular}{cc}
\includegraphics[width=0.48\textwidth]{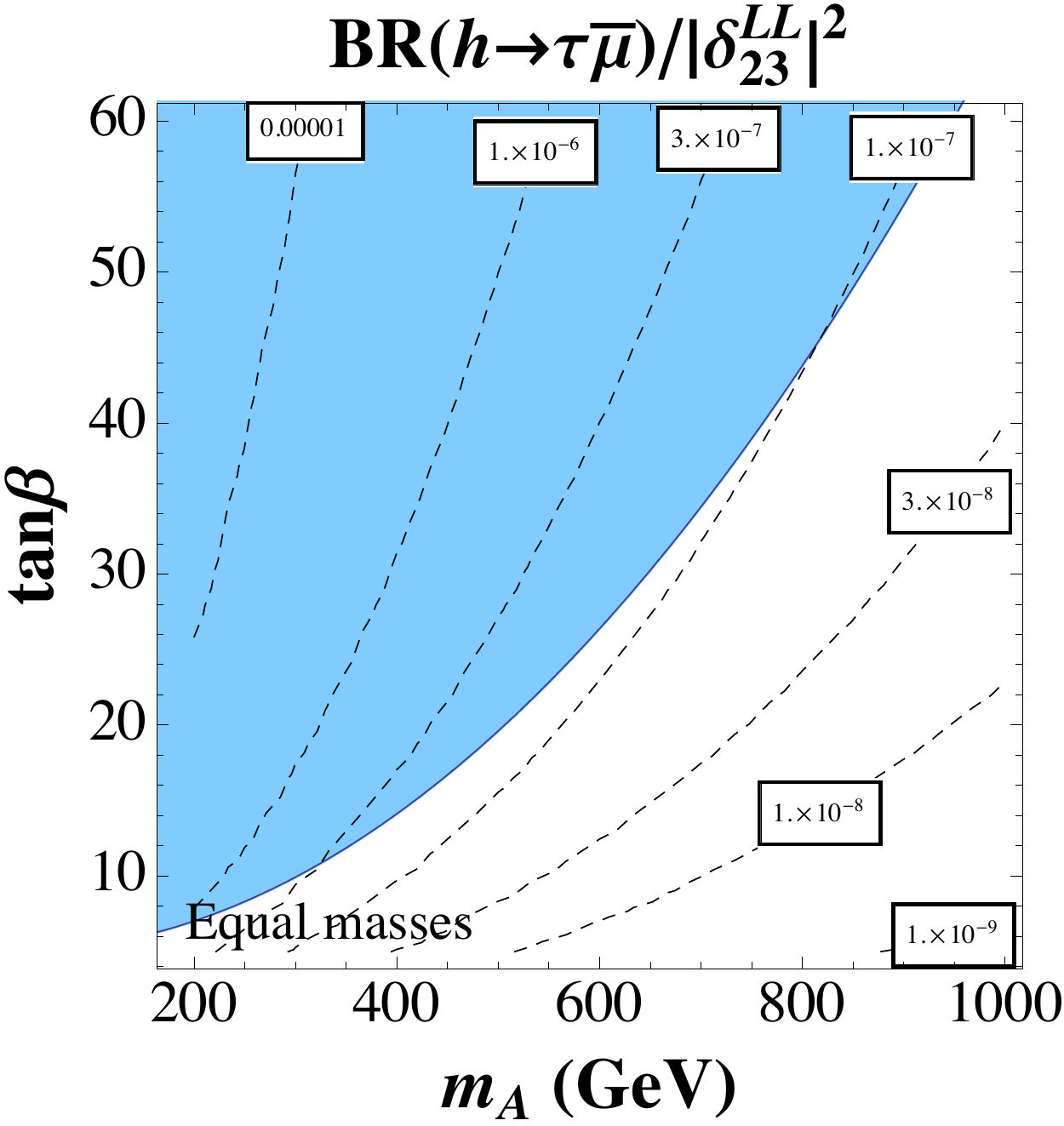} &
\includegraphics[width=0.48\textwidth]{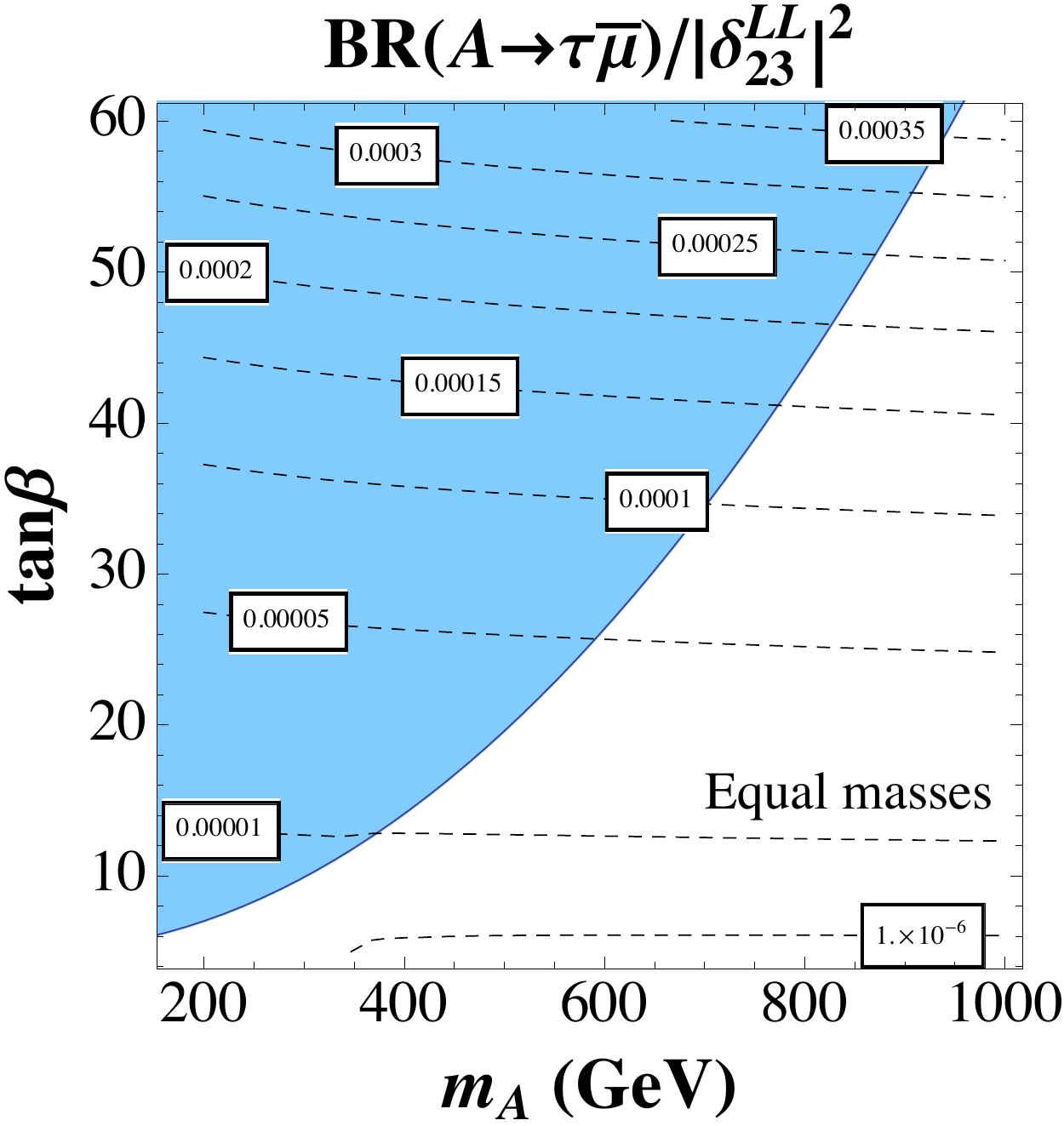}
\end{tabular}
\caption{Contour lines of BR($h \to \tau \bar \mu$)/$|\delta_{23}^{LL}|^2$ (left panel) and BR($A \to \tau \bar \mu$)/$|\delta_{23}^{LL}|^2$ (right panel) in the $[m_A, \tan\beta]$ plane within the {\it Equal masses} scenario with $m_\text{SUSY} =$ 4 TeV. The shaded blue area represents the 95\% C.L. excluded regions by the negative searches by ATLAS and CMS for neutral MSSM Higgs bosons decaying to a pair of $\tau$ leptons~\cite{Khachatryan:2014wca,Aad:2014vgg}. The results for the heavy scalar $H$ (not shown) are nearly equal to these ones for the pseudoscalar $A$.
}\label{LFVHDdelta23LL2-mAtanb_eqM}
\end{center}
\end{figure}

Finally, we show in figure~\ref{LFVHDdelta23LL2-mAtanb_eqM} the contour lines of BR($h \to \tau \bar \mu$)/$|\delta_{23}^{LL}|^2$ (left panel) and BR($H, A \to \tau \bar \mu$)/$|\delta_{23}^{LL}|^2$ (right panel) in the $[m_A, \tan\beta]$ plane predicted in the MIA within the {\it Equal masses} scenario with $m_\text{SUSY} =$ 4 TeV, being the shaded blue area the 95\% C.L. excluded regions by the negative searches by ATLAS and CMS for neutral MSSM Higgs bosons decaying to a pair of $\tau$ leptons~\cite{Khachatryan:2014wca,Aad:2014vgg}. The fact of fixing $m_\text{SUSY} =$ 4 TeV ensures us that the predictions are in agreement with the $\tau \to \mu \gamma$ upper limit, as can be inferred from figure~\ref{LFVHDdelta23LL2-mSUSYtanb_eqM}. In this case, the known decoupling behavior of BR($h \to \tau \bar \mu$) in the large $m_A$ limit is manifest on the left panel. The largest value for BR($h \to \tau \bar \mu$)/$|\delta_{23}^{LL}|^2$ is $1 \times 10^{-5}$, however it is again excluded by the ATLAS and CMS searches for neutral MSSM Higgs bosons. The largest $h \to \tau \bar \mu$ rates allowed by data are of ${\cal O} (10^{-7})$, out of the reach of the present and next future LHC experiments. Fortunately, the prospects for $H \to \tau \bar \mu$ and $A \to \tau \bar \mu$ are much more promising, as we can see on the right panel of figure~\ref{LFVHDdelta23LL2-mAtanb_eqM}. The MIA predictions for BR($H, A \to \tau \bar \mu$)/$|\delta_{23}^{LL}|^2$ are practically independent on $m_A$ and increase quadratically with $\tan\beta$ as expected. It reaches values, allowed by data, up to $3.5 \times 10^{-4}$ for large $m_A$ and $\tan\beta$, not very far from the current LHC sensitivity. It is important to mention that our predictions of the LFVHD rates are identical for the $\tau \bar \mu$ and $\bar \tau \mu$ final states, since we are assuming real $\delta^{LL}_{23}$, and in order to compare our results with the ATLAS and CMS reported data, we have to multiply our rates by a factor of 2. Our maximum branching ratio is then of ${\cal O}(10^{-3})$, only one order of magnitude lower than the current percent-level sensitivity achieved at the LHC~\cite{Khachatryan:2015kon,Aad:2015gha}.

\section{Conclusions}
\label{conclusions}

In this work we have analyzed in full detail, both analytically and numerically, the decay rates of the neutral MSSM Higgs bosons into a lepton and an anti-lepton with different flavor: $h,A,H \to l_k \bar l_m$ ($m \neq k$). Our computation of the LFV partial widths $\Gamma(h,A,H \to l_k \bar l_m)$ is a one-loop diagrammatic one, but different to previous analytic computations in the literature. Here it has been performed by the first time using the simple approximation provided by the MIA, which works with the electroweak interaction slepton and sneutrino eigenstates, $\tilde l^{L,R}_{i}$ and $\tilde \nu^L_i$, with $i=1,2,3$, and treats perturbatively the mass insertions changing lepton flavor, $\Delta^{AB}_{ij}$ with $AB=LL,LR,RL,RR$ and $i \neq j$. By using the MIA at the first order in the dimensionless parameters expansion $\delta^{AB}_{ij}$, we have found compact analytic results for all the form factors involved in the LFVHD amplitudes in terms of the well known 3- and 4-point scalar one-loop integrals, and the relevant MSSM parameters, namely, the soft masses $m_{\widetilde{L}_{i}}$, $m_{\widetilde{R}_{i}}$, $M_1$, and $M_2$, the Higgs sector input mass $m_A$, $\tan \beta$, and the $\mu$ parameter. Then, by performing an expansion of the loop integrals in powers of the external momenta and keeping just the leading and next-to-leading terms, we have been able to find a set of simple analytic formulas, both for each contributing diagram and for the total sum, with all the relevant contributions explicit. These relevant contributions consist of two qualitative different parts that we have analyzed and presented separately: The leading non-decoupling contributions of ${\cal O} ((m_{h,H,A}/m_{\rm SUSY})^0)$ that tend to a constant value for asymptotically large  $m_{\rm SUSY}$, and the next-to-leading decoupling contributions of  ${\cal O} (m_{h,H,A}^2/m_{\rm SUSY}^2)$.
At this point, we would like to emphasize that an alternative analytic computation to ours could be done by starting instead with the full analytic results of the form factors of~\cite{Arganda:2004bz}, given in terms of the physical sparticle masses and rotation matrices, then performing a Taylor expansion in powers of $\Delta^{AB}_{mk}$ and keeping the first order in this expansion. However, this is not an easy task since such a computation would involve a systematic Taylor expansion of all the physical slepton masses and rotation matrices elements, keeping all the relevant terms that will contribute to ${\cal O}(\Delta^{AB}_{mk})$ in the form factors, and expressing them in terms of the EW basis parameters like the soft masses, etc. This kind of computation has not been completed yet, to our knowledge, for the LFV form factors of the three neutral Higgs bosons to a comparable level of our  MIA computation, i.e. dealing with all the four slepton mixing cases $LL$, $LR$, $RL$, and $RR$, and keeping in the final results both the leading non-decoupling contributions of ${\cal O} ((m_{h,H,A}/m_{\rm SUSY})^0)$ and the next-to-leading decoupling contributions of  ${\cal O} (m_{h,H,A}^2/m_{\rm SUSY}^2)$.

We have also analyzed numerically the MIA results for the most interesting case of $h,H$, and $A$ decays into $\tau$ and $\mu$ leptons. After an exhaustive comparison with the full one-loop results, we have concluded that the MIA provides indeed quite accurate predictions for the explored mixing parameters range, $|\delta^{AB}_{23}|<1$. We have detected only a few cases, for specific choices of the model parameters, in which there occur strong cancellations among contributing diagrams, mainly due to some degree of degeneracy in the mass parameters, where the MIA does not provide a good result as compared to the full one-loop computation. This happens for instance in the  case of the {\it Equal masses} scenario with the non-vanishing flavor mixing input given by $\delta^{RR}_{23}$. In this case, we have checked by an explicit computation that to achieve a better convergence of the MIA with the full results one must include in addition the next-to-leading decoupling contributions of ${\cal O} (M_W^2/m_{\rm SUSY}^2)$ which we have not taken into account generically in this work. Nevertheless, we wish to emphasize that this detected mismatch MIA/full is not important at all for phenomenological purposes since the predicted rates in those cases are very tiny and therefore irrelevant.
Furthermore, it should be noticed that, for the heavy $m_A \gg M_W$ values considered here, it is only in the case of the lightest Higgs boson where generically the two types of next-to-leading corrections of ${\cal O}(M_W^2/m_\text{SUSY}^2)$ and ${\cal O}(m_{H_x}^2/m_\text{SUSY}^2)$ could be comparable in size and therefore, a priori, equally relevant. However, we have found that in the heavy SUSY masses scenario of our interest here, with $m_{\rm SUSY}>$ 1 TeV, these corrections are below ${\cal O}(10^{-13})$, and the maximum rates found for the lightest Higgs decays, allowed by data, are experimentally unreachable, being at most of ${\cal O}(10^{-7})$. Hence, we have focused our interest here on the LFV heavy Higgs bosons decays.

In summary, we have presented in this work a set of simple analytic formulas for the form factors and the associated effective vertices, computed within the MIA, that we think may be  very useful for future phenomenological studies of LFVHD and for their comparison with data. Finally, we have also concluded from our numerical results of the LFVHD rates, presented in contour plots in the $[m_A, \tan\beta]$ and $[m_{\rm SUSY}, \tan\beta]$ planes, that for the most promising case of $\delta^{LL}_{23}$ mixing, one can obtain maximum allowed values (by $\tau \to \mu \gamma$ experimental constraints and MSSM Higgs boson searches at the LHC) of up to BR$(H,A \to \tau \mu) \sim 10^{-3}$ (adding both final state $\tau \bar \mu$ and $\bar \tau \mu$ rates), not far from the present experimental sensitivity accomplished at the LHC. In the case of the lightest MSSM Higgs boson $h$, the rates are much smaller and clearly not reachable at the LHC.

\section*{Acknowledgments}

E.~A. would like to thank the hospitality of the IFLP (CONICET), where the final part of this work was done.
This work is supported by the European Union Grant No. FP7 ITN
INVISIBLES (Marie Curie Actions, Grant No. PITN- GA-2011- 289442), by the CICYT through Grant No. FPA2012-31880,  
by the Spanish Consolider-Ingenio 2010 Programme CPAN (Grant No. CSD2007-00042), 
and by the Spanish MINECO's ``Centro de Excelencia Severo Ochoa'' Programme under Grant No. SEV-2012-0249.
E.~A. is financially supported by the Spanish DGIID-DGA Grant No. 2013-E24/2 and the Spanish MICINN Grants No. FPA2012-35453 and No. CPAN-CSD2007-00042. This work has been also supported by CONICET (R.~M., A.~S.).

\section*{Appendices}

\appendix

\section{Relevant Feynman Rules}
\label{FeynRules}

\begin{figure}[t!]
\begin{center}
\begin{tabular}{c}
\includegraphics[width=0.75\textwidth]{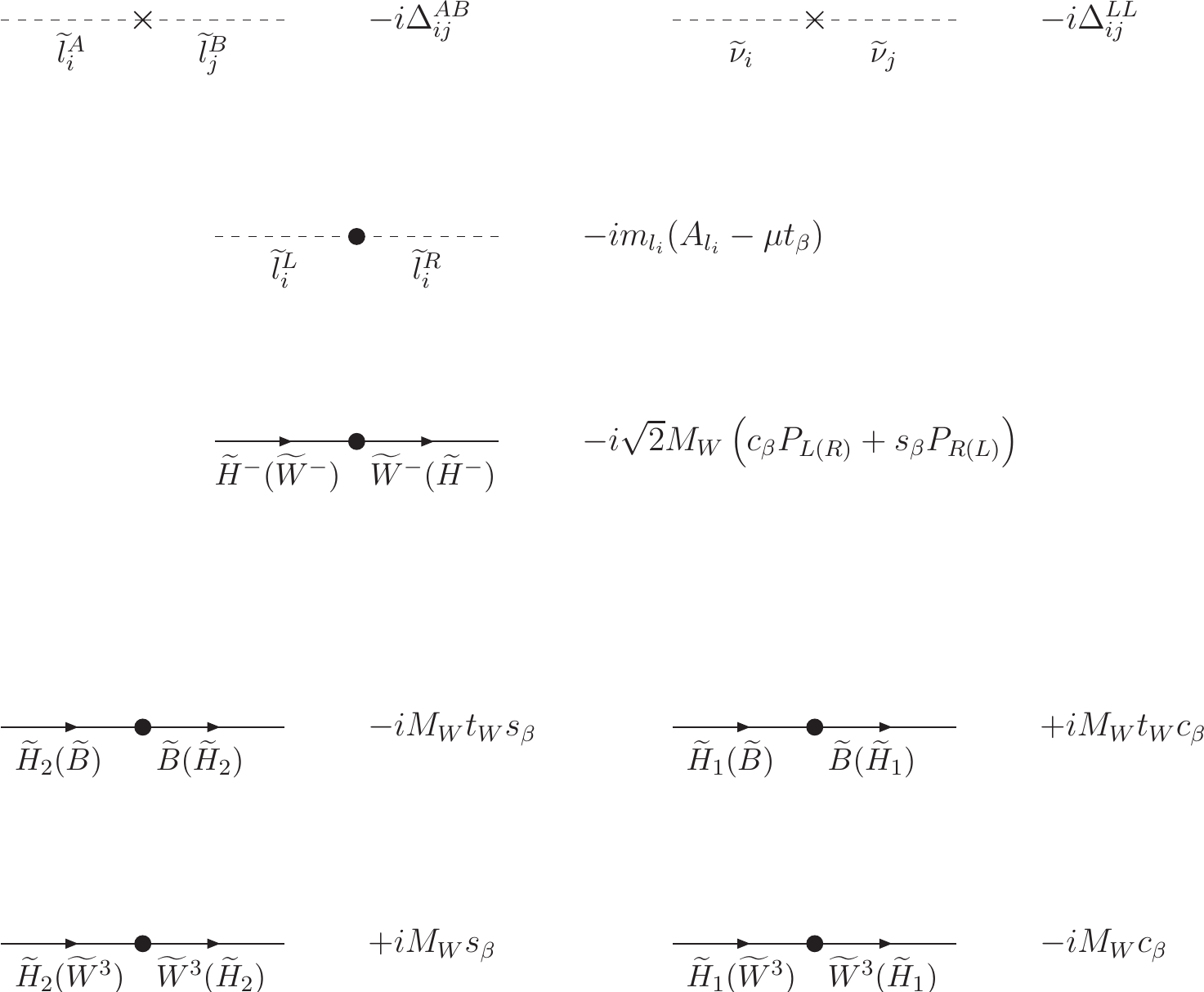}
\end{tabular}
\caption{Feynman rules for the relevant insertions. Insertions changing (non-changing) flavor are denoted by a cross (point).}
\label{LFVinsertions}
\end{center}
\end{figure}

\begin{figure}[t!]
\begin{center}
\begin{tabular}{c}
\includegraphics[width=0.75\textwidth]{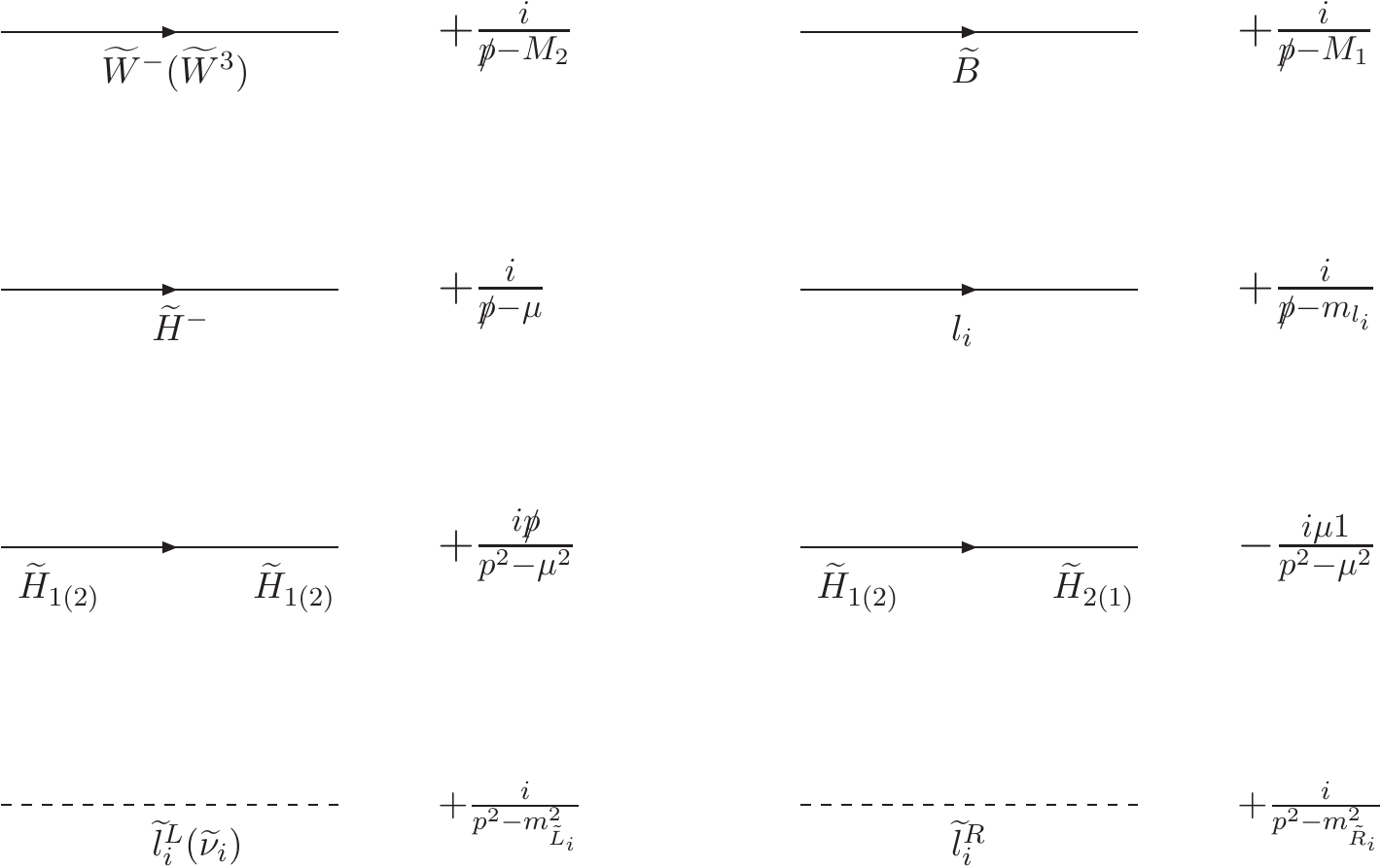}
\end{tabular}
\caption{Feynman rules for the relevant propagators.}
\label{propagators}
\end{center}
\end{figure}

\begin{figure}[t!]
\begin{center}
\begin{tabular}{c}
\includegraphics[width=0.75\textwidth]{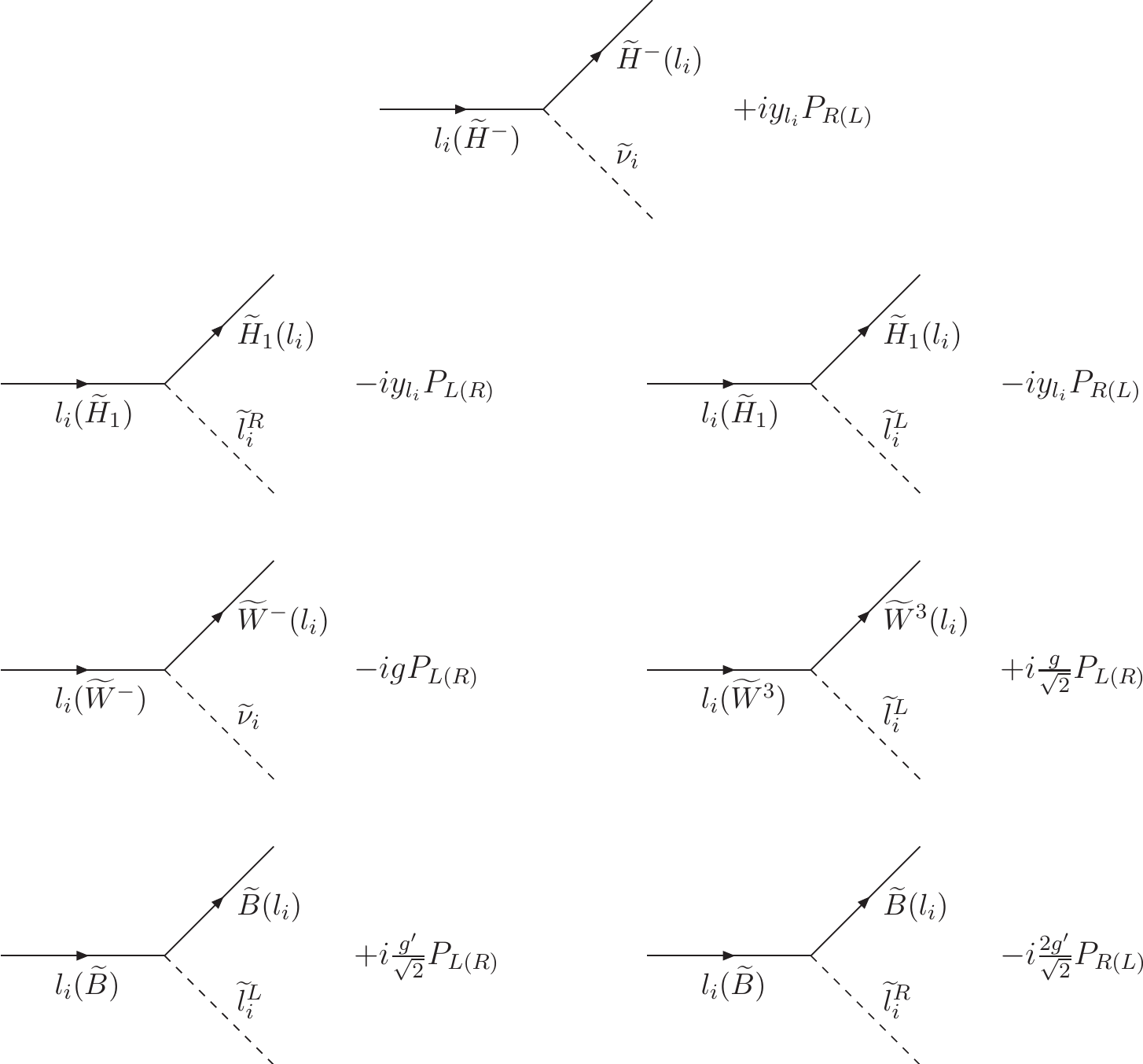}
\end{tabular}
\caption{Feynman rules for the relevant lepton-ino-slepton vertices.}
\label{l-ino-sl-vertices}
\end{center}
\end{figure}

\begin{figure}[t!]
\begin{center}
\begin{tabular}{c}
\includegraphics[width=0.55\textwidth]{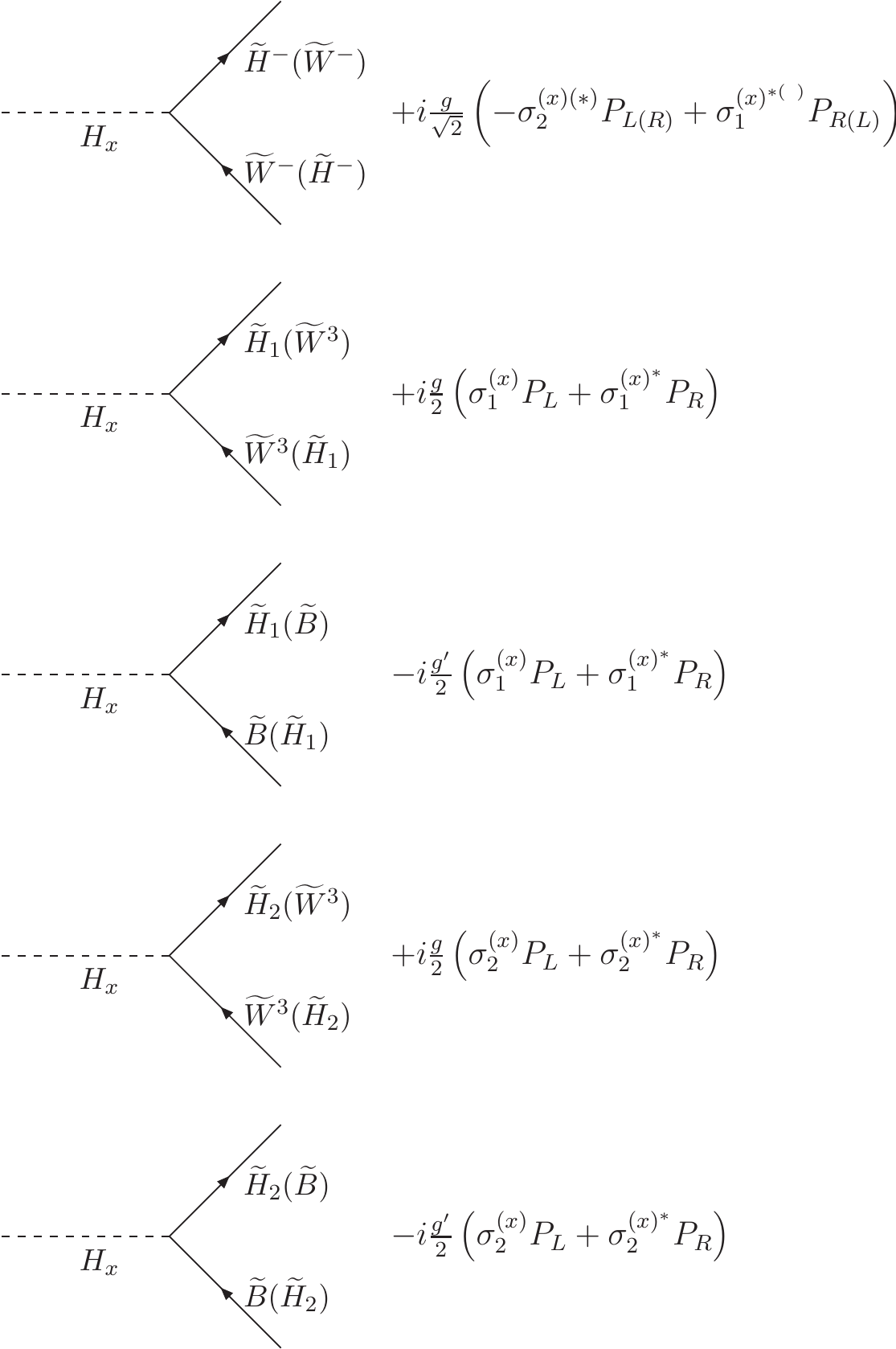}
\end{tabular}
\caption{Feynman rules for the relevant Higgs-ino-ino vertices.}
\label{H-ino-ino-vertices}
\end{center}
\end{figure}

\begin{figure}[t!]
\begin{center}
\begin{tabular}{c}
\includegraphics[width=0.75\textwidth]{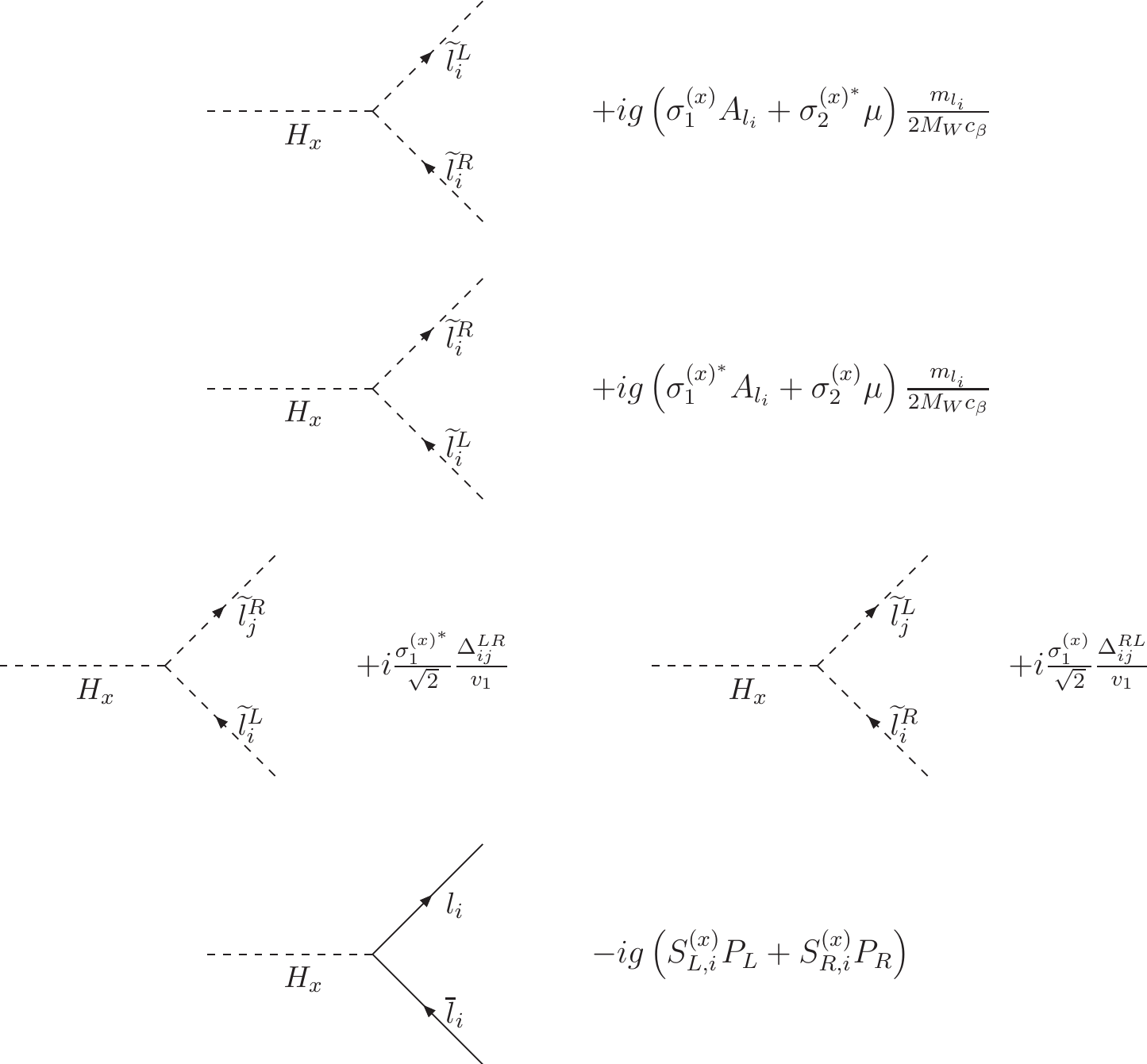}
\end{tabular}
\caption{Feynman rules for the relevant Higgs-slepton-slepton and Higgs-lepton-lepton vertices.}
\label{H-sl-sl-vertices}
\end{center}
\end{figure}

The relevant Feynman rules for the present computation are collected in figures~\ref{LFVinsertions}-\ref{H-sl-sl-vertices}.

The notation and conventions are:
\be
H_x = \left( \begin{array}{c}
h \\
H \\
A \end{array} \right) \,,
\ee
\be
\sigma_{1}^{(x)} = \left( \begin{array}{c}
s_{\alpha} \\
-c_{\alpha} \\
i s_{\beta} \end{array} \right)   \qquad   ,  \qquad
\sigma_{2}^{(x)} = \left( \begin{array}{c}
c_{\alpha} \\
s_{\alpha} \\
-i c_{\beta} \end{array} \right)   \qquad   ,  \qquad
\sigma_{3}^{(x)} = \left( \begin{array}{c}
s_{\alpha+\beta} \\
-c_{\alpha+\beta} \\
0 \end{array} \right) \,,
\label{sigma1y2}
\ee
\be
S_{L,i}^{(x)}=-\frac{m_{l_i}}{2M_{W}c_{\beta}} \sigma_{1}^{(x)^{*}}  \qquad  \ \ , \ \  \qquad   S_{R,i}^{(x)}=S_{L,i}^{(x)^{*}} \,,
\ee
\be 
y_{l_i}=\frac{g m_{l_i}}{\sqrt{2} M_W \cos \beta}=\frac{m_{l_i}}{v_1} \,.
\ee
Here and through the paper we use the short notation: $s_{\alpha}=\sin \alpha$, $c_{\alpha}=\cos \alpha$, $s_{\beta}= \sin \beta$, $c_{\beta}= \cos \beta$,
$t_{\beta}= \tan \beta$, $s_{\alpha+\beta}=\sin(\alpha+\beta)$, $c_{\alpha+\beta}=\cos(\alpha+\beta)$, and $t_W= \tan \theta_W$. $P_{L,R}=(1 \mp \gamma_5)/2$ are the usual $L,R$ projectors. $M_W$ and $M_Z$ are the $W^{\pm}$ and $Z$ gauge boson masses, respectively. $g$ and $g'$ are the gauge coupling constants of $SU(2)_L$ and $U(1)_Y$, respectively. In the propagators, $p$ denotes the flowing momentum and $1$ denotes the identity in spinor space.  


\section{Analytic expressions of the form factors}
\label{AnalyticFormFactors}

Here we present the analytic results of the form factors, $F_{L,R}^{(x)AB}$ with $AB=LL,LR,RL,RR$, in eq.~(\ref{formfactors}),   
from all the diagrams in figures~\ref{diagsMIALL}, \ref{diagsMIALR}, \ref{diagsMIARL}, and \ref{diagsMIARR}.  The contributions from each diagram are explicitly separated (with an obvious notation by a subscript referring to the corresponding diagram) and expressed in terms of the relevant one-loop functions, $C_0$, $C_2$, $D_0$, and ${\tilde D}_0$. These functions are given in Appendix~\ref{LoopIntegrals}.  
\ba
F_{L}^{(x)LL} &=& \frac{g^2}{16 \pi^{2}} \frac{m_{l_k}}{2 M_{W} c_{\beta}}  \left[  
- \left( -\mu M_{2}\sigma_{2}^{(x)}D_{0} + \sigma_{1}^{(x)^{*}}\widetilde{D}_{0} \right)_{(1a)} 
+ \left( \sigma_{1}^{(x)^{*}} (\mu M_{2} t_{\beta}D_{0} + \widetilde{D}_{0}) \right)_{(4a)} \right. \nonumber \\
&&\left. +\frac{t_{W}^{2}}{2} \left( -\mu M_{1}\sigma_{2}^{(x)}D_{0} \right)_{(5a)}
-\frac{1}{2}  \left( -\mu M_{2}\sigma_{2}^{(x)}D_{0} \right)_{(5b)}  \right. \nonumber \\
&&\left. +\frac{t_{W}^{2}}{2} \left( \sigma_{1}^{(x)^{*}}\widetilde{D}_{0} \right)_{(5c)}
-\frac{1}{2}  \left( \sigma_{1}^{(x)^{*}}\widetilde{D}_{0} \right)_{(5d)}  \right. \nonumber \\
&&\left. -t_{W}^{2} \left( M_{1}(\sigma_{1}^{(x)^{*}}A_{l_k}+\sigma_{2}^{(x)}\mu)D_{0} \right)_{(6a)} 
-\frac{t_{W}^{2}}{2}  \left( \sigma_{1}^{(x)^{*}} \mu M_{1} t_{\beta} D_{0} \right)_{(8h)}  \right. \nonumber \\
&&\left. +\frac{1}{2} \left( \sigma_{1}^{(x)^{*}} \mu M_{2} t_{\beta} D_{0} \right)_{(8i)}
-\frac{t_{W}^{2}}{2}  \left( \sigma_{1}^{(x)^{*}} \widetilde{D}_{0} \right)_{(8j)}  \right. \nonumber \\ 
&&\left. +\frac{1}{2} \left( \sigma_{1}^{(x)^{*}} \widetilde{D}_{0} \right)_{(8k)}  
+t_{W}^{2} \left( \sigma_{1}^{(x)^{*}} M_{1}(A_{l_k}-\mu\ t_{\beta}) D_{0} \right)_{(8l)} \right] \,,
\label{FLLL}
\ea

\ba
F_{R}^{(x)LL} &=& \frac{g^2}{16 \pi^{2}} \frac{m_{l_m}}{2 M_{W} c_{\beta}} \left[ 
- \left( -\mu M_{2}\sigma_{2}^{(x)^{*}}D_{0} + \sigma_{1}^{(x)}\widetilde{D}_{0} \right)_{(1b)}
- \left( \sigma_{1}^{(x)} (\mu M_{2} t_{\beta}D_{0} + \widetilde{D}_{0}) \right)_{(3a)}  \right. \nonumber \\
&&\left. +  \left( \sigma_{1}^{(x)} (C_{0}+C_{2}) \right)_{(3b)}
+  \left(  \sigma_{1}^{(x)} (\mu M_{2} t_{\beta}D_{0} + \widetilde{D}_{0}) \right)_{(4a)}  \right. \nonumber \\
&&\left. + \left(  \sigma_{1}^{(x)} (\mu M_{2} t_{\beta}D_{0} + \widetilde{D}_{0}) \right)_{(4b)}
-  \left( \sigma_{1}^{(x)} (C_{0}+C_{2}) \right)_{(4c)}  \right. \nonumber \\
&&\left. +\frac{t_{W}^{2}}{2} \left( -\mu M_{1}\sigma_{2}^{(x)^{*}}D_{0} \right)_{(5e)}
-\frac{1}{2}  \left( -\mu M_{2}\sigma_{2}^{(x)^{*}}D_{0} \right)_{(5f)}  \right. \nonumber \\
&&\left. +\frac{t_{W}^{2}}{2} \left( \sigma_{1}^{(x)} \widetilde{D}_{0} \right)_{(5g)}
-\frac{1}{2}  \left( \sigma_{1}^{(x)} \widetilde{D}_{0} \right)_{(5h)}  \right. \nonumber \\
&&\left. -t_{W}^{2} \left( M_{1}(\sigma_{1}^{(x)}A_{l_m}+\sigma_{2}^{(x)^{*}}\mu)D_{0} \right)_{(6b)}
+\frac{t_{W}^{2}}{2}  \left( \sigma_{1}^{(x)} \mu M_{1} t_{\beta} D_{0} \right)_{(7a)}  \right. \nonumber \\
&&\left. -\frac{1}{2} \left( \sigma_{1}^{(x)} \mu M_{2} t_{\beta} D_{0} \right)_{(7b)}
+\frac{t_{W}^{2}}{2}  \left( \sigma_{1}^{(x)} \widetilde{D}_{0} \right)_{(7c)}  \right. \nonumber \\
&&\left. -\frac{1}{2} \left( \sigma_{1}^{(x)} \widetilde{D}_{0} \right)_{(7d)}
+\frac{t_{W}^{2}}{2} \left( \sigma_{1}^{(x)} (C_{0}+C_{2}) \right)_{(7e)}  \right. \nonumber \\
&&\left. +\frac{1}{2}  \left( \sigma_{1}^{(x)} (C_{0}+C_{2}) \right)_{(7f)}  
-t_{W}^{2}  \left( \sigma_{1}^{(x)} M_{1}(A_{l_k}-\mu\ t_{\beta}) D_{0} \right)_{(7g)}  \right. \nonumber \\
&&\left. -\frac{t_{W}^{2}}{2} \left( \sigma_{1}^{(x)} \mu M_{1} t_{\beta} D_{0} \right)_{(8a)}
+\frac{1}{2}  \left( \sigma_{1}^{(x)} \mu M_{2} t_{\beta} D_{0} \right)_{(8b)}  \right. \nonumber \\
&&\left. -\frac{t_{W}^{2}}{2} \left( \sigma_{1}^{(x)} \widetilde{D}_{0} \right)_{(8c)}
+\frac{1}{2}  \left( \sigma_{1}^{(x)} \widetilde{D}_{0} \right)_{(8d)}  \right. \nonumber \\
&&\left. -\frac{t_{W}^{2}}{2} \left( \sigma_{1}^{(x)} (C_{0}+C_{2}) \right)_{(8e)} 
-\frac{1}{2}  \left( \sigma_{1}^{(x)} (C_{0}+C_{2}) \right)_{(8f)}  \right. \nonumber\\
&&\left. +t_{W}^{2}  \left( M_{1}(A_{l_m}-\mu\ t_{\beta})\sigma_{1}^{(x)} D_{0} \right)_{(8g)}
-\frac{t_{W}^{2}}{2}  \left( \sigma_{1}^{(x)} \mu M_{1} t_{\beta} D_{0} \right)_{(8h)}  \right. \nonumber \\
&&\left. +\frac{1}{2} \left( \sigma_{1}^{(x)} \mu M_{2} t_{\beta} D_{0} \right)_{(8i)}
-\frac{t_{W}^{2}}{2}  \left( \sigma_{1}^{(x)} \widetilde{D}_{0} \right)_{(8j)}  \right. \nonumber \\
&&\left. +\frac{1}{2} \left( \sigma_{1}^{(x)} \widetilde{D}_{0} \right)_{(8k)}
+t_{W}^{2} \left( M_{1}(A_{l_k}-\mu\ t_{\beta})\sigma_{1}^{(x)} D_{0} \right)_{(8l)}  \right] \,,
\label{FRLL}
\ea
 
\be
F_{L}^{(x)LR} = \frac{g^2 t_{W}^{2}}{16 \pi^{2}} \frac{M_{1} \sigma_{1}^{(x)^{*}}}{2 M_{W} c_{\beta}} \left[ 
-\left( C_{0} \right)_{(6c)} 
+\left( C_{0} \right)_{(8m)} \right] \,,
\label{FLLR}
\ee

\be
F_{R}^{(x)LR} = 0 \,,
\label{FRLR}
\ee

\be
F_{L}^{(x)RL} = 0 \,,
\label{FLRL}
\ee

\be
F_{R}^{(x)RL} = \frac{g^2 t_{W}^{2}}{16 \pi^{2}} \frac{M_{1} \sigma_{1}^{(x)}}{2 M_{W} c_{\beta}} 
\left[ 
-\left( C_{0} \right)_{(6d)} 
+\left( C_{0} \right)_{(8n)} 
\right] \,,
\label{FRRL}
\ee

\ba
F_{L}^{(x)RR} &=& \frac{g^2 t_{W}^{2}}{16 \pi^{2}} \frac{m_{l_m}}{2 M_{W} c_{\beta}} \left[ 
 \left( \mu M_{1}\sigma_{2}^{(x)}D_{0} \right)_{(5k)}
-  \left( \sigma_{1}^{(x)^{*}}\widetilde{D}_{0} \right)_{(5l)}  \right. \nonumber \\
&&\left. - \left( M_{1}(\sigma_{1}^{(x)^{*}}A_{l_m}+\sigma_{2}^{(x)}\mu)D_{0} \right)_{(6f)}
-  \left( \sigma_{1}^{(x)^{*}} \mu M_{1} t_{\beta} D_{0}  \right)_{(7h)}  \right. \nonumber \\
&&\left. - \left( \sigma_{1}^{(x)^{*}} \widetilde{D}_{0} \right)_{(7i)}
+ \left(  2\sigma_{1}^{(x)^{*}} (C_{0}+C_{2}) \right)_{(7j)}  \right. \nonumber \\
&&\left. -  \left( \sigma_{1}^{(x)^{*}} M_{1}(A_{l_k}-\mu\ t_{\beta}) D_{0} \right)_{(7k)}
+  \left( \sigma_{1}^{(x)^{*}} \mu M_{1} t_{\beta} D_{0} \right)_{(8o)}  \right. \nonumber \\
&&\left. + \left( \sigma_{1}^{(x)^{*}} \widetilde{D}_{0} \right)_{(8p)}
+ \left( \sigma_{1}^{(x)^{*}} \mu M_{1} t_{\beta} D_{0} \right)_{(8q)}  \right. \nonumber \\
&&\left. +  \left( \sigma_{1}^{(x)^{*}} \widetilde{D}_{0} \right)_{(8r)}
-  \left(  2 \sigma_{1}^{(x)^{*}} (C_{0}+C_{2}) \right)_{(8s)}  \right. \nonumber \\
&&\left. +  \left( \sigma_{1}^{(x)^{*}} M_{1}(A_{l_k}-\mu\ t_{\beta}) D_{0} \right)_{(8t)}
+ \left( \sigma_{1}^{(x)^{*}} M_{1}(A_{l_m}-\mu\ t_{\beta})  D_{0} \right)_{(8u)} \right] \,,
\label{FLRR}
\ea
 
 \ba
F_{R}^{(x)RR} &=& \frac{g^2 t_{W}^{2}}{16 \pi^{2}} \frac{m_{l_k}}{2 M_{W} c_{\beta}} \left[ 
 \left( \mu M_{1}\sigma_{2}^{(x)^{*}}D_{0} \right)_{(5i)}
-  \left( \sigma_{1}^{(x)} \widetilde{D}_{0} \right)_{(5j)}  \right. \nonumber \\
&&\left. - \left( M_{1}(\sigma_{1}^{(x)}A_{l_k}+ \sigma_{2}^{(x)^{*}}\mu)D_{0} \right)_{(6e)}
+  \left( \sigma_{1}^{(x)} \mu M_{1} t_{\beta} D_{0}  \right)_{(8o)}  \right. \nonumber \\
&&\left. + \left( \sigma_{1}^{(x)} \widetilde{D}_{0} \right)_{(8p)}
+  \left( \sigma_{1}^{(x)} M_{1}(A_{l_k}-\mu\ t_{\beta}) D_{0} \right)_{(8t)} \right] \,.
\label{FRRR}
\ea 

The arguments of the above loop integrals are the following: 
\ba
& D_{0}, \widetilde{D}_{0} = D_{0}, \widetilde{D}_{0} (0,p_{2},p_{1},m_{\widetilde{L}_{m}},m_{\widetilde{L}_{k}},\mu,M_{2}) & \text{in } (1a) \nonumber \\
& D_{0}, \widetilde{D}_{0} = D_{0}, \widetilde{D}_{0} (0,p_{2},p_{1},m_{\widetilde{L}_{m}},m_{\widetilde{L}_{k}},M_{2},\mu) & \text{in } (1b) \nonumber \\
& D_{0}, \widetilde{D}_{0} = D_{0}, \widetilde{D}_{0} (0,p_{2},0,m_{\widetilde{L}_{k}},m_{\widetilde{L}_{m}},\mu,M_{2}) & \text{in } (3a) \nonumber \\
& \hspace{-7mm} C_{0,2} = C_{0,2}(0,p_{2},m_{\widetilde{L}_{k}},m_{\widetilde{L}_{m}},M_{2})   & \text{in } (3b) \nonumber \\
& D_{0}, \widetilde{D}_{0} = D_{0}, \widetilde{D}_{0} (0,p_{3},0,m_{\widetilde{L}_{m}},m_{\widetilde{L}_{k}},\mu,M_{2}) & \text{in } (4a),(4b) \nonumber \\
& \hspace{-7mm} C_{0,2} = C_{0,2}(0,p_{3},m_{\widetilde{L}_{m}},m_{\widetilde{L}_{k}},M_{2})   & \text{in } (4c) \nonumber \\
& \ D_{0}, \widetilde{D}_{0} = D_{0}, \widetilde{D}_{0} (0,p_{2},p_{1},m_{\widetilde{L}_{m}},m_{\widetilde{L}_{k}},\mu,M_{1}) & \text{in } (5a),(5c) \nonumber \\
& \ D_{0}, \widetilde{D}_{0} = D_{0}, \widetilde{D}_{0} (0,p_{2},p_{1},m_{\widetilde{L}_{m}},m_{\widetilde{L}_{k}},\mu,M_{2}) & \text{in } (5b),(5d) \nonumber \\
& \ D_{0}, \widetilde{D}_{0} = D_{0}, \widetilde{D}_{0} (0,p_{2},p_{1},m_{\widetilde{L}_{m}},m_{\widetilde{L}_{k}},M_{1},\mu) & \text{in } (5e),(5g) \nonumber \\
& \ D_{0}, \widetilde{D}_{0} = D_{0}, \widetilde{D}_{0} (0,p_{2},p_{1},m_{\widetilde{L}_{m}},m_{\widetilde{L}_{k}},M_{2},\mu) & \text{in } (5f),(5h) \nonumber \\
& D_{0}, \widetilde{D}_{0} = D_{0}, \widetilde{D}_{0} (0,p_{2},p_{1},m_{\widetilde{R}_{m}},m_{\widetilde{R}_{k}},\mu,M_{1}) & \text{in } (5i), (5j) \nonumber \\
& D_{0}, \widetilde{D}_{0} = D_{0}, \widetilde{D}_{0} (0,p_{2},p_{1},m_{\widetilde{R}_{k}},m_{\widetilde{R}_{m}},M_{1},\mu) & \text{in } (5k), (5l) \nonumber \\
& \ \ \ \ \ D_{0} = D_{0} (p_{2},p_{1},0,M_{1},m_{\widetilde{R}_{k}},m_{\widetilde{L}_{k}},m_{\widetilde{L}_{m}})   & \text{in } (6a) \nonumber \\
& \ \ \ \ \ D_{0} = D_{0} (p_{2},0,p_{1},M_{1},m_{\widetilde{L}_{k}},m_{\widetilde{L}_{m}},m_{\widetilde{R}_{m}})   & \text{in } (6b) \nonumber \\
& \hspace{-7mm} C_{0} = C_{0}(p_{2},p_{1},M_{1},m_{\widetilde{R}_{k}},m_{\widetilde{L}_{m}})   & \text{in } (6c) \nonumber \\
& \hspace{-7mm} C_{0} = C_{0}(p_{2},p_{1},M_{1},m_{\widetilde{L}_{k}},m_{\widetilde{R}_{m}})   & \text{in } (6d) \nonumber \\
& \ \ \ \ \ D_{0} = D_{0} (p_{2},p_{1},0,M_{1},m_{\widetilde{L}_{k}},m_{\widetilde{R}_{k}},m_{\widetilde{R}_{m}})   & \text{in } (6e) \nonumber \\
& \ \ \ \ \ \ D_{0} = D_{0} (p_{2},0,p_{1},M_{1},m_{\widetilde{R}_{k}},m_{\widetilde{R}_{m}},m_{\widetilde{L}_{m}})   & \text{in } (6f) \nonumber \\
& D_{0}, \widetilde{D}_{0} = D_{0}, \widetilde{D}_{0} (0,p_{2},0,m_{\widetilde{L}_{k}},m_{\widetilde{L}_{m}},\mu,M_{1}) & \text{in } (7a),(7c) \nonumber \\
& D_{0}, \widetilde{D}_{0} = D_{0}, \widetilde{D}_{0} (0,p_{2},0,m_{\widetilde{L}_{k}},m_{\widetilde{L}_{m}},\mu,M_{2}) & \text{in } (7b),(7d) \nonumber \\
& \hspace{-7mm} C_{0,2} = C_{0,2}(0,p_{2},m_{\widetilde{L}_{k}},m_{\widetilde{L}_{m}},M_{1})   & \text{in } (7e) \nonumber \\
& \hspace{-7mm} C_{0,2} = C_{0,2}(0,p_{2},m_{\widetilde{L}_{k}},m_{\widetilde{L}_{m}},M_{2})   & \text{in } (7f) \nonumber \\
& \ \ \ \ D_{0} = D_{0} (0,0,p_{2},m_{\widetilde{R}_{k}},m_{\widetilde{L}_{k}},m_{\widetilde{L}_{m}},M_{1})   & \text{in } (7g) \nonumber \\
& D_{0}, \widetilde{D}_{0} = D_{0}, \widetilde{D}_{0} (0,p_{2},0,m_{\widetilde{R}_{k}},m_{\widetilde{R}_{m}},\mu,M_{1}) & \text{in } (7h), (7i) \nonumber \\
& \hspace{-7mm} C_{0,2} = C_{0,2} (0,p_{2},m_{\widetilde{R}_{k}},m_{\widetilde{R}_{m}},M_{1})   & \text{in } (7j) \nonumber \\
& \ \ \ \ D_{0} = D_{0} (0,0,p_{2},m_{\widetilde{L}_{k}},m_{\widetilde{R}_{k}},m_{\widetilde{R}_{m}},M_{1})   & \text{in } (7k) \nonumber \\
& D_{0}, \widetilde{D}_{0} = D_{0}, \widetilde{D}_{0} (0,p_{3},0,m_{\widetilde{L}_{m}},m_{\widetilde{L}_{k}},M_{1},\mu) & \text{in } (8a),(8c) \nonumber \\
& D_{0}, \widetilde{D}_{0} = D_{0}, \widetilde{D}_{0} (0,p_{3},0,m_{\widetilde{L}_{m}},m_{\widetilde{L}_{k}},M_{2},\mu) & \text{in } (8b),(8d) \nonumber \\
& \hspace{-7mm} C_{0,2} = C_{0,2}(0,p_{3},m_{\widetilde{L}_{k}},m_{\widetilde{L}_{m}},M_{1})   & \text{in } (8e) \nonumber \\
& \hspace{-7mm} C_{0,2} = C_{0,2}(0,p_{3},m_{\widetilde{L}_{k}},m_{\widetilde{L}_{m}},M_{2})   & \text{in } (8f) \nonumber \\
& \ \ \ \ \ D_{0} = D_{0} (0,0,p_{3},m_{\widetilde{R}_{m}},m_{\widetilde{L}_{m}},m_{\widetilde{L}_{k}},M_{1})   & \text{in } (8g) \nonumber \\    
& D_{0}, \widetilde{D}_{0} = D_{0}, \widetilde{D}_{0} (0,p_{3},0,m_{\widetilde{L}_{m}},m_{\widetilde{L}_{k}},\mu,M_{1}) & \text{in } (8h),(8j) \nonumber \\
& D_{0}, \widetilde{D}_{0} = D_{0}, \widetilde{D}_{0} (0,p_{3},0,m_{\widetilde{L}_{m}},m_{\widetilde{L}_{k}},\mu,M_{2}) & \text{in } (8i),(8k) \nonumber \\ 
& \ \ \ \ D_{0} = D_{0} (0,0,p_{3},m_{\widetilde{L}_{m}},m_{\widetilde{L}_{k}},m_{\widetilde{R}_{k}},M_{1})   & \text{in } (8l) \nonumber \\
& \hspace{-9mm} C_{0} = C_{0}(p_{3},0,M_{1},m_{\widetilde{L}_{m}},m_{\widetilde{R}_{k}})   & \text{in } (8m) \nonumber \\
& \hspace{-9mm} C_{0} = C_{0}(p_{3},0,M_{1},m_{\widetilde{R}_{m}},m_{\widetilde{L}_{k}})   & \text{in } (8n) \nonumber \\
& D_{0}, \widetilde{D}_{0} = D_{0}, \widetilde{D}_{0} (0,p_{3},0,m_{\widetilde{R}_{m}},m_{\widetilde{R}_{k}},\mu,M_{1}) & \text{in } (8o), (8p) \nonumber \\
& D_{0}, \widetilde{D}_{0} = D_{0}, \widetilde{D}_{0} (0,p_{3},0,m_{\widetilde{R}_{k}},m_{\widetilde{R}_{m}},M_{1},\mu) & \text{in } (8q), (8r) \nonumber \\
& \hspace{-7mm} C_{0,2} = C_{0,2} (0,p_{3},m_{\widetilde{R}_{k}},m_{\widetilde{R}_{m}},M_{1})   & \text{in } (8s) \nonumber \\
& \ \ \ \ D_{0} = D_{0} (0,0,p_{3},m_{\widetilde{R}_{m}},m_{\widetilde{R}_{k}},m_{\widetilde{L}_{k}},M_{1})   & \text{in } (8t) \nonumber \\
& \ \ \ \ \ D_{0} = D_{0} (0,0,p_{3},m_{\widetilde{R}_{k}},m_{\widetilde{R}_{m}},m_{\widetilde{L}_{m}},M_{1})   & \text{in } (8u) \nonumber 
\ea
For the particular case of the {\it Equal masses} scenario,  the analytic results of the form factors are considerably simplified. We include here the results for the ${\hat F}_{L,R}^{(x)AB}$ of eq.~(\ref{hatFF}), specifying the contributions from each diagram: 
 \ba
{\hat F}_{L}^{(x)LL}&=&\frac{g^2}{16 \pi^{2}}  \frac{m_{\tau}}{2 M_{W} c_{\beta}} \left[
 \sigma_{1}^{(x)^{*}} \left(\left(\frac{1}{3}+\frac{m_{H_{x}}^{2}}{40 m_{S}^{2}}\right)_{(1)}+
 \left(\frac{1}{6}t_\beta-\frac{1}{3} \right)_{(4)} + \frac{(1-t_W^2)}{2} \left(\frac{1}{3}+\frac{m_{H_{x}}^{2}}{40 m_{S}^{2}}\right)_{(5)} \right. \right.
 \nonumber  \\  
 &&\left. \left. \ \   -t_W^2 \left(\frac{1}{6}+\frac{m_{H_{x}}^{2}}{30 m_{S}^{2}}\right)_{(6)}  +
 \left(\frac{1}{12}t_\beta-\frac{1}{6}-\frac{1}{4} t_W^2 t_\beta +\frac{1}{3} t_W^2 \right)_{(8)}  \right) \right.
  \nonumber  \\  
 &&\left. \ \  + \sigma_{2}^{(x)} \left( \left(\frac{1}{6}+\frac{m_{H_{x}}^{2}}{60 m_{S}^{2}}\right)_{(1)}  + \frac{(1-t_W^2)}{2} \left(\frac{1}{6}+
 \frac{m_{H_{x}}^{2}}{60 m_{S}^{2}}\right)_{(5)} 
   -t_W^2\left(\frac{1}{6}+\frac{m_{H_{x}}^{2}}{30 m_{S}^{2}}\right)_{(6)} 
 \right)
 \right] \,,
\ea
 
 \ba
 {\hat F}_{R}^{(x)LL}&=&\frac{g^2}{16 \pi^{2}}  \frac{m_{\mu}}{2 M_{W} c_{\beta}}
 \left[
 \sigma_{1}^{(x)} \left(
 \left(\frac{1}{3}+\frac{m_{H_{x}}^{2}}{40 m_{S}^{2}}\right)_{(1)}
 -\left(\frac{1}{6} t_\beta \right)_{(3)}+
 \left(\frac{1}{3} t_\beta-\frac{1}{3} \right)_{(4)}  \right. \right. \nonumber \\
 &&\left.  \left. \ \    
 +\frac{(1-t_W^2)}{2} \left(\frac{1}{3}+\frac{m_{H_{x}}^{2}}{40 m_{S}^{2}}\right)_{(5)}
 -t_W^2 \left(\frac{1}{6}+\frac{m_{H_{x}}^{2}}{30 m_{S}^{2}}\right)_{(6)}  +
  \left(\frac{-1}{12}t_\beta+\frac{1}{4} t_W^2 t_\beta +\frac{-1}{2} t_W^2 \right)_{(7)} 
\right.  \right. \nonumber \\
 && \left. \left.\ \   
+
 \left(\frac{1}{6}t_\beta+\frac{-1}{6}-\frac{1}{2} t_W^2 t_\beta +\frac{5}{6} t_W^2 \right)_{(8)}  \right) 
  + \sigma_{2}^{(x)^{*}} \left(  
  \left(\frac{1}{6}+\frac{m_{H_{x}}^{2}}{60 m_{S}^{2}}\right)_{(1)} \right. \right. \nonumber \\
 &&  \left. \left. \ \   
  + \frac{(1-t_W^2)}{2} \left(\frac{1}{6}+
 \frac{m_{H_{x}}^{2}}{60 m_{S}^{2}}\right)_{(5)}
 -t_W^2\left(\frac{1}{6}+\frac{m_{H_{x}}^{2}}{30 m_{S}^{2}}\right)_{(6)} 
 \right)
 \right] \,,
\ea
 
\ba
{\hat F}_{L}^{(x)LR}&=&\frac{g}{16 \pi^{2}} \frac{1}{\sqrt{2}} t_W^2 \sigma_{1}^{(x)^{*}} \left(\left(\frac{1}{2}+\frac{m_{H_{x}}^{2}}{24 m_{S}^{2}}\right)_{(6)}
 -\left( \frac{1}{2} \right)_{(8)} \right) \,,
\ea
\ba
{\hat F}_{R}^{(x)RL}& =&  {+}{\hat F}_{L}^{(x)LR*}  \ \ ; \ \ \   {\hat F}_{R}^{(x)LR}={\hat F}_{L}^{(x)RL}=0 \,,
\ea
\ba
 {\hat F}_{L}^{(x)RR}&=&\frac{g^2}{16 \pi^{2}}  t_W^2\frac{m_{\mu}}{2 M_{W} c_{\beta}}
 \left[
 \sigma_{1}^{(x)^{*}} \left(
 \left(\frac{1}{3}+\frac{m_{H_{x}}^{2}}{40 m_{S}^{2}}\right)_{(5)}-
 \left(\frac{1}{6}+\frac{m_{H_{x}}^{2}}{30 m_{S}^{2}}\right)_{(6)} -\left( \frac{1}{2} \right)_{(7)} + \left( \frac{1}{3} \right)_{(8)} \right)
\right.  \nonumber \\
  && \left. \ \  + \sigma_{2}^{(x)} \left( \left(\frac{1}{6}+\frac{m_{H_{x}}^{2}}{60 m_{S}^{2}}\right)_{(5)}  - \left(\frac{1}{6}+
 \frac{m_{H_{x}}^{2}}{30 m_{S}^{2}}\right)_{(6)} 
 \right)
 \right] \,,
\ea
\ba
{\hat F}_{R}^{(x)RR}&=&\frac{g^2}{16 \pi^{2}}  t_W^2\frac{m_{\tau}}{2 M_{W} c_{\beta}}
 \left[
 \sigma_{1}^{(x)} \left(
 \left(\frac{1}{3}+\frac{m_{H_{x}}^{2}}{40 m_{S}^{2}}\right)_{(5)}-
 \left(\frac{1}{6}+\frac{m_{H_{x}}^{2}}{30 m_{S}^{2}}\right)_{(6)} - \left( \frac{1}{6} \right)_{(8)} \right) \right.
  \nonumber \\
  && \left. \ \  + \sigma_{2}^{(x)^{*}} \left( \left(\frac{1}{6}+\frac{m_{H_{x}}^{2}}{60 m_{S}^{2}}\right)_{(5)}  - \left(\frac{1}{6}+
 \frac{m_{H_{x}}^{2}}{30 m_{S}^{2}}\right)_{(6)} 
 \right)
 \right] \,.
\ea
Finally, we have also computed for this {\it Equal masses} scenario the subleading decoupling contributions of ${\cal O}(M_W^2/m_{\rm SUSY}^2)$ to the specific form factor 
${\hat F}_{R}^{(x)RR}$, where we have detected that there are strong cancellations among diagrams and these contributions play a relevant role in obtaining a better convergence between the MIA and the full results.
The main contributions at this order come from diagrams with two extra gaugino-Higgsino insertions in the internal fermion propagators of diagrams $5i,5j,6e,8o,8p,8t$; or one extra insertion gaugino-Higgsino and one extra of type $\tilde{l}_{k}^{L}-\tilde{l}_{k}^{R}$ in diagrams $5i,5j$; or only one extra insertion of type $\tilde{l}_{k}^{L}-\tilde{l}_{k}^{R}$ in diagram $6e$; or considering a new ``type $6$ like'' diagram -pure Bino exchange- with vertex $H_{x}-\tilde{l}_{k(m)}^{R}-\tilde{l}_{k(m)}^{R}$ (no chirality flip). 
 After this computation we have found that to include these new ${\cal O}(M_W^2/m_{\rm SUSY}^2)$ contributions into this $RR$ form factor one should replace $({\hat F}_{R}^{(x)RR})$ by $({\hat F}_{R}^{(x)RR}+{\tilde F}_{R}^{(x)RR})$, where:
 \ba
\tilde{F}_{R}^{(x)RR} &=& \frac{g^2 t_{W}^{2}}{16\pi^{2}} \frac{m_{\tau}}{2 M_{W} c_{\beta}} \frac{M_{W}^{2}}{m_S^{2}} \frac{t_{\beta}^{2}}{1+t_{\beta}^{2}} \left[ \left( \frac{\sigma_{1}^{(x)}}{60} \left( 3t_{W}^{2} + 13  - 4t_{W}^{2}t_{\beta} -12t_{\beta} \right) \right. \right. \nonumber\\
&& \left. \left. -\frac{\sigma_{1}^{(x)^{*}}}{5} -\frac{4\sigma_{2}^{(x)}}{15} -\frac{2\sigma_{2}^{(x)^{*}}}{15} +\frac{\sigma_{3}^{(x)}\sqrt{1+t_{\beta}^{2}}}{12t_{\beta}}\left( 1+t_{W}^{2} \right) \right) \right. \nonumber\\
&& \left. +\left( \frac{1+t_{W}^{2}}{60 t_{\beta}} \left( -8\sigma_{1}^{(x)} +4\sigma_{1}^{(x)^{*}} +\sigma_{2}^{(x)} +\sigma_{2}^{(x)^{*}} \right) +\frac{\sigma_{3}^{(x)}\sqrt{1+t_{\beta}^{2}}}{12t_{\beta}^{2}}\left( -1+5t_{W}^{2} \right) \right) \right. \nonumber\\
&& \left. +\left( \frac{1+t_{W}^{2}}{30 t_{\beta}^{2}} \left( -\sigma_{1}^{(x)} +\sigma_{1}^{(x)^{*}} +\sigma_{2}^{(x)} -\sigma_{2}^{(x)^{*}} \right) \right) \right] \,.
\label{FRRReqmassesMW2}
\ea
In the large $\tan\beta$ limit we obtain that this correction in eq.~(\ref{FRRReqmassesMW2}) grows linearly with $\tan\beta$ for $h$ and quadratically for $H$ and $A$. More specifically, we get for the heavy Higgs boson $H$ (and similarly for $A$):
\ba
 \tilde{F}_{R}^{(H)RR}\vert_{t_{\beta} \gg 1} &=& \frac{g^2 t_{W}^{2}}{16\pi^{2}} \frac{m_{\tau}}{2 M_{W}} \frac{M_{W}^{2}}{m_S^{2}} \frac{ 3 + t_{W}^{2}}{15} t_{\beta}^{2} \,.
\label{FRRReqmassesMW2largetanb}
\ea
 

\section{Relevant loop integrals and their expansions for heavy SUSY}
\label{LoopIntegrals}

The loop integrals that are relevant for the present computation are the following:
\ba
&\frac{i}{16\pi^{2}} C_{0},C^{\mu}(q_{1},q_{2},m_{1},m_{2},m_{3})= \nonumber \\
&\int  d \tilde k  \frac{1,k^{\mu}}{(k^{2}-m_{1}^{2})((k+q_{1})^{2}-m_{2}^{2})((k+q_{1}+q_{2})^{2}-m_{3}^{2})} \,, 
\ea 
and 
\ba
&\frac{i}{16 \pi^2} D_{0},\tilde D_{0} (q_{1},q_{2},q_{3},m_{1},m_{2},m_{3},m_{4})= \nonumber \\
&\int d \tilde k  \frac{1,k^2}{(k^{2}-m_{1}^{2})((k+q_{1})^{2}-m_{2}^{2})((k+q_{1}+q_{2})^{2}-m_{3}^{2})((k+q_{1}+q_{2}+q_{3})^{2}-m_{4}^{2})} \,, 
\ea
where 
\be
d \tilde k \equiv \frac{\mu_{0}^{4-D}d^{D}k}{(2\pi)^{D}} \,,
\ee  
and 
\be
C^{\mu}(q_{1},q_{2},m_{1},m_{2},m_{3})=\sum_{i=1}^{2} q_{i}^{\mu}C_{i}(q_{1},q_{2},m_{1},m_{2},m_{3}) \,.
\ee
The particular values of the relevant loop functions for zero external momenta are the following:
\be
C_{0}(0,0,m_{1},m_{2},m_{3})=\frac{a (b-c) \text{log}(a)+b (c-a) \text{log}(b)+c (a-b) \text{log}(c)}{(a-b) (a-c) (c-b)} \,,
\ee

\be
C_{2}(0,0,m_{1},m_{2},m_{3})= \nonumber 
\ee
\ba
\frac{a^2 \log (a) (b-c)^2-b^2 (a-c)^2 \log (b)+c (a-b) ((a-c) (b-c)+\log (c) (2 a b-c (a+b)))}{2 (a-b) (a-c)^2 (b-c)^2} \,, &&
\ea
where $a=m_{1}^{2}$, $b=m_{2}^{2}$, and $c=m_{3}^{2}$.
\ba
D_{0}(0,0,0,m_{1},m_{2},m_{3},m_{4})&=& \frac{1}{(a-b) (a-c) (b-c)} \left[ \frac{(-b+c) (-a+d+a \text{log}(a)-d \text{log}(d))}{a-d} \right. \nonumber \\
&&+\frac{(a-c) (-b+d+b \text{log}(b)-d \text{log}(d))}{b-d}   \nonumber \\
&& \left. +\frac{(-a+b) (-c+d+c \text{log}(c)-d \text{log}(d))}{c-d} \right] \,,
\label{D0zeromom}
\ea
where $a=m_{1}^{2}$, $b=m_{2}^{2}$, $c=m_{3}^{2}$, and $d=m_{4}^{2}$.
The $\tilde{D}_{0}$ function can be derived from $C_0$ and $D_0$ by:
\be
\tilde{D}_{0}(0,0,0,m_{1},m_{2},m_{3},m_{4}) = C_{0}(0,0,m_{2},m_{3},m_{4}) + m_{1}^{2}D_{0}(0,0,0,m_{1},m_{2},m_{3},m_{4}) \,.
\ee

At non-zero external momenta all these integrals can be Taylor expanded for heavy internal particle masses as compared to the external momenta, $m_i^2 \gg q_j^2$, and expressed generically as their values at zero external momenta plus corrections given by functions with extra powers of the small ${\cal O}(q_j^2/m_i^2)$ quantities.

For instance, by keeping just the ${\cal O}(p_1^2/m_i^2)$ corrections 
in $C_{0}(p_{2},p_{1},m_{1},m_{2},m_{3})$ we get:
\ba
C_{0}(p_{2},p_{1},m_{1},m_{2},m_{3})&=& C_{0}(0,0,m_{1},m_{2},m_{3}) \nonumber \\
&& + \frac{d}{2 (a-b)^2 (a-c)^2 (c-b)^3} [(a-b) (a-c) (b-c) (-2 b c+a (b+c)) \nonumber \\
&& \ \ \ \ \ \ \ \ \ \ \ \ -a^2 (b-c)^3 \text{log}(a) \nonumber \\
&& \ \ \ \ \ \ \ \ \ \ \ \ + b (a-c)^2 (-2 a c+b (b+c)) \text{log}(b) \nonumber \\
&& \ \ \ \ \ \ \ \ \ \ \ \ + (a-b)^2 c (2 a b-c (b+c)) \text{log}(c)] \,,
\ea
with $a=m_{1}^{2}$, $b=m_{2}^{2}$, $c=m_{3}^{2}$, and $d=p_1^2$.
And similarly for other loop functions. 

For the present computation we have computed all the relevant Taylor expansions including the ${\cal O}(p_1^2)$ corrections with $p_1^2=m_{H_x}^2$, for $C_{0,2}$, $D_0$, and ${\tilde D}_0$, but we omit to show them here for shortness.
Here we include instead just the simplest case, for illustrative purposes, that corresponds to taking all the involved SUSY masses to be equal, the so-called {\it Equal masses} scenario, keeping just the dominant and the leading subdominant contributions in the previously commented Taylor expansions. In this case, we get the following simple formulas: 
\ba
&& C_{0}(0,p_{2},m_S,m_S,m_S)\approx C_{0}(0,p_{3},m_S,m_S,m_S)\approx -\frac{1}{2 m_S^{2}}  \,,  \nonumber \\
&& C_{0}(p_{2},0,m_S,m_S,m_S)\approx C_{0}(p_{3},0,m_S,m_S,m_S)\approx -\frac{1}{2 m_S^{2}}  \,,  \nonumber \\
&& C_{2}(0,p_{2},m_S,m_S,m_S)\approx C_{2}(0,p_{3},m_S,m_S,m_S)\approx \frac{1}{6 m_S^{2}}  \,,  \nonumber \\
&& C_{0}(p_{2},p_{1},m_S,m_S,m_S)\approx -\frac{1}{2 m_S^{2}}-\frac{m_{H_{x}}^{2}}{24 m_S^{4}} \,, \nonumber \\
&& D_{0}(0,p_{2},0,m_S,m_S,m_S,m_S)\approx D_{0}(0,p_{3},0,m_S,m_S,m_S,m_S) \approx \frac{1}{6 m_S^{4}}  \,, \nonumber \\
&& \tilde{D}_{0}(0,p_{2},0,m_S,m_S,m_S,m_S)\approx \tilde{D}_{0}(0,p_{3},0,m_S,m_S,m_S,m_S)\approx -\frac{1}{3 m_S^{2}} \,,  \nonumber \\
&& D_{0}(0,0,p_{2},m_S,m_S,m_S,m_S)\approx D_{0}(0,0,p_{3},m_S,m_S,m_S,m_S)\approx \frac{1}{6 m_S^{4}}  \,, \nonumber \\
&& D_{0}(0,p_{2},p_{1},m_{\rm S},m_S,m_S,m_S)\approx \frac{1}{6 m_S^{4}}+\frac{m_{H_{x}}^{2}}{60 m_S^{6}} \,,  \nonumber \\
&& \tilde{D}_{0}(0,p_{2},p_{1},m_S,m_S,m_S,m_S)\approx -\frac{1}{3 m_S^{2}}-\frac{m_{H_{x}}^{2}}{40 m_S^{4}} \,,  \nonumber \\
&& D_{0}(p_{2},p_{1},0,m_S,m_S,m_S,m_S)\approx \frac{1}{6 m_S^{4}}+\frac{m_{H_{x}}^{2}}{30 m_S^{6}} \,, \nonumber \\
&& D_{0}(p_{2},0,p_{1},m_S,m_S,m_S,m_S)\approx \frac{1}{6 m_S^{4}}+\frac{m_{H_{x}}^{2}}{30 m_S^{6}} \,.
\label{eqmass}
\ea


\bibliographystyle{unsrt}

\end{document}